\documentclass[aps,prd,twocolumn,amsfonts,amsmath,amssymb,floatfix,superscriptaddress]{revtex4-2}
\usepackage{url}
\usepackage{bm}		% bold math
\usepackage{graphicx}	% figures
\usepackage{cancel}	% cancel
\usepackage{mathrsfs}
\usepackage[pdftex,pdfpagemode={UseOutlines},bookmarks,bookmarksopen,colorlinks,linkcolor={blue},citecolor={green},urlcolor={red}]{hyperref}
\usepackage{xcolor}
\definecolor{darkgreen}{hsb}{.333,1,.5}
\definecolor{darkblue}{hsb}{.667,1,.75}
\usepackage{hyperref}
\hypersetup{colorlinks=true,
        linkcolor=darkblue,
        linktoc=all,
        citecolor=darkgreen,
        urlcolor=darkblue}
\allowdisplaybreaks[1]

\DeclareMathOperator{\sgn}{\textrm{sgn}}
\DeclareMathOperator{\wedgie}{\wedge}

\begin{document}

%\font\tenssf = cmss10
%\input{wasyfont}		% bold math

\title{Wave equations in conformally separable, accreting, rotating black holes}

\author{Andrew J.\ S.\ Hamilton}
\email{Andrew.Hamilton@colorado.edu}
\affiliation{JILA, Box 440, U. Colorado Boulder, CO 80309, USA}
\affiliation{Dept.\ Astrophysical \& Planetary Sciences, U. Colorado Boulder}
\author{Tyler McMaken}
\email{Tyler.McMaken@colorado.edu}
\affiliation{JILA, Box 440, U. Colorado Boulder, CO 80309, USA}
\affiliation{Dept.\ Physics, U. Colorado Boulder}

\newcommand{\simpropto}{\raisebox{-0.8ex}[1.5ex][0ex]{$
		\begin{array}[b]{@{}c@{\;}} \propto \\
		[-1.4ex] \sim \end{array}$}}

\newcommand{\overbar}[1]{\mkern 1.5mu\overline{\mkern-1.5mu#1\mkern-1.5mu}\mkern 1.5mu}

\newlength{\lenminus}
%\settowidth{\lenminus}{0}
\settowidth{\lenminus}{}
\newcommand{\hminus}{\hspace{\lenminus}}

\newcommand{\dd}{d}
\newcommand{\ddi}[1]{\dd^#1\mkern-1.5mu}
\newcommand{\DD}{D}
\newcommand{\Dext}{\textrm{D}}
\newcommand{\ee}{e}
\newcommand{\im}{i}
\newcommand{\nn}{\nonumber\\}
\renewcommand{\Re}{{\rm Re \,}}
\renewcommand{\Im}{{\rm Im \,}}
\newcommand{\diag}{\textrm{diag}}

\newcommand{\bDD}{\bm{\DD}}
\newcommand{\be}{\bm{e}}
\newcommand{\bF}{\bm{F}}
\newcommand{\bg}{\bm{g}}
\newcommand{\bj}{\bm{j}}
\newcommand{\bl}{\bm{l}}
\newcommand{\bn}{\bm{n}}
\newcommand{\bmm}{\bm{m}}
\newcommand{\bp}{\bm{p}}
\newcommand{\bR}{\bm{R}}
\newcommand{\starbR}{{}^\ast\!\bm{R}}
\newcommand{\bRz}{\mathbf{\tilde R}}
\newcommand{\bv}{\bm{v}}
\newcommand{\bx}{\bm{x}}
\newcommand{\bgamma}{\bm{\gamma}}
\newcommand{\bepsilon}{\bm{\epsilon}}
\newcommand{\bGamma}{\bm{\Gamma}}
\newcommand{\bvarphi}{\bm{\varphi}}
\newcommand{\spinordot}{{\mkern2mu \cdot}}

\newcommand{\col}{{\rm col}}
\newcommand{\Cz}{{\tilde C}}
\newcommand{\eff}{{\rm eff}}
\newcommand{\ddsq}{\dd^2\mkern-1.5mu}
\newcommand{\ethsi}[2]{\eth_{#2}}
\newcommand{\ethSi}[2]{{}_{#1}\eth_{#2}}
\newcommand{\Dsi}[2]{{\cal D}_{#2}}
\newcommand{\DSi}[2]{{}_{#1}{\cal D}_{#2}}
\newcommand{\Dpsi}[2]{{\cal D}^\prime_{#2}}
\newcommand{\DpSi}[2]{{}_{#1}{\cal D}^\prime_{#2}}
\newcommand{\psiz}{{\tilde \psi}}
\newcommand{\Fz}{{\tilde F}}
\newcommand{\bFz}{\bm{\tilde{F}}}
\newcommand{\ARS}{\psi}
\newcommand{\ARSz}{\tilde{\ARS}}
\newcommand{\bARS}{\bm{\ARS}}
\newcommand{\FRS}{\Psi}
\newcommand{\FRSz}{\tilde{\FRS}}
\newcommand{\bFRS}{\bm{\FRS}}
\newcommand{\bFRSz}{\bm{\FRSz}}
\newcommand{\Gammaz}{{\tilde\Gamma}}
\newcommand{\jz}{{\tilde j}}
\newcommand{\Jz}{{\tilde J}}
\newcommand{\Mbh}{M_\bullet}
\newcommand{\rhoi}{\rho_{\rm i}}
\newcommand{\rhom}{\bar\rho}
\newcommand{\rhop}{\bar\rho^\ast}
\newcommand{\rhosep}{\rho_{\rm s}}
\newcommand{\Ux}{U}
\newcommand{\uel}{u}
\newcommand{\vel}{v}
\newcommand{\wel}{w}
\newcommand{\xin}{x_-}
\newcommand{\nY}[1]{{}_{#1\,} \! Y}

\newcommand{\pmu}{\mathrel{\raisebox{-4pt}{${\overset{\displaystyle +}{\displaystyle \pm}}$}}}
\newcommand{\sq}[2]{\square_{#2}}
\newcommand{\Sq}[2]{{}_{#1}\square_{#2}}
\newcommand{\sqvu}{\square_{\overset{\scriptstyle vu}{\scriptstyle uv}}}
\newcommand{\squv}{\square_{\overset{\scriptstyle uv}{\scriptstyle vu}}}
\newcommand{\sqpm}{\square_{\overset{\raisebox{0pt}[.6ex][0pt]{$\scriptstyle +-$}}{\raisebox{0pt}[.6ex][0pt]{$\scriptstyle -+$}}}}
\newcommand{\sqmp}{\square_{\overset{\raisebox{0pt}[.6ex][0pt]{$\scriptstyle -+$}}{\raisebox{0pt}[.6ex][0pt]{$\scriptstyle +-$}}}}
\newcommand{\Sqvu}[1]{{}_{#1}\sqvu}
\newcommand{\Squv}[1]{{}_{#1}\squv}
\newcommand{\Sqpm}[1]{{}_{#1}\sqpm}
\newcommand{\Sqmp}[1]{{}_{#1}\sqmp}
\newcommand{\uv}{{\overset{\scriptstyle u}{\scriptstyle v}}}
\newcommand{\vu}{{\overset{\scriptstyle v}{\scriptstyle u}}}
\newcommand{\hatpsi}{{\hat \psi}}

\newcommand{\Upup}{{\Uparrow}{\uparrow}}
\newcommand{\Updown}{{\Uparrow}{\downarrow}}
\newcommand{\Downup}{{\Downarrow}{\uparrow}}
\newcommand{\Downdown}{{\Downarrow}{\downarrow}}

\newcommand{\smallzero}{{\scriptscriptstyle 0}}
\newcommand{\smallminus}{{\scriptscriptstyle -}}
\newcommand{\smallplus}{{\scriptscriptstyle +}}
\newcommand{\smallpm}{{\scriptscriptstyle \pm}}
\newcommand{\smallup}{\scriptscriptstyle \uparrow}
\newcommand{\smalldown}{\scriptscriptstyle \downarrow}
\newcommand{\smallUp}{\scriptscriptstyle \Uparrow}
\newcommand{\smallDown}{\scriptscriptstyle \Downarrow}
\newcommand{\smallUpup}{\scriptscriptstyle \Upup}
\newcommand{\smallUpdown}{\scriptscriptstyle \Updown}
\newcommand{\smallDownup}{\scriptscriptstyle \Downup}
\newcommand{\smallDowndown}{\scriptscriptstyle \Downdown}

\newcommand{\pback}{\mkern-1.5mu}	% backspace after prime

\hyphenpenalty=3000

\begin{abstract}
The Teukolsky wave equations governing fields
of spin 0, $\tfrac{1}{2}$, 1, $\tfrac{3}{2}$, and 2
are generalized to the case of conformally separable solutions
for accreting, rotating black holes.
\end{abstract}

\pacs{04.20.-q}	% Classical general relativity

\date{\today}

\maketitle

\section{Introduction}

While the exterior geometry of an astrophysically realistic black hole is well-modeled by the Kerr metric, the region below the event horizon remains more elusive.
The inner horizon of a Kerr black hole is subject to
the Poisson-Israel \cite{Poisson:1990eh,Barrabes:1990}
``mass inflation'' instability,
in which incident outgoing and ingoing streams of accreting matter or radiation
drive an exponentially growing curvature.
The outcome of the inflationary instability,
from both a classical and quantum perspective, is a subject of active research
\cite{
Ottewill:2000,
Dafermos:2017dbw,
Casals:2017,Casals:2019a,Casals:2019b,
Lanir:2019,
Chesler:2019a,Chesler:2019b,Chesler:2021,
Barcelo:2019,Barcelo:2021,Barcelo:2022,
Taylor:2020,
Hollands:2020a,Hollands:2020b,
Zilberman:2020,Zilberman:2021,Zilberman:2022a,Zilberman:2022b,
Arrechea:2021,
Klein:2021,
McMaken:2021,McMaken:2022}.

The Kerr solution \cite{Kerr:1963,Teukolsky:2015}
and its electrovac cousins \cite{Newman:1965,Stephani:2003}
are strictly stationary, strictly separable, and axisymmetric.
Refs.~\cite{Hamilton:2010a,Hamilton:2010b,Hamilton:2010c}
have generalized these spacetimes to allow for self-similar accretion,
leading to conformally stationary, conformally separable, axisymmetric
solutions for the interior structure of accreting, rotating black holes.
Whereas strict separability posits that all geodesics are Hamilton-Jacobi separable, conformal separability posits only that null geodesics are separable.
The hypothesis of conformal separability imposes all the usual constraints on the possible form of the separable line-element \cite{Carter:1968c,Walker:1970} except that the overall conformal factor is permitted to be arbitrary.
The further hypothesis of conformal stationarity posits that the spacetime has a conformal timelike Killing vector, or in other words that the spacetime grows self-similarly, by accretion, and that the accretion rate is small.

The energy-momentum tensor that sources the conformally stationary, conformally separable solutions fits that of a collisionless fluid containing a combination of outgoing and ingoing components.
It is remarkable that the entire system of Einstein equations (and Maxwell equations, if the black hole is charged), coupled to the equations of a collisionless fluid, are jointly separable and solvable.

The conformally separable solutions are approximate,
holding in the asymptotic limit of small accretion rate.
It is helpful to emphasize the precise sense in which the conformally
separable solutions are approximate.
The standard $\Lambda$-Kerr-Newman line-element can be derived
from three assumptions:
time translation symmetry, azimuthal symmetry, and the separability of
the Hamilton-Jacobi equation for geodesics of particles of rest mass $m$.
More specifically, the $\Lambda$-Kerr-Newman line-element follows from
assuming that the Hamilton-Jacobi equation separates
``in the simplest possible way'' \cite{Carter:1968c},
which requires that the particle action $S$ be a separated sum
(this is eq.~(28) of \cite{Hamilton:2010b})
\begin{equation}
\label{HJaction}
  S =
  \tfrac{1}{2} m^2 \lambda
  -
  E t
  +
  L \phi
  +
  S_x ( x )
  +
  S_y ( y)
  \ ,
\end{equation}
in which $\lambda$ is an affine parameter,
$E$ and $L$ are a conserved energy and angular momentum
associated with time $t$ translation and azimuthal $\phi$ symmetry,
and $S_x(x)$ and $S_y(y)$ are respectively functions
of two other coordinates $x$ and $y$.
The conformally separable line-element~(\ref{lineelement}) follows
from relaxing the assumption of time-translation symmetry
to conformal time-translation symmetry,
meaning that the metric depends on a conformal time coordinate $t$
only through an overall conformal factor $\ee^{v t}$ with accretion rate $v$,
and from relaxing the assumption of strict Hamilton-Jacobi separability
to conformal separability,
meaning that equation~(\ref{HJaction}) is required to hold
only for massless particles, $m = 0$.
With the line-element~(\ref{lineelement}) in hand, it is then a laborsome
matter to separate the Einstein (and Maxwell) equations systematically.
That separation works only if certain terms are treated as negligible, which
\cite{Hamilton:2010b} shows is valid in the asymptotic limit of small
accretion rate $v$.  The proof that the negligible terms can in fact be
neglected is again laborsome.

Since the accretion rate is small, the geometry of the conformal solutions is well-approximated by the Kerr (or more general electrovac) geometry
down to just above the inner horizon.
However, even in the limit of a tiny accretion rate, hyper-relativistic counterstreaming between outgoing and ingoing streams just above the inner horizon ignites and then drives the Poisson-Israel mass inflation instability.
During inflation, the proper counter-streaming density and pressure, along with the Weyl curvature, exponentiate to huge (super-Planckian, if quantum gravity did not intervene) values.
The inflationary instability is the nonlinear realization of the infinite blueshift at the inner horizon first pointed out by \cite{Penrose:1968}.

The end result of mass inflation is the formation of a singularity near where the inner horizon should have been.
Perturbation analyses in the Kerr geometry
find this singularity to be either weak and null \cite{Dafermos:2017dbw}
or strong and spacelike \cite{Brady:1995un,Chesler:2021}.
The conformally stationary, conformally separable solutions predict a singularity of the latter type.
Whereas a weak, null singularity is derived from the assumption that the black hole remains isolated into the indefinite future
after an initial collapse or accretion event,
real astronomical black holes are not isolated,
with ongoing accretion from
the cosmic microwave background, dark matter,
gas, and ambient astronomical bodies
\cite{Hamilton:2008zz}.
As a result of the continued accretion,
mass inflation stalls at exponentially large curvature,
at which point the spacetime collapses
(in the sense that the conformal factor shrinks to an exponentially tiny scale)
toward a spacelike singularity.
This is the general behavior predicted by the nonlinear, dynamical solutions of \cite{Hamilton:2010a,Hamilton:2010b,Hamilton:2010c}.

The conformally separable solutions are unrealistic in the sense that they require a special incident accretion flow, which is constant in time and constant in angle over the inner horizon.
This would seem to diminish the physical relevance of the solutions.
It appears however that the solutions have more general application to arbitrary accretion flows, as long as the accretion rate is small, as is true most of the time in most astronomical black holes.
The reason is that during the initial, inflationary phase, the Kerr (or more general electrovac) geometry remains essentially unchanged
while outgoing and ingoing streams focus and blueshift exponentially along the principal null directions.
During this inflationary phase, tidal forces (second gradients of the metric) grow (exponentially) large, but the metric itself barely budges.
Eventually the tidal force starts to backreact on the spacetime,
leading to collapse in the transverse directions, and stretching along the principal directions.
Inflation and collapse occur over such a tiny proper time that what happens at one point on the inner horizon is causally disconnected from what happens at another point.
Different causal patches on the inner horizon evolve essentially independently of each other.
The conformally separable solutions \cite{Hamilton:2010a,Hamilton:2010b,Hamilton:2010c} fail deep into the collapse regime, where rotational motions reassert themselves.
\cite{Hamilton:2017qls} has explored numerically what happens at that point, finding that inflation and collapse is followed by Belinskii-Khalatnikov-Lifshitz (BKL) \cite{Belinskii:1970,Belinskii:1971,Belinskii:1972,Belinskii:1982,Garfinkle:2020} oscillatory collapse to a spacelike singular surface.
The conformally separable inflation and collapse regimes prove to be just the first two Kasner epochs in a succession of Kasner epochs separated by BKL bounces.

The purpose of the present paper is to derive the wave equations of massless, neutral fields of various spin (0, $\frac12$, 1, $\frac32$, and 2) in conformally separable black-hole spacetimes, extending the well-known Teukolsky solutions for stationary rotating black holes
\cite{Teukolsky:1972my,Teukolsky:1973ha,Teukolsky:2015,Chandrasekhar:1983,Staicova:2015}.
The problem is by itself of some interest: do the wave equations in rotating black holes remain separable if the condition of strict separability is relaxed to conformal separability? The answer is yes.
The generalization of the Teukolsky master equation
\cite{Teukolsky:1972my,Teukolsky:1973ha,Teukolsky:2015}
to the conformally separable solutions is given by
equations~(\ref{altwaveoperatorscoordinate})
with the potentials~(\ref{altpotential}).

The conformal wave equations are also relevant for analyzing quantum effects near the inner horizons of astronomically realistic black holes.
Semiclassically, the quantum backreaction to the metric is governed by the renormalized vacuum expectation value of a field's energy-momentum tensor, which is built out of solutions to the wave equation.
For Kerr black holes, this quantity has recently been calculated by
\cite{Zilberman:2022a} at the inner horizon,
revealing singular behavior at that null boundary.
The fully story of what happens at the inner horizon
of an astronomically realistic rotating black hole
is a topic of active research
\cite{
Dafermos:2017dbw,
Casals:2019a,
Chesler:2019a,
McMaken:2021,McMaken:2022,
Zilberman:2022a,Zilberman:2022b}.
One of the big outstanding questions is,
do quantum effects transform a weak, null singularity
into a strong, spacelike singularity,
even in the absence of accretion from outside?

This paper considers only uncharged black holes, even though conformally separable solutions exist also for charged black holes \cite{Hamilton:2010c}.
The problem is that electromagnetic and gravitational perturbations become inextricably linked in the presence of a finite electric field, preventing separation of the wave equations.
In his epic monograph, Chandrasekhar \cite{Chandrasekhar:1983} was able to separate the coupled electromagnetic and gravitational perturbations for the case of a Reissner-Nordstr{\"o}m (charged, nonrotating) black hole,
but not for a charged, rotating black
hole.

The plan of this paper is as follows.
Section~\ref{confsep-sec} reviews the conformally separable
black hole solutions,
and defines the Newman-Penrose and spinor frames
with respect to which the wave equations are separable.
Section~\ref{waveeqs-sec} summarizes the wave equations for massless
fields of arbitrary spin.
The summary in \S\ref{waveeqs-sec} is based on the wave equations
derived in \S\S\ref{scalar-sec}--\ref{spintwo-sec}
for each of spins
$s = 0$, $\tfrac{1}{2}$, $1$, $\tfrac{3}{2}$, and $2$.

Appendix~\ref{Dapp} takes a deeper dive into the derivation of
wave equations in general spacetimes.
Appendix~\ref{relation-app}
gives a relation between differential operators
needed to write down the expressions~(\ref{F0})
for the boost weight~0 component $\Fz_0$ of an electromagnetic wave.
Appendix~\ref{chandra-app} translates between the notation of Chandrasekhar
\cite{Chandrasekhar:1983} and the present paper.

In this paper, mid latin indices $k,l,\ldots$ run over tetrad vector indices, while early latin indices $a,b,\ldots$ run over chiral spinor indices.
Greek indices $\kappa, \lambda, \ldots$ are coordinate indices.
The units are such that the speed of light and Newton’s gravitational constant are unity, $c=G=1$.

\section{Conformally separable rotating black holes}
\label{confsep-sec}

\subsection{Line-element}
\label{lineelement-sec}

This section~\ref{lineelement-sec}
summarizes results from
\cite{Hamilton:2010a,Hamilton:2010b}.

The conformally separable line-element is,
in conformal coordinates $\{ x , t , y , \phi \}$
\begin{align}
\label{lineelement}
  \dd s^2
  =
  \rho^2
  \Bigl[
  &{\dd x^2 \over \Delta_x}
  -
  {\Delta_x \over ( 1 - \omega_x \omega_y )^2}
  \left( \dd t - \omega_y \, \dd \phi \right)^2
\nonumber
\\
  + \,
  &{\dd y^2 \over \Delta_y}
  +
  {\Delta_y \over ( 1 - \omega_x \omega_y )^2}
  \left( \dd \phi - \omega_x \, \dd t \right)^2
  \Bigr]
  \ .
\end{align}
The line-element~(\ref{lineelement}) defines not only a metric,
but also an orthonormal tetrad, \S\ref{tetrad-sec},
whose null directions are chosen to align with the principal null directions.
The line-element~(\ref{lineelement}) is Kerr (or more general electrovac),
with two differences elaborated below,
the conformal factor $\rho$, equation~(\ref{rho}),
and the radial horizon function $\Delta_x$,
equations~(\ref{DeltaxLKN}) and~(\ref{Deltaxinf}).
Separation of the Einstein equations is most natural with respect
to radial and angular coordinates $x$ and $y$, which are
related to the usual Boyer-Lindquist radius $r$ and polar angle $\theta$ by
\begin{equation}
\label{rtheta}
  r
  \equiv
  a \cot ( a x )
  \ , \quad
  \cos\theta
  \equiv
  - y
  \ ,
\end{equation}
where $a$ is the angular momentum per unit mass of the black hole,
with $a$ positive for right-handed rotation about the axis of rotation.
Note that
\begin{equation}
\label{Rdef}
  {\partial \over \partial x}
  =
  - R^2 {\partial \over \partial r}
  \ , \quad
  R \equiv \sqrt{r^2 + a^2}
  \ .
\end{equation}
The quantities
$\omega_x$ and $\omega_y$
in the conformally separable line-element~(\ref{lineelement})
take their usual Kerr forms,
and are functions respectively
only of the radial and angular coordinates,
\begin{equation}
\label{omegaLKN}
  \omega_x
  =
  {a \over R^2}
  \ , \quad
  \omega_y
  =
  a \sin^2\!\theta
  \ .
\end{equation}
Physically,
$\omega_x$
is the angular velocity of the principal null frame through the coordinates,
while $\omega_y$ is the specific angular momentum of lightrays
on the principal null congruence.

The first difference in the line-element~(\ref{lineelement})
between stationary ($\Lambda$-Kerr-Newman) and conformally separable
black hole spacetimes is in the conformal factor $\rho$.
In stationary black hole spacetimes,
the conformal factor $\rho$ is separable,
its square being a sum
$\rho^2 = \rhosep^2 = r^2 \,{+}\, a^2 \cos^2\!\theta$
of radial and angular parts.
In the conformally separable spacetimes,
the conformal factor $\rho$
is a product of the $\Lambda$-Kerr-Newman separable factor $\rhosep$,
a time-dependent factor $\ee^{\vel t}$,
and an inflationary factor $\ee^{- \xi(x)}$,
\begin{equation}
\label{rho}
  \rho
  =
  \rhosep \rhoi
  \ , \quad
  \rhosep
  =
  \sqrt{r^2 + a^2 \cos^2\!\theta}
  \ , \quad
  \rhoi
  =
  \ee^{\vel t - \xi(x)}
  \ .
\end{equation}
It is useful to define the complex combination
\begin{equation}
\label{rhopm}
  \rhom
  \equiv
  r - \im a \cos\theta
  \ ,
\end{equation}
whose complex conjugate is $\rhop = r + \im a \cos\theta$,
and whose absolute value is the separable conformal factor,
$| \rhom | = \sqrt{\rhom \rhop} = \rhosep$.
The conformally separable solutions apply
in the limit of small but nonvanishing accretion rate
\begin{equation}
\label{vel0}
  \vel
  \rightarrow
  0
  \ .
\end{equation}
%SAY MORE ABOUT SMALL ACCRETION RATE?
The conformally separable spacetimes
possess a Killing vector $\partial / \partial \phi$
associated with azimuthal symmetry,
a conformal Killing vector $\partial / \partial t$
associated with conformal time translation invariance,
and a traceless conformal Killing tensor $K^{mn}$
\cite{Hamilton:2010b}
\begin{equation}
  K^{mn}
  \equiv
  \tfrac{1}{2} \rho^2 \diag ( 1, -1, 1, 1 )
  \ ,
\end{equation}
which satisfies the conformal Killing equation
\begin{equation}
  \DD_{(k} K_{mn)}
  -
  \tfrac{1}{3} \eta_{(km} \DD^l K_{n)l}
  =
  0
  \ .
\end{equation}
The connection between the existence of a conformal Killing tensor
and the separability of wave equations is discussed by
\cite{Jeffryes:1984}.

There is a gauge freedom in the choice of zero point and scaling
of conformal time $t$.
The proper time far from the black hole yet well inside any
cosmological horizon defines the Kerr time
$t_{\rm Kerr}$,
\begin{equation}
  t_{\rm Kerr}
  \equiv
  \int \ee^{\vel t} \, \dd t
  \propto
  {\ee^{\vel t} \over \vel}
  \ .
\end{equation}
The proper mass $M$ of the black hole increases proportionally
to the conformal factor, thus linearly with Kerr time.
It is natural to scale conformal time, hence $\vel$, so that
\begin{equation}
  M = \vel t_{\rm Kerr}
  \ .
\end{equation}
With this choice,
the accretion rate $\vel$ is just equal to the dimensionless rate
at which the proper mass of the black hole increases,
as measured by a distant observer,
\begin{equation}
  \vel
  =
  \dot{M}
  \ .
\end{equation}

The second difference in the line-element~(\ref{lineelement})
between stationary ($\Lambda$-Kerr-Newman) and conformally separable
black hole spacetimes is in the radial horizon function $\Delta_x$.
The conformally separable spacetimes share with their separable cousins
the property that the radial horizon function $\Delta_x$
and the angular polar function $\Delta_y$
are functions only of respectively the radial coordinate $x$
and the angular coordinate $y$.
Indeed, the polar function $\Delta_y$ for the conformally separable spacetimes
is unchanged from the $\Lambda$-Kerr-Newman spacetimes,
and is, including a possible cosmological constant $\Lambda$,
\begin{equation}
\label{DeltayLKN}
  \Delta_y
  =
  \sin^2\!\theta
  \left( 1 + {\Lambda a^2 \cos^2\!\theta \over 3}
  %- {2 \NUTbh \cos\theta}
  %- {\Qmagbh^2 \over a^2}
  \right)
  \ .
\end{equation}
The analysis in this paper holds both outside and inside the outer horizon.
The conformally separable geometry is the standard
%$\Lambda$-Kerr-Newman
$\Lambda$-Kerr
geometry down to just above the inner horizon,
where the usual Teukolsky wave equations are recovered.
In the $\Lambda$-Kerr region outside the outer horizon
and down to just above the inner horizon,
the horizon function $\Delta_x$ is
\begin{equation}
\label{DeltaxLKN}
  \Delta_x
  \, \underset{r \gg r_-}{\parbox{2.2em}{\rightarrowfill\,}} \,
  {1 \over R^2}
  \left(
  1 - {2 \Mbh r \over R^2}
  %+ {\Qelecbh^2
  %+ \Qmagbh^2
  %\over R^2}
  - {\Lambda r^2 \over 3}
  \right)
  \ ,
\end{equation}
where $\Mbh$ is the conformal mass of the black hole, a constant,
%$\Qelecbh$ its electric charge,
%$\Qelecbh$ and $\Qmagbh$ are its electric and magnetic charge,
and $\Lambda$ is the cosmological constant.
%The
%charge and
%cosmological constant is included
%not because it is are required,
%but rather
%because its inclusion is a trivial generalization.
%Their only role in the solutions is to set the boundary
%condition for the horizon function at (just above) the inner horizon.
%However, all the fields considered in this paper are neutral.
%for simplicity.
%The inflationary instability is quite generic,
%occuring independently of.

However, the conformally separable solutions undergo violent
inflation from just above the inner horizon inward.
Just above the inner horizon, the horizon function
is negative and tending to zero, $\Delta_x \rightarrow - 0$.
The radial coordinate $x$ is then timelike,
and increasing inward, the direction of increasing proper time,
while the time coordinate $t$ is spacelike,
and increasing in the outgoing direction
(at the inner horizon, outgoing particles accreted in the past
encounter ingoing particles accreted in the future,
so it is natural to choose outgoing particles inside the black hole
to be moving forward in $t$,
while ingoing particles move backward in $t$).
The conformally separable solutions are characterized by
two small parameters $\uel$, $\vel$
which specify the rates
$\uel \pm \vel$
of accretion of outgoing $(+)$ and ingoing ($-$) collisionless streams
incident on the inner horizon
(do not confuse these accretion rates with the ingoing and outgoing
tetrad indices $u$ and $v$ introduced in \S\ref{tetrad-sec}).
Positivity of both accretion rates requires
\begin{equation}
  \uel > \vel > 0
  \ .
\end{equation}
The conformally separable solutions are valid in the limit of small
but finite $\uel$ and $\vel$.

Solution of the Einstein equations for the conformally
separable line element~(\ref{lineelement})
driven by outgoing and ingoing null streams
leads to evolutionary equations for the inflationary exponent
$\xi$ and the horizon function $\Delta_x$
near the inner horizon
\cite{Hamilton:2010a,Hamilton:2010b}.
First, define radial and angular tortoise coordinates $x^\ast$ and $y^\ast$ by
\begin{equation}
\label{xytortoise}
  x^\ast
  \equiv
  \int {\dd x \over -\Delta_x}
  \ , \quad
  y^\ast
  \equiv
  \int
  {\dd y \over \Delta_y}
  \ .
\end{equation}
Outside the outer horizon $r_+$,
the radial tortoise coordinate $x^\ast$ increases outward,
from $- \infty$ at the horizon $r_+$, to
%NO
%to $+ \infty$ either at the cosmological horizon $r_\Lambda$ if there is one,
%or at infinite radius if not,
%YES
\begin{equation}
  x^\ast \rightarrow \left\{
  \begin{array}{ll}
%  - \infty
%  &
%  r \rightarrow r_+ \ ,
%  \\
  + \infty
  &
  \mbox{at~} r \rightarrow \infty \mbox{~if~} \Lambda = 0 \ ,
  \\
  + \infty
  &
  \mbox{at~} r \rightarrow r_\Lambda \mbox{~if~} \Lambda > 0 \ ,
  \\
  \mbox{finite}
  &
  \mbox{at~} r \rightarrow \infty \mbox{~if~} \Lambda < 0 \ .
  \end{array}
  \right.
\end{equation}
Inside the outer horizon, the tortoise coordinate $x^\ast$
increases inward,
from $- \infty$ at the horizon $r_+$.
%For $\Lambda$-Kerr-Newman,
%\begin{subequations}
%\begin{align}
%  x^\ast
%  &=
%  \sum_{{\rm roots}~i}
%  \left.
%  {1 \over \dd ( R^4 \Delta_x ) / \dd x}
%  \right|_{r = r_i}
%  \ln | r - r_i |
%  \ ,
%\\
%  y^\ast
%  &=
%  \sum_{{\rm roots}~y_i}
%  \left.
%  {1 \over \dd \Delta_y / \dd y}
%  \right|_{y = y_i}
%  \ln | y - y_i |
%  \ .
%\end{align}
%\end{subequations}
Define the dimensionless quantity $\Ux(x)$ by
\begin{equation}
\label{Ux}
  \Ux
  \equiv
  {\dd \xi \over \dd x^\ast}
  \ .
\end{equation}
The Einstein equations near the inner horizon lead to
\cite{Hamilton:2010b}
%\begin{subequations}
%\label{DDUx}
%\begin{align}
%\label{DUx}
%  {\dd \Ux \over \dd x}
%  +
%  2
%  {\Ux^2 - \vel^2 \over \Delta_x}
%  &=
%  0
%  \ ,
%\\
%\label{DDx}
%  {\dd \Delta_x \over \dd x}
%  +
%  3 \Ux
%  &=
%  \Delta_x^\prime
%  \ ,
%\end{align}
%\end{subequations}
\begin{subequations}
\label{DDUx}
\begin{align}
\label{DUx}
  {\dd \Ux \over \dd x^\ast}
  &=
  2
  ( \Ux^2 - \vel^2 )
  \ ,
\\
\label{DDx}
  {\dd \ln \Delta_x \over \dd x^\ast}
  &=
  3 \Ux
  -
  \Delta_x^\prime
  \ ,
\end{align}
\end{subequations}
where
$\Delta_x^\prime \equiv \left. \dd \Delta_x / \dd x \right|_{\xin}$,
a constant,
is the (positive) derivative of the electrovac
horizon function at the inner horizon
$x = \xin$.
The initial condition for $\Ux$ well above the inner horizon is
\begin{equation}
\label{Uxinit}
  \Ux = \uel
  \ .
\end{equation}
The differential equations~(\ref{DDUx})
with the initial condition~(\ref{Uxinit}) solve to
give the inflationary exponent $\xi$, horizon function $\Delta_x$,
and radial coordinate $x$ near the inner horizon:
\begin{subequations}
\label{UDxinf}
\begin{align}
  \ee^\xi
  &=
  \left( {\Ux^2 - \vel^2 \over \uel^2 - \vel^2} \right)^{1/4}
  \ ,
\\
  \ee^{\vel x^\ast}
  &=
  \left[
  {
  ( \Ux - \vel )
  ( \uel + \vel )
  \over
  ( \Ux + \vel )
  ( \uel - \vel )
  }
  \right]^{1/4}
  \ ,
\\
\label{Deltaxinf}
  \Delta_x
  &=
  - \ee^{3 \xi - \Delta_x^\prime x^\ast}
  \ ,
%\\
%  \Delta_x
%  &=
%  -
%  \ee^{3 \xi}
%  \left[
%  {
%  ( \Ux + \vel )
%  ( \uel - \vel )
%  \over
%  ( \Ux - \vel )
%  ( \uel + \vel )
%  }
%  \right]^{\Delta_x^\prime / ( 4 \vel )}
%  \ ,
\\
\label{xinf}
  x - \xin
  &=
  -
  \int
  {\Delta_x \, \dd \Ux \over 2 ( \Ux^2 - \vel^2 )}
  \ .
\end{align}
\end{subequations}
Equation~(\ref{xinf})
says that the radial coordinate $x$
is essentially frozen at its inner horizon value $\xin$
throughout inflation and collapse.

\subsection{Tetrad}
\label{tetrad-sec}

The line-element~(\ref{lineelement})
defines not only
a metric
$g_{\kappa\lambda}$,
which is an inner product of coordinate tangent vectors
$\be_\kappa$,
but also,
through
\begin{equation}
  \be_\kappa \cdot \be_\lambda
  =
  g_{\kappa\lambda}
  =
  \eta_{kl} e^k{}_\kappa e^l{}_\lambda
  =
  e^k{}_\kappa \bgamma_k \cdot e^l{}_\lambda \bgamma_l
  \ ,
\end{equation}
a vierbein matrix
$e^k{}_\kappa$,
and a corresponding orthonormal tetrad
$\{ \bgamma_0 , \bgamma_1 , \bgamma_2 , \bgamma_3 \}$,
%(which could be written
%$\{ \bgamma_x , \bgamma_t , \bgamma_\theta , \bgamma_\phi \}$
%if one preferred),
whose inner products form the Minkowski metric,
$\bgamma_k \cdot \bgamma_l = \eta_{kl}$.
The orthonormal tetrad behaves smoothly across horizons
provided that the time axis $\bgamma_0$ is chosen timelike and future-pointing
both outside ($\Delta_x > 0$) and inside ($\Delta_x < 0$) the horizon,
while the radial axis $\bgamma_1$ is outgoing
both outside and inside the horizon:
\begin{equation}
\label{gamma0123}
  \{ \bgamma_0 , \bgamma_1 , \bgamma_2 , \bgamma_3 \}
  =
  \left\{
  \begin{array}{ll}
  \{ \bgamma_t , \bgamma_x , \bgamma_\theta , \bgamma_\phi \}
  &
  \Delta_x > 0
  \\
  \{ \bgamma_x , \bgamma_t , \bgamma_\theta , \bgamma_\phi \}
  &
  \Delta_x < 0
  \end{array}
  \right.
  \ .
\end{equation}
The sign of the vierbein coefficient $e^1{}_x$ is negative outside the horizon
(because the spacelike radial coordinate $x$ decreases outwards,
equation~(\ref{Rdef})),
while $e^0{}_x$ is positive inside the horizon
(because the timelike radial coordinate $x$ increases inwards).

Whenever dealing with massless fields,
it is advantageous to use the Newman-Penrose double-null tetrad formalism.
The double-null Newman-Penrose basis tetrad
$\{ \bgamma_v , \bgamma_u , \bgamma_+ , \bgamma_- \}$
corresponding to the orthonormal tetrad
$\{ \bgamma_0 , \bgamma_1 , \bgamma_2 , \bgamma_3 \}$,
is
\begin{equation}
\label{gammavupm}
  \bgamma_\vu
  \equiv
  \tfrac{1}{\sqrt{2}}
  ( \bgamma_0 \pm \bgamma_1 )
  \ , \quad
  \bgamma_\pm
  \equiv
  \tfrac{1}{\sqrt{2}}
  ( \bgamma_2 \pm \im \bgamma_3 )
  \ ,
\end{equation}
%where the $\pm$ in $\gamma_u$ is $+$ outside,
%$-$ inside, the outer horizon $r_+$.
The choice~(\ref{gamma0123}) of orthonormal axes ensures that
the Newman-Penrose basis vectors~(\ref{gammavupm})
behave smoothly across horizons,
with $\bgamma_v$ outgoing and $\bgamma_u$ ingoing
both outside and inside the outer horizon.
The tetrad metric of the Newman-Penrose basis vectors is
($k,l = v,u,+,-$)
\begin{equation}
  \gamma_{kl}
  \equiv
  \bgamma_k \cdot \bgamma_l
  =
  \left(
  \begin{array}{cccc}
  0 & -1 & 0 & 0 \\
  -1 & 0 & 0 & 0 \\
  0 & 0 & 0 & 1 \\
  0 & 0 & 1 & 0
  \end{array}
  \right)
  \ .
\end{equation}

The transverse Newman-Penrose basis vectors $\bgamma_\pm$
defined by equations~(\ref{gammavupm})
are complex conjugates of each other,
\begin{equation}
  \bgamma_- = \bgamma_+^\ast
  \ .
\end{equation}
Consequently the directed derivatives $\partial_\pm$
in the $\bgamma_\pm$ directions are complex conjugates,
\begin{equation}
  \partial_- = \partial_+^\ast
  \ .
\end{equation}
More generally, covariant tetrad-frame derivatives $\DD_\pm$
in the $\bgamma_\pm$ directions are complex conjugates,
\begin{equation}
  \DD_- = \DD_+^\ast
  \ .
\end{equation}

Boosting the tetrad frame by rapidity $\eta$ in the $vu$-plane
boosts the outgoing $\bgamma_v$ and ingoing $\bgamma_u$ basis vectors by
\begin{equation}
\label{gammavuboost}
  \bgamma_v \rightarrow \ee^{\eta} \bgamma_v
  \ , \quad
  \bgamma_u \rightarrow \ee^{- \eta} \bgamma_u
  \ ,
\end{equation}
while spatially rotating the tetrad frame by angle $\zeta$ in the $+-$-plane
rotates the $\bgamma_+$ and $\bgamma_-$ basis vectors by
\begin{equation}
\label{gammapmspin}
  \bgamma_+ \rightarrow \ee^{- \im \zeta} \bgamma_+
  \ , \quad
  \bgamma_- \rightarrow \ee^{\im \zeta} \bgamma_-
  \ .
\end{equation}

Tetrad-frame tensors inherit their transformation properties
from the tetrad-frame basis vectors $\bgamma_k$.
A tensor of boost weight $\sigma$ is multiplied by
$\ee^{\sigma\eta}$
under a Lorentz boost by rapidity $\eta$ in the $vu$-plane,
while a tensor of spin weight $\varsigma$ is multiplied by
$\ee^{- \im \varsigma \zeta}$
under a right-handed spatial rotation by angle $\zeta$ in the $+-$-plane
\cite{Geroch:1973}.
The boost and spin weights of a tetrad-frame tensor
can be determined by inspection, according to the rules
\begin{subequations}
\label{boostspinrules}
\begin{align}
  &
  \mbox{boost weight}
\\
\nonumber
  &
  =
  \mbox{number of $v$ minus number of $u$ covariant indices}
  \ ,
\\
  &
  \mbox{spin weight}
\\
\nonumber
  &
  =
  \mbox{number of $+$ minus number of $-$ covariant indices}
  \ .
\end{align}
\end{subequations}
Contravariant indices count oppositely to covariant indices.

\subsection{Spinors}
\label{spinor-sec}

To deal with fields of half-integral spin,
it is necessary to introduce a matrix representation
of the tetrad $\bgamma_k$.
These matrices are commonly called Dirac $\gamma$-matrices
(which accounts for the notation $\bgamma_k$ for the tetrad basis vectors).
In the chiral representation
where the chiral operator is diagonal,
equation~(\ref{gamma5}) below,
the Dirac $\gamma$-matrices are the
$4 \times 4$ real unitary ($\bgamma^k = \bgamma_k^\dagger$) matrices
\begin{subequations}
\label{diracgammaI}
\begin{align}
  \bgamma_v =
  \left(
  \begin{array}{cc}
     0 & \sigma_v \\
     -\sigma_u & 0
  \end{array}
  \right)
  \ &, \quad
  \bgamma_u =
  \left(
  \begin{array}{cc}
     0 & \sigma_u \\
     -\sigma_v & 0
  \end{array}
  \right)
  \ ,
\\
  \bgamma_+ =
  \left(
  \begin{array}{cc}
     0 & \sigma_+ \\
     \sigma_+ & 0
  \end{array}
  \right)
  \ &, \quad
  \bgamma_- =
  \left(
  \begin{array}{cc}
     0 & \sigma_- \\
     \sigma_- & 0
  \end{array}
  \right)
  \ ,
\end{align}
\end{subequations}
where $\sigma_k$ are the Newman-Penrose Pauli matrices
\begin{subequations}
\begin{align}
\label{PauliI}
  \sigma_v \equiv
  \sqrt{2}
  \left(
    \begin{array}{cc}
      1 & 0 \\
      0 & 0
    \end{array}
  \right)
\ &, \quad
  \sigma_u \equiv
  \sqrt{2}
  \left(
    \begin{array}{cc}
      0 & 0 \\
      0 & 1
    \end{array}
  \right)
\ ,
\\
  \sigma_+ \equiv
  \sqrt{2}
  \left(
    \begin{array}{cc}
      0 & 1 \\
      0 & 0
    \end{array}
  \right)
\ &, \quad
  \sigma_- \equiv
  \sqrt{2}
  \left(
    \begin{array}{cc}
      0 & 0 \\
      1 & 0
    \end{array}
  \right)
  \ .
\end{align}
\end{subequations}
The Lorentz-invariant chiral operator $\gamma_5$ is $-\im$
times the pseudoscalar $I$,
\begin{align}
\label{gamma5}
  \gamma_5
  &\equiv
  - \im I
  \equiv
  - \im \bgamma_0 \bgamma_1 \bgamma_2 \bgamma_3
  =
  \bgamma_v \wedgie \bgamma_u \wedgie \bgamma_+ \wedgie \bgamma_-
\nonumber
\\
  &=
  - \frac{\im}{4!} \,
  \varepsilon^{klmn} \,
  \bgamma_k \bgamma_l \bgamma_m \bgamma_n
  =
  \left(
  \begin{array}{cc}
  1 & 0 \\
  0 & -1
  \end{array}
  \right)
  \ ,
\end{align}
the sign convention for the totally antisymmetric tensor being
$\varepsilon^{klmn} = -\varepsilon_{klmn} = [klmn]$ in a locally inertial frame.

The Dirac $\gamma$-matrices act by matrix multiplication on Dirac spinors,
which are spin~$\tfrac{1}{2}$ spinors in 3+1 spacetime dimensions.
A Dirac spinor $\psi$ is a complex
linear combination of chiral basis spinors $\bepsilon_a$,
\begin{equation}
\label{diracspinor}
  \psi = \psi^a \bepsilon_a
  \ ,
\end{equation}
in which $\bepsilon_a$ is the foursome of chiral basis spinors
\begin{equation}
  \bepsilon_a
  =
  \{
  \bepsilon_{\smallUpup} ,
  \bepsilon_{\smallDowndown} ,
  \bepsilon_{\smallUpdown} ,
  \bepsilon_{\smallDownup}
  \}
  \ .
\end{equation}
In the chiral representation,
the chiral basis spinors $\bepsilon_a$ are column vectors with $1$ in the $a$th place,
zero elsewhere; for example
$\bepsilon_{\smallUpup} = \col ( 1 , 0 , 0 , 0 )$.
The basis spinors with boost and spin aligned,
$\bepsilon_{\smallUpup}$ and $\bepsilon_{\smallDowndown}$,
are right-handed, with positive chirality,
while spinors with boost and spin antialigned,
$\bepsilon_{\smallUpdown}$ and $\bepsilon_{\smallDownup}$,
are left-handed, with negative chirality.
The basis spinors satisfy a Lorentz-invariant antisymmetric Dirac scalar product
\begin{equation}
\label{spinormetric}
  \bepsilon_a \cdot \bepsilon_b
  =
  \varepsilon_{ab}
  \equiv
  \left(
  \begin{array}{cccc}
  0 & 1 & 0 & 0 \\
  -1 & 0 & 0 & 0 \\
  0 & 0 & 0 & -1 \\
  0 & 0 & 1 & 0
  \end{array}
  \right)
  \ .
\end{equation}

The spinor indices
$\{ \Uparrow , \Downarrow , \uparrow , \downarrow \}$
are spinor analogs of the Newman-Penrose vector indices
$\{ v , u , + , - \}$.
Boosting the tetrad frame by rapidity $\eta$ in the $vu$-plane
boosts basis spinors with boost index $\Uparrow$ and $\Downarrow$ by
(compare equation~(\ref{gammavuboost}))
\begin{equation}
  \bepsilon_{\smallUp} \rightarrow \ee^{\eta / 2} \bepsilon_{\smallUp}
  \ , \quad
  \bepsilon_{\smallDown} \rightarrow \ee^{- \eta / 2} \bepsilon_{\smallDown}
  \ ,
\end{equation}
while spatially rotating the tetrad frame by angle $\zeta$ in the $+-$-plane
rotates basis spinors with spin index $\uparrow$ and $\downarrow$ by
(compare equation~(\ref{gammapmspin}))
\begin{equation}
  \bepsilon_{\smallup} \rightarrow \ee^{- \im \zeta / 2} \bepsilon_{\smallup}
  \ , \quad
  \bepsilon_{\smalldown} \rightarrow \ee^{\im \zeta / 2} \bepsilon_{\smalldown}
  \ .
\end{equation}
%The boost weights of covariant $\Uparrow$ and $\Downarrow$ indices
%are $\frac{1}{2}$ and $- \frac{1}{2}$,
%while the spin weights of covariant $\uparrow$ and $\downarrow$ indices
%are $\frac{1}{2}$ and $- \frac{1}{2}$.
The boost and spin weights of a spinor tensor
can be determined by inspection, according to the rules
\begin{subequations}
\label{boostspinrulesspin}
\begin{align}
  &
  \mbox{boost weight}
\\
\nonumber
  &
  =
  \mbox{$\tfrac{1}{2}$ number of $\Uparrow$ minus number of $\Downarrow$ covariant indices}
  \ ,
\\
  &
  \mbox{spin weight}
\\
\nonumber
  &
  =
  \mbox{$\tfrac{1}{2}$ number of $\uparrow$ minus number of $\downarrow$ covariant indices}
  \ .
\end{align}
\end{subequations}
Contravariant indices count oppositely to covariant indices.

The spin~$\tfrac{3}{2}$ fields considered in \S\ref{spinthreehalves-sec}
carry both vector and spinor indices.
The boost (spin) weight of a vector-spinor tensor
is just the sum of the boost (spin) weights of the vector and spinor indices,
equations~(\ref{boostspinrules}) and~(\ref{boostspinrulesspin}).

\section{Wave equations}
\label{waveeqs-sec}

The bulk of this paper, \S\S\ref{scalar-sec}--\ref{spintwo-sec},
is devoted to deriving wave equations
for massless fields for each of spins
$s = 0$, $\tfrac{1}{2}$, $1$, $\tfrac{3}{2}$, and $2$
(in this paper, $s$ is always positive).
This section summarizes the results,
giving expressions valid for any of the spins
0, $\tfrac{1}{2}$, $1$, $\tfrac{3}{2}$, or $2$.

%The spin $s$ of a field characterizes how it transforms
%under local Lorentz transformations.

For brevity and clarity,
many of the equations in this paper
carry upper and lower indices,
such as \raisebox{-1pt}{$\vu$} and $\pm$ in equations~(\ref{partialvupm});
unless otherwise stated,
such equations should be interpreted in the `natural' way
as pairs of equations in which all upper indices apply to the upper equation,
and all lower indices to the lower equation.

Appendix~\ref{Dapp}
takes a deeper dive into the derivation of the wave equations
in a general spacetime,
clarifying the origin of the wave equations in the conformally separable
spacetimes considered in this paper.

\subsection{Petrov type D}

The conformally separable black-hole spacetimes are Petrov type~D, meaning that
the only nonvanishing component of the Weyl tensor is its
boost and spin weight zero component.
%In the conformally separable black-hole spacetimes,
The right-handed boost/spin weight zero Weyl component
$\Cz_0 \equiv \Cz_{vuvu}$ defined by equation~(\ref{Czcomponent}) is
\begin{align}
\label{Cz0}
  \rho^2 \Cz_0
  &
  =
  \frac{1}{12}
  \left(
  {\ddsq ( R^4 \Delta_x ) \over \dd r^2}
  +
  {\ddsq \Delta_y \over \dd y^2}
  \right)
\\
\nonumber
  &\quad
  +
  \frac{1}{2} \,
  \rhom
  \left[
  {\partial \over \partial r} \left(
  R^4 \Delta_x {\partial \over \partial r}
  \right)
  +
  {\partial \over \partial y} \left(
  \Delta_y {\partial \over \partial y}
  \right)
  \right]
  \rhom^{-1}
  \ .
\end{align}
The left-handed component is the same with
$\rhom \rightarrow \rhom^\ast$
on the second line,
so is the complex conjugate of the right-handed component.

\cite{TorresdelCastillo:1988}
has emphasized that all vacuum spacetimes of Petrov type~D
satisfy decoupled wave equations that can be solved by separation of variables.
The conformally separable black hole spacetimes are type~D,
but they are not vacuum;
rather the conformally separable spacetimes are sourced by
collisionless outgoing and ingoing streams
that produce an exponentially growing proper energy-momentum
in the center-of-mass frame.

\subsection{Frequency and azimuthal mode number}

Conformal time translation symmetry and axisymmetry
of the background
imply that wave amplitudes $\psi$ can be expanded in modes
of definite conformal frequency $\wel$ and azimuthal mode number $m$:
\begin{equation}
\label{psiwm}
  \psi
  =
  \ee^{- \im ( \wel t + m \phi )}
  \Psi ( x , y )
  \ .
\end{equation}
The signs of $\wel$ and $m$ accord with the convention
that positive frequency $\wel$ corresponds to positive energy,
and that a wave of azimuthal angular momentum $m$ varies as
$\ee^{- \im m \phi}$ under a right-handed spatial rotation.

Periodicity requires $m$ to be integral for integral spin $s$,
or half-integral for half-integral spin $s$.
Solutions with real frequency $\wel$ define normal modes.
The frequency $\wel$ can also be complex.
Of particular interest in gravitational wave astronomy
are quasinormal modes,
a discrete spectrum of complex frequencies for each angular mode $\ell m$,
corresponding to long-lived modes of decay of a perturbed black hole
\cite{Kokkotas:1999,Berti:2009,Yang:2012,Teukolsky:2015}.

Acting on modes $\psi$, equation~(\ref{psiwm}),
of definite (possibly complex) frequency $\wel$ and azimuthal number $m$,
the Newman-Penrose directed derivatives $\partial_k$ are related to
coordinate derivatives by
\begin{subequations}
\label{partialvupm}
\begin{align}
\label{partialvu}
  \sqrt{2}
  \rho \,
  \partial_\vu
  \psi
  &=
  \mkern4mu\pmu\mkern-4mu
  {1 \over \sqrt{|\Delta_x|}}
  \left( \mp \Delta_x {\partial \over \partial x} + \im \alpha_x \right)
  \psi
  \ ,
\\
  \sqrt{2}
  \rho \,
  \partial_\pm
  \psi
  &=
  {1 \over \sqrt{\Delta_y}}
  \left( \Delta_y {\partial \over \partial y} \pm \alpha_y \right)
  \psi
  \ ,
\end{align}
\end{subequations}
where the initial $\pmu$ sign in equation~(\ref{partialvu})
is $+$ for $\partial_v$,
and $+$ outside, $-$ inside,
the horizon for $\partial_u$,
and where $\alpha_x$ and $\alpha_y$ are defined by
\begin{equation}
\label{alphaxy}
  \alpha_x
  \equiv
  \wel + m \, \omega_x
  \ , \quad
  \alpha_y
  \equiv
  \wel \, \omega_y + m
  \ .
\end{equation}
It is evident that the radial differential operators
$\rho \partial_\vu$ are purely radial,
in the sense that they involve a derivative with respect to
the radial coordinate $x$ and not the angular coordinate $y$,
while the angular differential operators
$\rho \partial_\pm$ are purely angular,
in the sense that they involve a derivative with respect to
the angular coordinate $y$ and not the radial coordinate $x$.

\subsection{Chirality}

The chiral operator $\gamma_5$, equation~(\ref{gamma5}),
is a Lorentz-invariant pseudoscalar.
Waves of massless fields of nonzero spin can be decomposed
into a sum of independently evolving
right- and left-handed chiral components,
which are eigenstates of $\gamma_5$ with
eigenvalues $+1$ and $-1$ respectively.
Physically, the right- and left-handed chiral components
correspond to waves in which the spin axis is
respectively aligned and antialigned with the boost axis.

%IN SENSE CONJUGATE ANGULAR INDICES $+ <-> -$ AND $f_s$, $\rhom$.
%As remarked following equation~(\ref{fs}),
%components of opposite chirality satisfy complex conjugate equations,
%Thus it suffices to solve for one chiral component,
%and then its complex conjugate supplies
%a solution for the opposite chiral component.
%In \S\ref{spinhalf-sec} on spin~$\frac{1}{2}$ waves,
%both right-handed and left-handed chiralities are
%initially retained, to illustrate what happens,
%but in later sections for higher spin waves,
%the field is immediately projected into its right-handed chiral component.

A wave of given chirality and spin $s$
has $2s+1$ components, with boost weights
$\sigma = -s , -s{+}1 , ... , s$,
and spin weights $\varsigma$ either equal
($\varsigma = \sigma$, right-handed chirality)
or opposite
($\varsigma = -\sigma$, left-handed chirality)
to the boost weight.
The $2s+1$ different components are coupled by equations of motion,
so are not independent of each other,
but rather oscillate in harmony.
In flat space far from the black hole,
a component of spin $s$ and boost weight $\sigma$ falls off with radius as,
equation~(\ref{psifarflat}),
\begin{equation}
\label{psir}
  \psi_{\sigma} \sim r^{-1-s\mp\sigma}
  \ ,
\end{equation}
where the $\mp$ sign is $-$ for outgoing waves, $+$ for ingoing waves,
that is, the sign of $\sigma$ is opposite to the direction of motion
(the boost direction).
Only the largest component survives far from the black hole,
satisfying $\psi_{\sigma} \sim r^{-1}$.
The large component is called the propagating component of the wave.
Each of the outgoing and ingoing components
has either of two chiralities,
with spin weight $\varsigma$ equal (right-handed)
or opposite (left-handed) to the boost weight $\sigma$.

\subsection{Boost and spin raising and lowering operators}

The boost and spin weight of a field
can be read off from its covariant chiral indices,
equations~(\ref{boostspinrules}) and~(\ref{boostspinrulesspin}).
Operating on a field with one of the Newman-Penrose directed derivatives
$\partial_k$ yields an object
whose boost weight (if $k = v$ or $u$) or spin weight (if $k = +$ or $-$)
differs by $\pm 1$ from that of the field.
In effect, the Newman-Penrose directed derivatives
`raise' and `lower' the boost and spin weights of a field.
However,
the directed radial derivatives 
$\partial_v$ and $\partial_u$
do not commute with the directed angular derivatives
$\partial_+$ and $\partial_-$.

It is advantageous to define modified versions of the
Newman-Penrose directed derivatives
with the property that the radial (boost) derivatives
commute with the angular (spin) derivatives,
besides having the property that,
like the directed derivatives $\rho \partial_k$, equations~(\ref{partialvupm}),
when acting on modes of definite frequency $\wel$ and azimuthal number $m$,
the radial derivatives are purely radial,
and the angular derivatives are purely angular.
Define therefore boost and spin raising and lowering operators
$\ethSi{\sigma}{k}$
by
\cite{Newman:1962,Goldberg:1967,Geroch:1973}
(see Appendix~\ref{Dapp} for an exposition of
how these operators are defined in a general spacetime)
\begin{subequations}
\label{ethsi}
\begin{align}
\label{ethsivu}
  \ethSi{\sigma}{\vu}
  &\equiv
  \mkern4mu\pmu\mkern-4mu
  \sqrt{2}
  %\left( \sqrt{| \Delta_x |} \over \omega_x \right)^{\pm \sigma}
  \left( R^2 \sqrt{| \Delta_x |} \right)^{\pm \sigma} \!
  \rho \,
  \partial_{\overset{\scriptstyle v}{\scriptstyle u}}
  %\left( \sqrt{| \Delta_x |} \over \omega_x \right)^{\mp \sigma}
  \left( R^2 \sqrt{| \Delta_x |} \right)^{\mp \sigma}
  \ ,
\\
\label{ethsipm}
  \ethSi{\varsigma}{\pm}
  &\equiv
  \pm
  \sqrt{2}
  \left( \sqrt{\Delta_y} \right)^{\pm \varsigma} \!
  \rho \,
  \partial_\pm
  \left( \sqrt{\Delta_y} \right)^{\mp \varsigma}
  \ ,
\end{align}
\end{subequations}
where the initial $\pmu$ sign in~(\ref{ethsivu})
is $+$ for the raising operator $\ethSi{\sigma}{v}$,
and $+$ outside, $-$ inside,
the horizon for the lowering operator $\ethSi{\sigma}{u}$.
The operators~(\ref{ethsi}) are constructed so that
the boost raising and lowering operators
$\ethSi{\sigma}{v}$ and
$\ethSi{\sigma}{u}$
commute with the spin raising and lowering operators
$\ethSi{\varsigma}{+}$ and
$\ethSi{\varsigma}{-}$,
for arbitrary boosts $\sigma$ and spins $\varsigma$.
%The raising and lowering operators~(\ref{ethsi})
%transform the $2s+1$ components of a spin~$s$ wave among each other,
%increasing and decreasing by 1
%the boost weight $\sigma$ (eq.~(\ref{ethsivu}))
%and the spin weight $\varsigma$ (eq.~(\ref{ethsipm}))
%of the component.
%(as remarked above,
%$\sigma = \varsigma$ for right-handed,
%$\sigma = -\varsigma$ for left-handed chirality).
The initial signs in equations~(\ref{ethsi})
ensure that the raising and lowering operators
are Hermitian conjugates of each other,
equations~(\ref{hermitianconjugate}).
The $\sqrt{2}$ factor brings the operators to conventional normalization.
The definition~(\ref{ethsivu}) of the boost raising and lowering operators
holds both outside the horizon,
where the horizon function is positive,
$\Delta_x > 0$,
and inside the horizon,
where the horizon function is negative,
$\Delta_x < 0$.
%The contravariant components of the boost and spin raising and lowering operators are
%$\ethusi{\sigma}{\vu} = - \ethsi{\sigma}{\uv}$.

The boost/spin index $\sigma$ or $\varsigma$ on
the boost/spin raising and lowering operators
%$\ethSi{\sigma}{k}$
$\ethSi{\sigma}{\vu}$ or $\ethSi{\varsigma}{\pm}$
is often suppressed in this paper,
since it
%The exponent $s$ in the boost or spin
%raising and lowering operators~(\ref{ethsi})
equals the boost/spin weight of the object being operated on.
The boost and spin raising and lowering operators
respectively
raise and lower by one the boost and spin weight
of the object they are operating on.

Acting on modes~(\ref{psiwm}) of given frequency $\wel$ and azimuthal mode $m$,
the boost raising and lowering operators~(\ref{ethsivu})
connecting adjacent boost weights are Hermitian conjugates
with respect to the integration measure $\dd r$
over a suitable integration interval;
and likewise
the spin raising and lowering operators~(\ref{ethsipm})
connecting adjacent spin weights are Hermitian conjugates
with respect to the integration measure $\dd y$
over the interval $[ -1 , 1 ]$:
\begin{subequations}
\label{hermitianconjugate}
\begin{align}
\label{hermitianconjugateboost}
  \ethSi{\sigma}{v}^\dagger
  &=
  \ethSi{\sigma+1}{u}
  \ ,
\\
\label{hermitianconjugatespin}
  \ethSi{\varsigma}{+}^\dagger
  &=
  \ethSi{\varsigma+1}{-}
  \ .
\end{align}
\end{subequations}
Hermitian conjugacy of the spin raising and lowering operators
follows from the fact that for any
differentiable functions $\chi$ and $\psi$
of the same definite (possibly complex) frequency $\wel$ and azimuthal mode $m$,
an integration by parts shows that
\begin{equation}
  \int
  \chi
  ( \ethSi{\varsigma}{+}
  \psi ) \,
  \dd y
  =
  \left[
  \sqrt{\Delta_y} \, \chi \psi
  \right]
  +
  \int
  ( \ethSi{\varsigma+1}{-}
  \chi )
  \psi \,
  \dd y
  \ .
\end{equation}
The surface term vanishes
if the integration interval is $[ -1 , 1 ]$,
since the polar function $\Delta_y$ vanishes at these limits.
Similarly,
Hermitian conjugacy of the boost raising and lowering operators
follows from an analogous integration by parts,
\begin{equation}
  \int
  \chi
  ( \ethSi{\sigma}{v}
  \psi ) \,
  \dd r
  =
  \mp
  \left[
  R^2 \sqrt{| \Delta_x |} \chi \psi
  \right]
  +
  \int
  ( \ethSi{\sigma+1}{u}
  \chi )
  \psi \,
  \dd r
  \ ,
\end{equation}
where the sign $\mp$ on the surface term is
$-$ outside the horizon ($\Delta_x > 0$),
$+$ inside the horizon ($\Delta_x < 0$).
The surface term vanishes at horizons,
where $\Delta_x = 0$,
and also at infinity provided that the functions
$\chi$ and $\psi$ decrease sufficiently rapidly at infinity.

\subsection{Wave operators}

As remarked after equation~(\ref{psir}),
the propagating component of a spin~$s$ wave
is the component with the most negative boost weight
$\sigma$ along the direction of motion,
$\sigma = -s$ in the outgoing direction,
and $\sigma = +s$ in the ingoing direction.
For propagating waves of right-handed chirality,
the spin weight equals the boost weight,
$\varsigma = \sigma$,
and the spin~$s$ wave equations derived
in \S\ref{scalar-sec}-\S\ref{spintwo-sec}
are of the form
\begin{equation}
\label{waveequations}
  \left(
  \Sqvu{\pm s}
  -
  \Sqpm{\pm s}
  \right)
  \hatpsi_{\pm s}
  =
  0
  \ ,
\end{equation}
where the wave operators $\Sq{\sigma}{kl}$
are defined by equations~(\ref{waveoperatorsraiselower}).
%The top row of indices in equation~(\ref{waveequations}) is for ingoing,
%%($\sigma = \varsigma = s$),
%the bottom row for outgoing waves.
%($\sigma = \varsigma = -s$).
For left-handed chirality,
the spin weight is minus the boost weight,
$\varsigma = -\sigma$,
and $\Sqpm{\pm s} \rightarrow \Sqmp{\mp s}$
in the wave equation~(\ref{waveequations}).
Components with general boost weights $\sigma$ satisfy wave equations similar
to~(\ref{waveequations}),
but with the addition of a term proportional to the boost/spin weight zero
component $\Cz_0$ of the Weyl tensor, equation~(\ref{Cz0}),
\begin{equation}
\label{waveequationsCz}
  \left(
  \Sqvu{\sigma}
  -
  \Sqpm{\sigma}
  +
  c_{s|\sigma|} \rho^2 \Cz_0
  \right)
  \hatpsi_{\sigma}
  =
  0
  \ ,
\end{equation}
in which $c_{s|\sigma|}$ is a constant that depends on the spin $s$
and on the absolute value $|\sigma|$ of the boost weight of the component
(see equations~(\ref{spinonewavesq0a}), (\ref{spinonewavesq0b}),
(\ref{spinthreehalfwavesqp12a})--(\ref{spinthreehalfwavesqm12b}),
and
(\ref{spintwowavesq0a}), (\ref{spintwowavesq0b})).
The constant $c_{s|\sigma|}$ vanishes if $|\sigma| = s$,
as is true for propagating waves and their complements of opposite boost,
in which case the wave equation reduces to~(\ref{waveequations}).
If $\rho^2 \Cz_0$
were a separated sum of radial and angular coordinates $x$ and $y$,
then the wave equations~(\ref{waveequationsCz}) for general boost weights
would indeed become separable,
but the term on the second line of the expression~(\ref{Cz0}) for $\rho^2 \Cz_0$
is a mixed function of radial and angular coordinates, preventing separability.
%The wave equations~(\ref{waveequations}) hold
%not only for the propagating components but also for non-propagating
%components of the most extreme boost and spin weight.
%Thus the top row of indices in equation~(\ref{waveequations})
%holds with $\sigma = \varsigma = -s$
%for both propagating (outgoing) and non-propagating (ingoing) waves,
%while the bottom row holds with $\sigma = \varsigma = +s$
%for both propagating (ingoing) and non-propagating (outgoing) waves.

The scaled wave amplitude $\hatpsi_{\sigma}$ (with a hat)
in equation~(\ref{waveequations})
is related to the native wave amplitude $\psi_{\sigma}$ by
\begin{equation}
\label{psifs}
  \psi_{\sigma}
  =
  \left\{
  \begin{array}{ll}
  f_s
  \hatpsi_{\sigma}
  &
  \mbox{right chirality}
  \\[1ex]
  f_s^\ast
  \hatpsi_{\sigma}
  &
  \mbox{left chirality}
  \end{array}
  \right.
  \ ,
\end{equation}
where (note $s$ is positive)
\begin{equation}
\label{fs}
  f_s \equiv
  {1 \over
  \rhom{}^s
  \rhoi^{s+1}}
  \ ,
\end{equation}
with $\rhoi$ the inflationary conformal factor defined by
equation~(\ref{rho}),
and $\bar\rho$ the complex conformal factor
defined by equation~(\ref{rhopm}).
The scaling factors $f_s$ and $f_s^\ast$
for right- and left-handed chiralities
are complex conjugates of each other.
%NOT TRUE FOR QUASINORMAL MODES, WHERE $\wel$ IS COMPLEX
%As seen below,
%%equations~(\ref{waveoperatorscoordinate}) or~(\ref{altwaveoperatorscoordinate}),
%equation~(\ref{altwaveoperatorscoordinate}),
%the radial wave operators $\square_{uv}$ and $\square_{vu}$
%are complex conjugates of each other.

The wave equations~(\ref{waveequations}) admit separated solutions
($\varsigma = \pm\sigma$ right/left-handed)
\begin{equation}
\label{psis}
  \hatpsi_{\sigma}
  =
  \ee^{- \im ( \wel t + m \phi )}
  X_\sigma (x) Y_\varsigma(y)
  \ ,
\end{equation}
the equations~(\ref{waveequations}) separating as
%(top row outgoing, bottom row ingoing)
\begin{equation}
\label{waveequationssep}
  \bigl(
  \Sqvu{\sigma}
  -
  \lambda_{\sigma}
  \bigr)
  \hatpsi_{\sigma}
  =
  0
  \ , \ \ 
  \bigl(
  \Sqpm{\sigma}
  -
  \lambda_{\sigma}
  \bigr)
  \hatpsi_{\sigma}
  =
  0
  \ ,
\end{equation}
for some eigenvalues $\lambda_{\sigma}$.
For normal modes the frequency $\wel$ is real,
in which case the angular wave operators $\square_{+-}$ and $\square_{-+}$,
equation~(\ref{waveoperatorsraiselowermp}),
are Hermitian,
so the eigenvalues $\lambda_{\sigma}$ are real.
For quasinormal modes the frequency $\wel$ is complex,
and the eigenvalues $\lambda_{\sigma}$ are complex.
%(the sign in front of the eigenvalues
%in equations~(\ref{waveequationssep})
%is chosen so that the eigenvalues $\lambda_\sigma$ are generally positive).
As demonstrated in \S\ref{scalar-sec}--\S\ref{spintwo-sec}
for spins $s = 0$, $\tfrac{1}{2}$, $1$, $\tfrac{3}{2}$, and $2$,
in terms of the boost and spin raising and lowering operators
$\ethsi{\sigma}{k}$ defined by equations~(\ref{ethsi}),
radial and angular spin~$s$ wave operators $\square_{kl}$
acting on waves $\hat{\psi}_{\sigma}$
of boost $\sigma$ are
(spin weight $\varsigma = +\sigma$ right-handed,
$\varsigma = -\sigma$ left-handed)
%\begin{widetext}
\begin{subequations}
\label{waveoperatorsraiselower}
\begin{align}
\label{waveoperatorsraiseloweruv}
  \Sqvu{\sigma}
  \hatpsi_{\sigma}
  &\equiv
  \biggl[
  - \sgn(\Delta_x)
  \ethSi{\sigma\mp1}{\overset{\scriptstyle v}{\scriptstyle u}}
  \,
  \ethSi{\sigma}{\overset{\scriptstyle u}{\scriptstyle v}}
  +
  4
  \im
  \left( \sigma \mp \tfrac{1}{2} \right)
  %( 2 \sigma \mp 1 )
  %\wel {\dd \ln \omega_x \over \dd x}
  \wel r
\nonumber
\\
  &
  \quad\ \ 
  +
  \tfrac{1}{3}
  \left( \sigma \mp \tfrac{1}{2} \right)
  \left( \sigma \mp 1 \right)
  {\ddsq ( R^4 \Delta_x ) \over \dd r^2}
  \biggr]
  \hatpsi_{\sigma}
  \ ,
\\
\label{waveoperatorsraiselowermp}
  \Sqpm{\varsigma}
  \hatpsi_{\varsigma}
  &\equiv
  \biggl[
  \ethSi{\varsigma\mp1}{\pm}
  \,
  \ethSi{\varsigma}{\mp}
  -
  4
  \left( \varsigma \mp \tfrac{1}{2} \right)
  %( 2 \varsigma \mp 1 )
  %\wel {\dd \omega_y \over \dd y}
  \wel a y
\nonumber
\\
  &
  \quad\ \ 
  -
  \tfrac{1}{3}
  \left( \varsigma \mp \tfrac{1}{2} \right)
  \left( \varsigma \mp 1 \right)
  {\ddsq \Delta_y \over \dd y^2}
  \biggr]
  \hatpsi_{\varsigma}
  \ .
\end{align}
\end{subequations}
Note that the indices $\sigma$ or $\varsigma$ on the radial and angular
wave operators $\Sq{\sigma}{}$ and $\Sq{\varsigma}{}$
are always equal to the boost/spin weights $\sigma$ and $\varsigma$
of the field $\hatpsi$ that they operate on,
so could be omitted for brevity.
%NO
%Since $\sigma$ and $\varsigma$ are taken positive
%in these equations~(\ref{waveoperatorsraiselower}),
%the factors involving $\sigma$ and $\varsigma$ on the right hand sides
%should be understood as involving the absolute values $|\sigma\|$
%and $|\varsigma|$ of the boost and spin weight.
For normal modes, for which by definition the frequency $\wel$ is real,
each of the angular operators $\square_{+-}$ and $\square_{-+}$ is Hermitian
thanks to the Hermitian conjugacy of the angular operators
$\ethsi{}{+}$ and $\ethsi{}{-}$, equation~(\ref{hermitianconjugatespin}).
The radial operators $\square_{vu}$ and $\square_{uv}$
on the other hand are Hermitian only if the imaginary term
$4 \im \left( \sigma \mp \tfrac{1}{2} \right)$
in equation~(\ref{waveoperatorsraiseloweruv}) vanishes;
thus $\square_{vu}$ is Hermitian only for
$\sigma = +\tfrac{1}{2}$,
while $\square_{uv}$ is Hermitian only for
$\sigma = -\tfrac{1}{2}$.

Equations~(\ref{sqsR}) and~(\ref{sqsL}) in Appendix~\ref{Dapp}
give expressions for the difference
$\sqvu - \sqpm$
of radial and angular wave operators~(\ref{waveoperatorsraiselower})
which enter the wave equations~(\ref{waveequations})
and~(\ref{waveequationsCz}).

Operating on separated solutions~(\ref{psis}),
the boost and spin raising and lowering operators~(\ref{ethsi}) yield
\begin{subequations}
\label{ethsired}
\begin{align}
\label{ethsiredvu}
  \ethSi{\sigma}{\overset{\scriptstyle u}{\scriptstyle v}}
  \hatpsi_{\sigma}
  &=
  \ee^{- \im ( \wel t + m \phi )}
  Y_\varsigma(y)
  \left(
  \ethSi{\sigma}{\overset{\scriptstyle u}{\scriptstyle v}}
  -
  {\im \alpha_x \over \sqrt{| \Delta_x |}}
  \right)
  X_\sigma(x)
  \ ,
\\
\label{ethsiredpm}
  \ethSi{\varsigma}{\mp}
  \hatpsi_{\varsigma}
  &=
  \ee^{- \im ( \wel t + m \phi )}
  X_\sigma(x)
  \left(
  \ethSi{\varsigma}{\mp}
  +
  {\alpha_y \over \sqrt{\Delta_y}}
  \right)
  Y_\varsigma(y)
  \ ,
\end{align}
\end{subequations}
where $\alpha_x$ and $\alpha_y$ are defined by equation~(\ref{alphaxy}).
In terms of coordinate derivatives,
the boost and spin raising and lowering operators
acting on functions $X_\sigma(x)$ of boost weight $\sigma$
and $Y_\varsigma(y)$ of spin weight $\varsigma$
are, with $x^\ast$ and $y^\ast$ the tortoise coordinates
defined by equations~(\ref{xytortoise}),
\begin{subequations}
\label{ethsisep}
\begin{align}
\label{ethsisepvu}
  \mkern-1mu
  \ethsi{\sigma}{\overset{\scriptstyle v}{\scriptstyle u}}
  X_\sigma(x)
  &
  =
  {1 \over \sqrt{| \Delta_x |}}
  \left(
  \pm
  {\dd \over \dd x^\ast}
  -
  {\sigma \over 2}
  {\dd \ln \left( R^4 | \Delta_x | \right)
  \over \dd x^\ast}
  \right)
  X_\sigma(x)
  \ ,
%\\
%\nonumber
%  &
%  =
%  \mkern4mu\pmu\mkern-4mu
%  \pm
%  {1 \over \sqrt{| \Delta_x |}}
%  \left( R^4 | \Delta_x | \right)^{\sigma/2}
%  {\dd \over \dd x^\ast}
%  \left( R^4 | \Delta_x | \right)^{- \sigma/2}
%  X_\sigma(x)
%  \ ,
\\
\label{ethsiseppm}
  \mkern-1mu
  \ethsi{\varsigma}{\pm}
  Y_\varsigma(y)
  &
  =
  {1 \over \sqrt{\Delta_y}}
  \left(
  \pm
  {\dd \over \dd y^\ast}
  -
  {\varsigma \over 2}
  {\dd \ln \Delta_y \over \dd y^\ast}
  \right)
  \,
  Y_\varsigma(y)
  \ .
%\nn
%  &
%  =
%  \pm
%  {1 \over \sqrt{\Delta_y}}
%  \Delta_y^{\varsigma/2}
%  {\dd \over \dd y^\ast}
%  \Delta_y^{- \varsigma/2}
%  \,
%  Y_\varsigma(y)
%  \ ,
\end{align}
\end{subequations}
%Outside the outer horizon $r_+$,
%the tortoise coordinate $x^\ast$ increases outward,
%from $- \infty$ at the horizon $r_+$.
%Inside the outer horizon, the tortoise coordinate $x^\ast$
%increases inward,
%from $- \infty$ at the horizon $r_+$.
In terms of coordinate derivatives,
the wave equations~(\ref{waveequationssep})
acting on separated solutions~(\ref{psis}) for
propagating
outgoing ($\sigma = -s$) and ingoing ($\sigma = s$)
right-handed
($\varsigma = \sigma$)
and left-handed
($\varsigma = -\sigma$)
waves reduce to
\begin{subequations}
\label{altwaveoperatorscoordinate}
\begin{align}
\label{altwaveoperatorscoordinatex}
%  \left(
%  \squv
%  -
%  \lambda_{\sigma\varsigma}
%  \right)
%  \hat{\psi}_{\sigma\varsigma}
%  &=
%  \ee^{- \im ( \wel t + m \phi )}
%  Y_\varsigma(y)
  {1 \over R \Delta_x}
  \left[
  {\ddi{2} \over \dd x^{\ast 2}}
  +
  \left(
  %\alpha_x + {\im \sigma \over 2} {\dd \Delta_x \over \dd x}
  \alpha_x - {\im \sigma \over 2} {\dd \ln|\Delta_x| \over \dd x^\ast}
  \right)^2
  \!
  +
  V_\sigma
  \right]
  R X_\sigma
  &=
  0
  \ ,
\\
\label{altwaveoperatorscoordinatey}
%  \left(
%  \sqmp
%  -
%  \lambda_{\sigma\varsigma}
%  \right)
%  \hat{\psi}_{\sigma\varsigma}
%  &=
  -
%  \ee^{- \im ( \wel t + m \phi )}
%  X_\sigma(x)
  {1 \over \Delta_y}
  \left[
  {\ddi{2} \over \dd y^{\ast 2}}
  -
  \left(
  %\alpha_y - {\varsigma \over 2} {\dd \Delta_y \over \dd y}
  \alpha_y - {\varsigma \over 2} {\dd \ln\Delta_y \over \dd y^\ast}
  \right)^2
  \!
  +
  W_\varsigma
  \right]
  Y_\varsigma
  &=
  0
  \ ,
\end{align}
\end{subequations}
where the radial and angular potentials $V_\sigma$ and $W_\varsigma$ are
\begin{subequations}
\label{altpotential}
\begin{align}
\label{altpotentialx}
  V_\sigma
  &=
  \biggl[
  \tfrac{1}{6}
  ( 1 + 2 \sigma^2 )
  {\ddi{2} \Delta_x \over \dd x^2}
  -
  \tfrac{1}{3}
  ( 1 - 4 \sigma^2 )
  a^2 \Delta_x
\\
\nonumber
  &
  \mkern154mu
  - 2 \im \sigma {\dd \alpha_x \over \dd x}
  %- {4 \im \sigma m a r \over R^2}
  -
  \lambda_{\sigma\varsigma}
  \biggr]
  \Delta_x
  \ ,
\\
\label{altpotentialy}
  W_\varsigma
  &=
  \left[
  \tfrac{1}{6}
  ( 1 + 2 \varsigma^2 )
  {\ddi{2} \Delta_y \over \dd y^2}
  - 2 \varsigma {\dd \alpha_y \over \dd y}
  %+ 4 \varsigma \wel a y
  +
  \lambda_{\sigma\varsigma}
  \right]
  \Delta_y
  \ ,
\end{align}
\end{subequations}
the derivatives of $\alpha_x$ and $\alpha_y$
defined by equations~(\ref{alphaxy}) being
\begin{equation}
  {\dd \alpha_x \over \dd x}
  %=
  %m {\dd \omega_x \over \dd x}
  =
  %2 m r \omega_x
  {2 m a r \over R^2}
  \ , \quad
  {\dd \alpha_y \over \dd y}
  %=
  %\wel {\dd \omega_y \over \dd y}
  =
  - 2 \wel a y
  \ .
\end{equation}
Equations~(\ref{altwaveoperatorscoordinate})
with the potentials~(\ref{altpotential})
constitute the generalization of the Teukolsky master equation
\cite{Teukolsky:1972my,Teukolsky:1973ha,Teukolsky:2015}
to the conformally separable solutions
for accreting, rotating black holes.
The angular eigenfunctions $Y_\varsigma$ are unchanged from those of
$\Lambda$-Kerr(-Newman) (stationary) black holes.

For Kerr(-Newman) (zero cosmological constant),
the angular eigenfunctions can be expressed
as spin-weighted spheroidal harmonics
\cite{Berti:2006} (with parameter $c = - \wel a$),
or
as confluent Heun functions
\cite{Fiziev:2010}.

For spherical black holes,
the angular eigenfunctions $Y_\varsigma$
reduce to spin-weighted spherical harmonics,
whose eigenvalues are
\begin{equation}
  \lambda_{\varsigma}
  =
  \ell ( \ell + 1 ) + \tfrac{1}{3} ( 1 - \varsigma^2 )
  \quad
  \mbox{(spherical)}
  \ ,
\end{equation}
with harmonic number $\ell = \ell_0 , \, \ell_0{+}1 , \, ...$
starting from $\ell_0 = \max(|\varsigma|,|m|)$.

The wave equations~(\ref{altwaveoperatorscoordinate})
with~(\ref{altpotential})
satisfy some discrete symmetries.
If the frequency $\wel$ is real
(or if $\wel$ is complex, but left unconjugated,
so that $\alpha_x$ and $\alpha_y$ are unconjugated),
then the wave equation~(\ref{altwaveoperatorscoordinatex}) for $X_{-\sigma}$
is the complex conjugate of that for $X_\sigma$.
The wave equations~(\ref{altwaveoperatorscoordinate})
are unchanged
if the boost and spin weights $\sigma$ and $\varsigma$ are flipped
and at the same time the signs of the frequency $\wel$
and azimuthal number $m$,
hence $\alpha_x$ and $\alpha_y$, are flipped,
%CONSISTENT WITH QFT
\begin{equation}
  \sigma \rightarrow - \sigma
  \ , \quad
  \varsigma \rightarrow - \varsigma
  \ , \quad
  \wel \rightarrow - \wel
  \ , \quad
  m \rightarrow - m
  \ .
\end{equation}
%reflecting the underlying symmetry of the conformal spacetime
%\begin{equation}
%  t \rightarrow - t
%  \ , \quad
%  \phi \rightarrow - \phi
%  \ .
%\end{equation}
%Thus if $\hatpsi_{s \wel m}$ is an eigenmode of the form~(\ref{psis}),
%then there is an eigenmode $\hatpsi_{-s}$ of opposite spin
%\begin{equation}
%\label{psims}
%  \hatpsi_{-s , -\wel, -m}
%  =
%  \ee^{\im ( \wel t + m \phi )}
%  X_\sigma(x) Y_\sigma(y)
%  \ ,
%\end{equation}
%with the same eigenfunctions $X_\sigma(x)$ and $Y_\sigma(y)$ as for $\hatpsi_s$,
%and the same eigenvalues,
%\begin{equation}
%  \lambda_{-s , -\wel , -m}
%  =
%  \lambda_{s , \wel , m}
%  \ .
%\end{equation}
%\end{widetext}

\subsection{Asymptotics}
\label{asymptotic-sec}

The radial wave equation~(\ref{altwaveoperatorscoordinatex}) simplifies
whenever the radial potential $V_\sigma$ is negligible,
\begin{equation}
\label{Vsmall}
  V_\sigma \sim 0
  \ .
\end{equation}
This happens if
the horizon function $\Delta_x$, equation~(\ref{DeltaxLKN}), is small,
and the angular momenta
(the azimuthal number $m$, and the eigenvalue $\lambda_{\varsigma}$ of
the angular wave equation~(\ref{altwaveoperatorscoordinatey}))
are not too large.
The horizon function $\Delta_x$ goes to zero
at horizons (inner, outer, or cosmological),
and at spatial infinity
in asymptotically flat space
(which happens if the cosmological constant is zero).
In the case of the inner horizon,
the term
$\Delta_x {\ddi{2} \Delta_x / \dd x^2}$
in the potential $V_\sigma$ grows large during inflation and collapse,
and must be retained;
this case is deferred to the end of this subsection~\ref{asymptotic-sec}.

When the potential $V_\sigma$ is negligible, WKB solution
of the radial wave equation~(\ref{altwaveoperatorscoordinatex}) gives
\begin{equation}
\label{XWKB}
  R X_\sigma
  \sim
  \ee^{\pm \im ( w x^\ast + m \omega_x^\ast )}
  | \Delta_x |^{\pm \sigma/2}
  \ ,
\end{equation}
where $\omega_x^\ast$ is an $\omega_x$-weighted tortoise coordinate,
\begin{equation}
  \omega_x^\ast
  \equiv
  - \int {\omega_x \, \dd x \over \Delta_x}
  \ .
\end{equation}

Consider first spatial infinity in asymptotically flat (Minkowski) space.
In this case the horizon function~(\ref{DeltaxLKN}) falls off as
$\Delta_x \sim 1/r^2$ as $r \rightarrow \infty$,
so that the WKB solution~(\ref{XWKB}) for the radial wave $X_\sigma$ is
\begin{equation}
\label{XWKBflat}
  X_\sigma
  \sim
  \ee^{\pm \im ( \wel x^\ast + m \omega_x^\ast )} r^{- 1 \mp \sigma}
  \quad
  r \rightarrow \infty
  \ .
\end{equation}
The Weyl tensor also goes to zero at infinity,
so the wave equations~(\ref{waveequationsCz})
for arbitrary boost weights $\sigma$ become separable,
and the WKB solution~(\ref{XWKBflat}) holds for arbitrary boost weight.
The upper sign in equation~(\ref{XWKBflat})
is for an outgoing wave,
where $\psi \sim \ee^{-\im \omega ( t - x^\ast )}$,
while the lower sign
is for an ingoing wave,
where $\psi \sim \ee^{-\im \omega ( t + x^\ast )}$.
The scaling factor $f_s$, equation~(\ref{psifs}),
contributes an additional factor of $r^{-s}$.
The net wavefunction $\psi_{\sigma}$, equation~(\ref{psis}),
of a component of boost weight $\sigma$ far from the black hole is
\begin{equation}
\label{psifarflat}
  \psi_{\sigma}
  \sim
  r^{-1-s}
  \left\{
  \begin{array}{ll}
  r^{-\sigma}
  \ee^{- \im \left[ \wel ( t - x^\ast ) + m ( \phi - \omega_x^\ast ) \right]}
  &
  \mbox{outgoing}
  \\
  r^{+\sigma}
  \ee^{- \im \left[ \wel ( t + x^\ast ) + m ( \phi + \omega_x^\ast ) \right]}
  &
  \mbox{ingoing}
  \end{array}
  \right.
  \ .
\end{equation}
The propagating component is the one that falls off most slowly at infinity,
so the propagating wave
has $\sigma = -s$ for an outgoing wave
and $\sigma = +s$ for an ingoing wave,
as already claimed following equation~(\ref{psir}).
For the propagating wave, the radial factor is $\psi_\sigma \sim r^{-1}$.
Equation~(\ref{psifarflat})
holds for both right-handed ($\varsigma = +\sigma$)
and left-handed ($\varsigma = -\sigma$) chiralities.

Now consider waves in the vicinity of the outer horizon,
or the cosmological horizon if the cosmological constant $\Lambda$ is positive.
Near a horizon, the propagating wavefunctions $\psi_{\sigma}$ are
(${\sigma = -s}$ outgoing, ${\sigma = +s}$ ingoing)
\begin{equation}
\label{psihor}
  \psi_{\sigma}
  \sim
  | \Delta_x |^{-s/2}
  \left\{
  \begin{array}{ll}
  \ee^{- \im \left[ \wel ( t - x^\ast ) + m ( \phi - \omega_x^\ast ) \right]}
  &
  \mbox{outgoing}
  \\
  \ee^{- \im \left[ \wel ( t + x^\ast ) + m ( \phi + \omega_x^\ast ) \right]}
  &
  \mbox{ingoing}
  \end{array}
  \right.
  \ ,
\end{equation}
which diverge as
$| \Delta_x |^{-s/2}$
for both outgoing and ingoing waves.
A tensor of boost weight $\sigma$ is multiplied by
$\ee^{\sigma \eta}$
under a boost by rapidity $\eta$ in the $vu$-plane.
The rapidity $\eta$ is positive for an outward boost,
negative for an inward boost.
The divergence of the propagating components at the outer horizon
can be removed by an outward boost of the outgoing wave
by boost factor $\ee^{\eta} = | \Delta_x |^{-1 / 2}$,
and by an inward boost of the ingoing wave
by boost factor $\ee^{\eta} = | \Delta_x |^{1 / 2}$,
\begin{align}
\label{psiboosted}
  &
  \psi_{\sigma}
  \sim
\\
\nonumber
  &
  \left\{
  \begin{array}{ll}
  \ee^{- \im \left[ \wel ( t - x^\ast ) + m ( \phi - \omega_x^\ast ) \right]}
  &
  \mbox{out wave, out frame}
  \\
  \ee^{- \im \left[ \wel ( t - x^\ast ) + m ( \phi - \omega_x^\ast ) \right]}
  | \Delta_x |^{-s}
  &
  \mbox{out wave, in frame}
  \\
  \ee^{- \im \left[ \wel ( t + x^\ast ) + m ( \phi + \omega_x^\ast ) \right]}
  &
  \mbox{in wave, in frame}
  \\
  \ee^{- \im \left[ \wel ( t + x^\ast ) + m ( \phi + \omega_x^\ast ) \right]}
  | \Delta_x |^{-s}
  &
  \mbox{in wave, out frame}
  \end{array}
  \right.
  \ .
\end{align}
Equation~(\ref{psiboosted}) says that
a (suitably boosted) outgoing observer sees an outgoing wave to have constant
amplitude near the horizon,
and similarly an ingoing observer sees an ingoing wave to have constant
amplitude;
but an outgoing observer sees an ingoing wave,
and an ingoing observer sees an outgoing wave,
boosted by a diverging factor $| \Delta_x |^{-s}$.

The apparent divergence of outgoing waves seen by an ingoer,
and of ingoing waves seen by an outgoer, should be interpreted with care.
Near the outer horizon of a black hole,
outgoing waves, whether outside or inside the horizon,
always move away from the horizon,
so an ingoer always sees the amplitude of an outgoing wave
near the outer horizon
to be smaller than when the outgoing wave was emitted.
Similarly, near a cosmological horizon,
ingoing waves, whether outside or inside the horizon,
always move away from the horizon,
so again an outgoer always sees the amplitude of an ingoing wave
near the cosmological horizon
to be smaller than when the outgoing wave was emitted.
Thus observers near the outer horizon or cosmological horizon
do not see any actual divergence.
%By contrast,
%outgoing waves near the inner horizon of a black hole
%always move towards the horizon,
%so in that case an ingoer sees the amplitude of an outgoing wave
%to be larger than when the outgoing wave was emitted.
%Thus observers near the inner horizon horizon do
%see an actual divergence.

An asymptotic analysis applies also
in the inflationary/collapse regime near the inner horizon,
where the horizon function $\Delta_x$ is small (negative)
and the angular momenta ($m$ and $\lambda_{\varsigma}$) are not too large;
but the ${\Delta_x \ddi{2} \Delta_x / \dd x^2}$ term
in the potential $V_\sigma$ grows large during inflation and collapse,
and must be retained.
The Einstein equations~(\ref{DDUx}) allow
the radial wave equation~(\ref{altwaveoperatorscoordinatex})
to be recast in terms of the inflationary variable $U$
defined by equation~(\ref{Ux}),
\begin{align}
\label{XWKBinf}
  \Biggl[
  \left(
  2 ( \Ux^2 - \vel^2 ) {\dd \over \dd \Ux}
  \right)^2
  +
  \left(
  \alpha_x
  -
  \frac{\im \sigma}{2} ( 3 \Ux - \Delta_x^\prime )
  \right)^2
  &
\nonumber
\\
  + \,
  ( 1 + 2 \sigma^2 )
  ( \Ux^2 - \vel^2 )
  \Biggr]
  X_\sigma
  =
  0
  &
  \ ,
\end{align}
the term on the second line
being the $\Delta_x {\ddi{2} \Delta_x / \dd x^2}$ term
in the potential $V_\sigma$,
equation~(\ref{altpotentialx}).
The solutions to equation~(\ref{XWKBinf})
can be expressed in terms of hypergeometric functions
%$_2F_1[a,b;c;z]$,
$_2F_1$,
but the expressions are unenlightening.
%and have the drawback that the arguments
%$c$ and $z \equiv (\Ux{+}\vel)/(2v)$
%are proportional to $1/v$,
%and are therefore generically huge since $v$ is asymptotically tiny.
A more insightful approach is to take the $v \rightarrow 0$ limit
of equation~(\ref{XWKBinf}),
in which case the solutions are Whittaker functions
$M_{\kappa,\lambda}(z)$ and $W_{\kappa,\lambda}(z)$,
which are scaled versions of Kummer confluent hypergeometric functions
$_1F_1$
\cite{Abramowitz:1970}.
The solutions to equation~(\ref{XWKBinf}) with $v \rightarrow 0$ are
($(M|W)$ in the following denotes either of the two Whittaker functions
$M$ or $W$)
\begin{equation}
\label{XWKBinfMW}
  X_\sigma
  =
  (M|W)_{\frac{3 \sigma}{4} , \frac{\sigma}{4}}
  \left(
  {- 2 \im \alpha_x + \sigma \Delta_x^\prime
  \over 2 U}
  \right)
  \ ,
\end{equation}
the tortoise coordinate $x^\ast$ in this case reducing to,
from equation~(\ref{DUx}),
\begin{equation}
  x^\ast
  =
  \mbox{constant}
  -
  {1 \over 2 U}
  \ .
\end{equation}

The radial wave solutions~(\ref{XWKBinfMW}) may seem abstruse,
but it is worth noting that Whittaker functions are
solutions to the radial wave equation
for fields of spin $s$ in flat (Minkowski) spacetime.
Specifically,
waves of spin $s$ and boost weight $\sigma$ in flat spacetime are
($\varsigma = +\sigma$ right-handed, $\varsigma = -\sigma$ left-handed)
\begin{align}
  &\psi_{\sigma}
\nonumber
\\
  &=
  r^{-1-s} \ee^{-\im ( \wel t + m \phi )}
  \, Y_\varsigma(y)
  \left\{
  \begin{array}{ll}
  M_{\sigma , \, \ell + \tfrac{1}{2}} ( 2 \im \wel r ) \,
  &
  \mbox{outgoing}
  \\
  W_{\sigma , \, \ell + \tfrac{1}{2}} ( 2 \im \wel r ) \,
  &
  \mbox{ingoing}
  \end{array}
  \right.
\nonumber
\\
  &\underset{r \rightarrow \infty}{\parbox{2.2em}{\rightarrowfill}} \,
  r^{-1-s}
  \, Y_\varsigma(y)
  \left\{
  \begin{array}{ll}
  r^{-\sigma}
  \ee^{- \im \left[ \wel ( t - r ) + m \phi \right]}
  &
  \mbox{outgoing}
  \\
  r^{+\sigma}
  \ee^{- \im \left[ \wel ( t + r ) + m \phi ) \right]}
  &
  \mbox{ingoing}
  \end{array}
  \right.
  \ .
\end{align}
Note that in flat spacetime,
$\alpha_x$ defined by equation~(\ref{alphaxy})
reduces to the temporal frequency $\alpha_x = \wel$,
and $Y_\varsigma(y) \ee^{-\im m \phi}$
reduce to standard spin-weighted spherical harmonics.

%\subsection{Angular eigenfunctions}
%
%\begin{widetext}
%In $\Lambda$-Kerr-Newmann,
%\begin{equation}
%  {\ddsq ( R^4 \Delta_x ) \over \dd r^2}
%  =
%  2
%  -
%  {\tfrac{2}{3}}
%  \Lambda
%  ( 6 r^2 + a^2 )
%  \ , \quad
%  {\ddsq \Delta_y \over \dd y^2}
%  =
%  - \,
%  2
%  -
%  {\tfrac{2}{3}}
%  a^2 \Lambda ( 6 y^2 - 1 )
%  \ .
%\end{equation}
%For Kerr, the spin~$s$ angular equations are
%\begin{equation}
%  \left[
%  ( 1 - y^2 ) {\ddsq \over \dd y^2}
%  - 2 y {\dd \over \dd y}
%  -
%  {\left[ a \wel ( 1 - y^2 ) - m \,\pm\, s y \right]^2
%  \over 1 - y^2}
%  -
%  4 a \wel m
%  -
%  \tfrac{1}{3}
%  ( 1 + 2 s^2 )
%  -
%  2 \lambda_{\pm \sigma}
%  \right]
%  Y_{\pm s}(y)
%  =
%  0
%  \ ,
%\end{equation}
%whose solutions are Teukolsky's spin-weighted spheroidal harmonics.
%
%Confluent Heun function Kerr-Newman angular solutions
%\cite{Staicova:2015}
%\begin{equation}
%  Y_s
%  =
%  \ee^{a w y}
%  \left\{
%  \begin{array}{l}
%  \HeunC
%  (
%  - \, \tfrac{1}{3} + 2 a w - 4 a m w + \lambda
%  , \,
%  4 a w , \, 1-m , \, 1+m , \, 4 a w , \sin^2\!\tfrac{1}{2}\theta
%  )
%  \\
%  (
%  \HeunC
%  (
%  - \, \tfrac{1}{3} + 2 a w - m (1+m) + \lambda
%  , \,
%  4 a ( 1 + m ) w , \, 1+m , \, 1+m , \, 4 a w , \sin^2\!\tfrac{1}{2}\theta
%  )
%  \end{array}
%  \right.
%\end{equation}
%\end{widetext}

\subsection{Cosmological constant}
\label{lambda-sec}

The outgoing and ingoing hyper-relativistic streams
that drive inflation near the inner horizon
in the conformally separable solutions
carry an exponentially growing proper energy-momentum.
But because the streams are null,
the trace of their energy-momentum is zero.
The only possible nonvanishing contribution to the Ricci scalar $R$
in the conformally separable solutions is from a cosmological constant
$\Lambda$, which contributes a Ricci scalar
\begin{equation}
  R = 4 \Lambda
  \ .
\end{equation}
%The conformally separable solutions are asymptotically exact
%in the limit of small accretion rates,
%and it is in this limit that a cosmological constant can be admitted.
In the real Universe, the cosmological constant $\Lambda$ is tiny compared to
the energy density of any astronomical black hole,
so is negligible in practice.
It is nevertheless useful for the sake of completeness
to consider the possibility of a cosmological constant in the wave equations.

For spin half, one, or two, $s = \tfrac{1}{2}$, 1, or 2,
the wave equations for propagating waves remain separable
in the presence of a cosmological constant,
the only effect of the cosmological constant being on the functional form
of the horizon function $\Delta_x$, equation~(\ref{DeltaxLKN}).
For spins $s = 0$ or $\tfrac{3}{2}$,
the effect of a cosmological constant
(besides modifying the horizon function $\Delta_x$)
is to replace the difference of radial and angular wave operators
in the wave equations by
\begin{equation}
  \sqvu
  -
  \sqpm
  \rightarrow
  \sqvu
  -
  \sqpm
  +
  \chi_s
  \rho^2 \Lambda
  \ ,
\end{equation}
for some spin-dependent constant $\chi_s$.
In the $\Lambda$-Kerr-(Newman) regime away from the inner horizon,
the conformal factor is separable,
$\rho^2 \rightarrow \rhosep^2 = r^2 + a^2 y^2$,
equation~(\ref{rho}),
and the wave equations remain separable
if the radial and angular wave operators $\sqvu$ and $\sqpm$
are adjusted by $+ \chi_s r^2 \Lambda$
and $- \chi_s a^2 y^2 \Lambda$ respectively.
This adjustment fails
in the inflationary regime near the inner horizon
because the conformal factor $\rho = \rhosep \rhoi$ ceases to be separable.
%However, a feature of the inflationary solutions
%is that during early inflation the inflationary exponent $\xi(x)$
%remains sensibly equal to zero even while its derivatives
%$\partial \xi / \partial x$ and
%$\partial^2 \xi / \partial x^2$
%grow exponentially huge.
%If inflation persisted, as opposed to being followed by collapse,
%the exponential growth of derivatives of $\xi$ would lead to
%a ``weak null singularity''
%\cite{Poisson:1990eh,Barrabes:1990,Dafermos:2017dbw}.
%In the conformally separable solutions,
%ongoing accretion in due course precipitates collapse.
However,
the conformally separable solutions hold in the asymptotic limit~(\ref{vel0})
of small accretion rate.
In this limit, the conformal factor can be effectively separated as
$\rho^2 = ( \rho^2 - a^2 y^2 ) + a^2 y^2$,
and the radial and angular wave operators $\sqvu$ and $\sqpm$
defined by equations~(\ref{waveoperatorsraiselower})
adjusted as
\begin{align}
\label{waveoperatorsLambda}
  \sqvu
  &\rightarrow
  \sqvu
  +
  \chi_s
  ( \rho^2 - a^2 y^2 ) \Lambda
  \ ,
\nonumber
\\
  \sqpm
  &\rightarrow
  \sqpm
  -
  \chi_s
  a^2 y^2 \Lambda
  \ ,
\end{align}
once again yielding separable wave equations.
The adjustment~(\ref{waveoperatorsLambda}) works:
in the $\Lambda$-Kerr(-Newman) regime where the conformal factor
$\rho$ is separable;
in the early inflationary regime,
where the derivatives
$\partial \xi / \partial x$ and $\partial^2 \xi / \partial x^2$
of the inflationary factor grow exponentially huge
but the inflationary factor $\xi(x)$ itself remains sensibly equal to zero,
so the inflationary conformal factor is still equal to one, $\rhoi = 1$;
and the late inflationary and collapse phases,
where $\xi(x)$ grows exponentially huge
but the angular coordinate $y$ is frozen at a constant value,
its value on the inner horizon.

A possible cosmological constant $\Lambda$ is retained for completeness
throughout this paper.
Throughout this paper,
the radial and angular wave operators $\sqvu$ and $\sqpm$
are as defined by equations~(\ref{waveoperatorsraiselower}),
without the adjustment~(\ref{waveoperatorsLambda})
from a cosmological constant.

\section{Spin~$0$ waves}
\label{scalar-sec}

%\cite{Duff:1993wm}
%It's wrong, it's trivial, it was my idea.

The wave equation for a minimally coupled massless scalar field $\varphi$ is
\begin{equation}
\label{DD0minimal}
  \DD^k \DD_k 
  \varphi
  =
  0
  \ .
\end{equation}
Denote the scalar field $\varphi \equiv \psi_0$,
and define a scaled scalar $\hat\varphi$ by,
equation~(\ref{psifs}),
\begin{equation}
  \varphi
  \equiv
  f_0
  \hat{\varphi}
  \ ,
\end{equation}
with $f_0 \equiv 1/\rhoi = \ee^{- \, \vel t + \xi(x)}$,
equation~(\ref{fs}).
The d'Alembertian $\DD^k \DD_k$
for the scalar field $\varphi$
can be written in terms of the spin $s = 0$ wave operators $\square_{kl}$ defined
by equations~(\ref{waveoperatorsraiselower}),
%or (\ref{altwaveoperatorscoordinate}),
\begin{equation}
\label{DD0square}
  \rho^2
  \DD^k \DD_k 
  \varphi
  =
  f_0
  \left(
  \sq{0}{vu} - \sq{0}{+-}
  +
  \tfrac{1}{6} \rho^2 R
  \right)
  \hat{\varphi}
  \ ,
\end{equation}
where $R$ is the Ricci scalar.
A more detailed exposition of the derivation of wave equations
is given in Appendix~\ref{Dapp};
the scalar d'Alembertian is given by equation~(\ref{npscalarconfsep}).
As discussed in \S\ref{lambda-sec},
the only possible contribution to the Ricci scalar in the conformally
separable black-hole spacetimes is from a cosmological constant $\Lambda$,
in which case $R = 4 \Lambda$.
If the cosmological constant is nonzero,
there are two possibilities that yield a separable spin~0 wave equation.
The first is to adjust the radial and angular wave operators $\square$
per~(\ref{waveoperatorsLambda}).
The second is to take the Ricci scalar over to the left hand side of
equation~(\ref{DD0square}),
in which case the equation becomes the conformally coupled scalar wave equation
%THIS PAPER PROVES CONFORMAL COUPLING BY MAKING SOME ASSUMPTIONS
%\cite{Sonega:1993}
\cite{Buck:2010sv}
\begin{equation}
\label{DD0squareconf}
  \rho^2
  \left(
  \DD^k \DD_k 
  -
  \tfrac{1}{6}
  R
  \right)
  \varphi
  =
  f_0
  \left(
  \sq{0}{vu} - \sq{0}{+-}
  \right)
  \hat{\varphi}
  =
  0
  \ .
\end{equation}
Equation~(\ref{DD0squareconf})
establishes the correctness of
the claimed wave equation~(\ref{waveequations})
for the case of zero spin, $s = 0$.

%Lagrangian is
%% Table 1, page 75, with f(varphi) = (1/3) varphi^2, eq (240) of
%\cite{Forger:2003ut}
%\begin{equation}
%  L_0
%  =
%  - \,
%  \tfrac{1}{2}
%  \partial^k \varphi \, \partial_k \varphi
%  -
%  \tfrac{1}{12}
%  R \varphi^2
%\end{equation}
%energy-momentum tensor is
%\begin{align}
%  4\pi
%  T_{kl}
%  &=
%  \partial_k \varphi \, \partial_l \varphi
%  -
%  \tfrac{1}{2}
%  \bgamma_{kl}
%  \partial^m \varphi \, \partial_m \varphi
%\\
%\nonumber
%  &\quad
%  + \,
%  \tfrac{1}{6}
%  \left(
%  \bgamma_{kl} \DD^m \DD_m - \DD_k \DD_l
%  + R_{kl}
%  -
%  \tfrac{1}{2}
%  \bgamma_{kl}
%  R
%  \right)
%  \varphi^2
%\end{align}
%whose trace is zero.

\section{Spin~$\frac{1}{2}$ waves}
\label{spinhalf-sec}

The wave equation for massless spin~$\frac{1}{2}$ waves is
the massless Dirac equation
\begin{equation}
\label{spinhalfwave}
  \bDD \psi
  =
  0
  \ ,
\end{equation}
where
$\psi$ is a Dirac spinor, equation~(\ref{diracspinor}),
$\bDD$ is the covariant derivative
\begin{equation}
\label{bDD}
  \bDD \equiv \bgamma^k \DD_k
  \ ,
\end{equation}
and
$\bgamma_k$ are Dirac $\gamma$-matrices
given by equations~(\ref{diracgammaI}).
To make the boost and spin weights transparent,
according to the rules~(\ref{boostspinrules}),
it is convenient to work with the covariant components
$\psi_a$
of the spinor,
\begin{equation}
  \psi =
  \psi_a \bepsilon^a
  \ .
\end{equation}
This is consistent with the convention that, for example,
$\bgamma^v \ethsi{}{v}$ with covariant $\ethsi{}{v}$,
and not $\bgamma_v \eth^v$ with contravariant $\eth^v$,
is a raising operator.
The covariant components $\psi_a$
are related to the contravariant components $\psi^a$
of the expansion~(\ref{diracspinor}) by
\begin{equation}
  \psi_a
  =
  \left(
  \begin{array}{c}
  \psi_{\smallUpup} \\
  \psi_{\smallDowndown} \\
  \psi_{\smallUpdown} \\
  \psi_{\smallDownup} \\
  \end{array}
  \right)
  =
  \varepsilon_{ab} \psi^b
  =
  \left(
  \begin{array}{c}
  \psi^{\smallDowndown} \\
  - \psi^{\smallUpup} \\
  - \psi^{\smallDownup} \\
  \psi^{\smallUpdown} \\
  \end{array}
  \right)
  \ .
\end{equation}
The top two components
$\psi_{\smallUpup}$ and $\psi_{\smallDowndown}$
are right-handed,
with boost and spin weights $\sigma = \varsigma = \pm \tfrac{1}{2}$,
while the bottom two components
$\psi_{\smallUpdown}$ and $\psi_{\smallDownup}$
are left-handed,
with boost and spin weights $\sigma = -\varsigma = \pm \tfrac{1}{2}$.

The massless spin~$\frac{1}{2}$ wave equation~(\ref{spinhalfwave})
can be written
\begin{equation}
\label{spinhalfwaveGamma}
  \bgamma^k \left( \partial_k + \tfrac{1}{2} \bGamma_k \right) \psi
  =
  0
  \ ,
\end{equation}
where the tetrad-frame connection
$\bGamma_k$ is the set of 4 bivectors
\begin{equation}
  \bGamma_k
  =
  %\tfrac{1}{2}
  \Gamma_{klm} \bgamma^l \wedgie \bgamma^m
\end{equation}
implicitly summed over distinct antisymmetric pairs of indices $lm$.
The tetrad-frame connections $\Gamma_{klm}$
are also called Lorentz connections,
or Ricci rotation coefficients,
or spin coefficients in the context of the Newman-Penrose formalism.

Massless spin~$\tfrac{1}{2}$ waves of opposite chirality do not mix.
The right- and left-handed chiral components
$\psiz$
of the Dirac spinor are
(the tilde on $\psiz$ signifies a right- or left-handed chiral component)
\begin{equation}
  \psiz
  \equiv
  \tfrac{1}{2}
  ( 1 \pm \gamma_5 ) \psi
  \ ,
\end{equation}
where the $\pm$ sign is $+$ for right-handed, $-$ for left-handed.
The right- and left-handed spinor fields
each have two distinct nonvanishing complex components,
with boost weights respectively $\sigma = +1/2$ and $\sigma = -1/2$
and spin weights
$\varsigma = \sigma$ for right-handed,
$\varsigma = -\sigma$ for left-handed:
\begin{equation}
\label{psizcomponents}
  \psiz_{\sigma}
  =
  \left(
  \begin{array}{l}
  \psiz_{+1/2} \\
  \psiz_{-1/2} \\
  \end{array}
  \right)
  \equiv
  \left\{
  \begin{array}{ll}
  \left(
  \begin{array}{l}
  \psiz_{\Upup} \\
  \psiz_{\Downdown} \\
  \end{array}
  \right)
  &
  \mbox{right}
  \\
  \left(
  \begin{array}{l}
  \psiz_{\Updown} \\
  \psiz_{\Downup} \\
  \end{array}
  \right)
  &
  \mbox{left}
  \end{array}
  \right.
  \ .
\end{equation}
Define the scaled right- and left-handed spinors
$\hat{\psi}_{\sigma}$
(with hats)
by
\begin{equation}
  \psiz_{\sigma}
  =
  \left\{
  \begin{array}{ll}
  f_\frac{1}{2} \,
  \hat\psi_{\sigma}
  & \mbox{right}
  \\
  f_\frac{1}{2}^\ast \,
  \hat\psi_{\sigma}
  & \mbox{left}
  \end{array}
  \right.
  \ ,
\end{equation}
in accordance with equation~(\ref{psifs}).
In terms of the scaled spinors $\hat{\psi}_{\sigma}$,
the spin~$\frac{1}{2}$ wave equation~(\ref{spinhalfwaveGamma}) is
(equations~(\ref{npDiracR}) in Appendix~\ref{Dapp}
give these equations in a general spacetime)
\begin{equation}
\label{waveequationspinhalf}
  \bDD \psi
  =
  {1
  \over \rho}
  \left(
  \begin{array}{r}
  \displaystyle
  -
  f_\frac{1}{2}^\ast
  (
  \ethsi{-\frac{1}{2}}{v} \hat{\psi}_{\smallDownup}
  +
  \ethsi{-\frac{1}{2}}{+} \hat{\psi}_{\smallUpdown}
  )
  \\[1ex]
  \displaystyle
  f_\frac{1}{2}^\ast
  (
  \ethsi{\hminus\frac{1}{2}}{-} \hat{\psi}_{\smallDownup}
  \mp
  \ethsi{\hminus\frac{1}{2}}{u} \hat{\psi}_{\smallUpdown}
  )
  \\[1ex]
  \displaystyle
  f_\frac{1}{2}
  (
  \ethsi{-\frac{1}{2}}{v} \hat{\psi}_{\smallDowndown}
  +
  \ethsi{\hminus\frac{1}{2}}{-} \hat{\psi}_{\smallUpup}
  )
  \\[1ex]
  \displaystyle
  -
  f_\frac{1}{2}
  (
  \ethsi{-\frac{1}{2}}{+} \hat{\psi}_{\smallDowndown}
  \mp
  \ethsi{\hminus\frac{1}{2}}{u} \hat{\psi}_{\smallUpup}
  )
  \end{array}
  \right)
  =
  0
  \ ,
\end{equation}
where the $\mp$ sign in the second and last rows is
$-$ outside the horizon ($\Delta_x > 0$),
$+$ inside the horizon ($\Delta_x < 0$).
The $\gamma$-matrices~(\ref{diracgammaI}),
hence the covariant derivative $\bDD \equiv \bgamma^k \DD_k$,
connect spinors of opposite chirality,
so the top two components of the wave equation~(\ref{waveequationspinhalf})
are equations for the left-handed spinor components,
while the bottom two are for the right-handed spinor components.
Equation~(\ref{waveequationspinhalf})
shows that the two left-handed spinor components are coupled to each other,
and the two right-handed spinor components are coupled to each other,
but the left-handed spinor is decoupled from the right-handed spinor.
Since the boost operators
$\ethsi{}{v}$
and
$\ethsi{}{u}$
commute with the spin operators
$\ethsi{}{+}$ and
$\ethsi{}{-}$,
the four components in the column vector
in the wave equation~(\ref{waveequationspinhalf})
can be combined in pairs
to yield wave equations for each spinor component separately:
\begin{subequations}
\label{spinhalfwavesep}
\begin{align}
  \left( \sq{1/2}{vu} - \sq{1/2}{+-} \right)
  \hat{\psi}_{\smallUpup}
  &=
  0
  \ ,
\\
  \left( \sq{-1/2}{uv} - \sq{-1/2}{-+} \right)
  \hat{\psi}_{\smallDowndown}
  &=
  0
  \ ,
\\
  \left( \sq{1/2}{vu} - \sq{-1/2}{-+} \right)
  \hat{\psi}_{\smallUpdown}
  &=
  0
  \ ,
\\
  \left( \sq{-1/2}{uv} - \sq{1/2}{+-} \right)
  \hat{\psi}_{\smallDownup}
  &=
  0
  \ ,
\end{align}
\end{subequations}
where the spin~$\frac{1}{2}$ wave operators $\square_{kl}$
are defined by equations~(\ref{waveoperatorsraiselower}).
Equations~(\ref{spinhalfwavesep})
establish the correctness of
the claimed wave equation~(\ref{waveequations})
for the case of spin half, $s = \tfrac{1}{2}$.
%Note that for spin $s = \frac{1}{2}$,
%the wave operators are simply
%$\square_{kl} = \ethsi{\mp \frac{1}{2}}{k} \ethsi{\pm \frac{1}{2}}{l}$.
The wave equations~(\ref{spinhalfwavesep})
admit separated solutions
of the form
\begin{equation}
\label{psihalf}
  \hat{\psi}_{\sigma}
  =
  \ee^{- \im ( \wel t + m \phi )}
  X_\sigma(x) Y_\varsigma(y)
  \ ,
\end{equation}
where $\sigma$ and $\varsigma$
each run over boost and spin weights $\pm\tfrac{1}{2}$.
The corresponding eigenvalues are
$\lambda_{\sigma\varsigma}$.

The Teukolsky-Starobinski
\cite{Press:1973,Starobinsky:1973}
identities follow immediately from the simple form
$\square_{kl} = \ethsi{\mp \frac{1}{2}}{k} \ethsi{\pm \frac{1}{2}}{l}$
modulo a sign,
equation~(\ref{waveoperatorsraiselower}),
of the spin~$\frac{1}{2}$ wave operators:
\begin{subequations}
\label{spinhalfteukolskystarobinski}
\begin{align}
\label{spinhalfteukolskystarobinski1}
  \left(
  \square_{uv}
  \,
  \ethsi{1/2}{u}
  -
  \ethsi{1/2}{u}
  \,
  \square_{vu}
  \right)
  \hat{\psi}_{\smallUpup}
  &=
  0
  \ ,
\\
\label{spinhalfteukolskystarobinski2}
  \left(
  \square_{vu}
  \,
  \ethsi{-1/2}{v}
  -
  \ethsi{-1/2}{v}
  \,
  \square_{uv}
  \right)
  \hat{\psi}_{\smallDowndown}
  &=
  0
  \ ,
\\
\label{spinhalfteukolskystarobinski3}
  \left(
  \square_{-+}
  \,
  \ethsi{1/2}{-}
  -
  \ethsi{1/2}{-}
  \,
  \square_{+-}
  \right)
  \hat{\psi}_{\smallUpup}
  &=
  0
  \ ,
\\
\label{spinhalfteukolskystarobinski4}
  \left(
  \square_{+-}
  \,
  \ethsi{-1/2}{+}
  -
  \ethsi{-1/2}{+}
  \,
  \square_{-+}
  \right)
  \hat{\psi}_{\smallDowndown}
  &=
  0
  \ .
\end{align}
\end{subequations}
The identities~(\ref{spinhalfteukolskystarobinski})
are for the right-handed components;
a similar set of identities holds for the left-handed components.
The first
Teukolsky-Starobinski identity~(\ref{spinhalfteukolskystarobinski1})
shows that
$\ethsi{1/2}{u} \, \hat{\psi}_{\smallUpup}$
is an eigenfunction of
$\square_{uv}$
with eigenvalue
$\lambda_{\smallUpup}$,
while the last
Teukolsky-Starobinski identity~(\ref{spinhalfteukolskystarobinski4})
shows that
$\ethsi{-1/2}{+} \, \hat{\psi}_{\smallDowndown}$
is an eigenfunction of
$\square_{+-}$
with eigenvalue
$\lambda_{\smallDowndown}$.
But the fourth row of~(\ref{waveequationspinhalf})
shows that
$\ethsi{1/2}{u} \hat{\psi}_{\smallUpup}$
%and
equals
$\ethsi{-1/2}{+} \hat{\psi}_{\smallDowndown}$ modulo a sign,
%are equal,
so it follows that the eigenvalues must be equal,
$\lambda_{\smallUpup} = \lambda_{\smallDowndown} = \lambda$.

The eigenvalue $\lambda$ is real and positive.
This follows from the fact that the spin~$\frac{1}{2}$
wave operators are positive definite,
being the product of an operator and its Hermitian conjugate,
%eigenvalues are $\lambda \lambda^\ast$, which is real positive
equations~(\ref{hermitianconjugate}):
\begin{equation}
  \square_{vu}
  =
  \ethsi{-1/2}{v} \, \ethsi{1/2}{u}
  =
  \ethsi{1/2}{u}^\dagger \, \ethsi{1/2}{u}
  \ , \quad
  \square_{+-}
  =
  \ethsi{-1/2}{+} \, \ethsi{1/2}{-}
  =
  \ethsi{1/2}{-}^\dagger \, \ethsi{1/2}{-}
  \ .
\end{equation}
Thus $\lambda$ can be written as the square of some real number $\mu$,
\begin{equation}
  \mu^2 = \lambda
  \ .
\end{equation}

The third row of~(\ref{waveequationspinhalf})
shows that
$X_{-1/2}$
raised once is proportional to
$X_{+1/2}$,
and that
$Y_{+1/2}$
lowered once is proportional to
$Y_{-1/2}$.
Similarly,
the last row of~(\ref{waveequationspinhalf})
shows that
$X_{+1/2}$
lowered once is proportional to
$X_{-1/2}$,
and that
$Y_{-1/2}$
raised once is proportional to
$Y_{+1/2}$.
With a convenient choice of relative normalization,
the eigenfunctions are related by
\begin{subequations}
\label{spinhalfXY}
\begin{align}
  \ethsi{\frac{1}{2}}{v}
  X_{-1/2}
  =
  -
  \mu \,
  X_{+1/2}
  \ &, \quad
  \ethsi{\frac{1}{2}}{u}
  X_{+1/2}
  =
  \sgn(\Delta_x)
  \mu \,
  X_{-1/2}
  \ ,
\\
  \ethsi{\frac{1}{2}}{+}
  Y_{-1/2}
  =
  \mu \,
  Y_{+1/2}
  \ &, \quad
  \ethsi{\frac{1}{2}}{-}
  Y_{+1/2}
  =
  \mu \,
  Y_{-1/2}
  \ .
\end{align}
\end{subequations}
Equations~(\ref{spinhalfXY})
specify the relation between the boost/spin weight $\pm\frac{1}{2}$
components of an eigenmode of given chirality, right or left.
Modes of opposite chirality evolve independently of each other.

\section{Spin~1 waves}
\label{spinone-sec}

Wave equations for massless, neutral spin~$1$ waves
--- electromagnetic waves ---
are provided by Maxwell's equations,
\begin{equation}
\label{maxwellbF}
  \bDD \bF
  =
  \bj
  \ ,
\end{equation}
where
$\bDD$ is the covariant derivative defined by equation~(\ref{bDD}),
$\bF$ is the electromagnetic field bivector
\begin{equation}
  \bF
  \equiv
  %\tfrac{1}{2}
  F_{kl} \bgamma^k \wedgie \bgamma^l
\end{equation}
implicitly summed over distinct antisymmetric pairs of indices $kl$,
and $\bj$ is the electric current.
The electromagnetic units here are Heaviside;
in Gaussian units the right hand side of Maxwell's equation~(\ref{maxwellbF})
would be $4\pi \bj$.
In the present case the source current is taken to vanish,
\begin{equation}
  \bj = 0
  \ .
\end{equation}
The assumption of vanishing source current
requires that the black hole be uncharged.
The single equation~(\ref{maxwellbF}) embodies both source-free (magnetic)
and source (electric) parts of Maxwell's equations.

Like massless Dirac spinors,
massless spin~$1$ waves
decompose into two distinct chiralities that do not mix.
The right- and left-handed chiral components
$\bFz$
of the electromagnetic field bivector are
(the tilde on $\bFz$ signifies a right- or left-handed chiral component)
\begin{equation}
\label{bFz}
  \bFz
  \equiv
  \tfrac{1}{2}
  ( 1 \pm \gamma_5 )
  \bF
  \ ,
\end{equation}
where the $\pm$ sign is $+$ for right-handed, $-$ for left-handed.
Equation~(\ref{bFz}) implies that
the components $\Fz_{kl}$ of the right and left-handed electromagnetic field
$\bFz$ are
\begin{equation}
  \Fz_{kl}
  =
  \tfrac{1}{2}
  %( F_{kl} \mp \im \, \starF_{kl} )
  \left( F_{kl} \mp \im \, \varepsilon_{kl}{}^{mn} F_{mn} \right)
  \ .
\end{equation}
(implicitly summed over distinct antisymmetric indices $mn$)
with $-$ for right-handed, $+$ for left-handed.
Maxwell's equations~(\ref{maxwellbF}) with zero source current are then
\begin{equation}
\label{maxwell}
  \DD^k \Fz_{kl}
  =
  0
  \ .
\end{equation}
The right- and left-handed electromagnetic field $\Fz_{kl}$ bivectors
each have three distinct nonvanishing complex components,
with boost weights respectively $\sigma = +1$, $0$, and $-1$,
and spin weights
$\varsigma = \sigma$ for right-handed,
$\varsigma = -\sigma$ for left-handed:
\begin{align}
\label{Fzcomponents}
  \Fz_\sigma
  &\equiv
  \left(
  \begin{array}{l}
  \Fz_{+1} \\
  \Fz_{0} \\
  \Fz_{-1}
  \end{array}
  \right)
\nonumber
\\
  &\equiv
  \left\{
  \begin{array}{ll}
  \left(
  \begin{array}{l}
  \Fz_{v+} \\
  \Fz_{vu} \\
  \Fz_{u-}
  \end{array}
  \right)
  =
  \left(
  \mkern-5mu
  \begin{array}{c}
  F_{v+} \\
  \frac{1}{2} \left( F_{vu} - F_{+-} \right) \\
  F_{u-}
  \end{array}
  \mkern-5mu
  \right)
  &
  \mbox{right}
  \\
  \left(
  \begin{array}{l}
  \Fz_{v-} \\
  \Fz_{vu} \\
  \Fz_{u+}
  \end{array}
  \right)
  =
  \left(
  \mkern-5mu
  \begin{array}{c}
  F_{v-} \\
  \frac{1}{2} \left( F_{vu} + F_{+-} \right) \\
  F_{u+}
  \end{array}
  \mkern-5mu
  \right)
  &
  \mbox{left}
  \end{array}
  \right.
  \ .
\end{align}
The three components $\Fz_\sigma$
with boost weights \mbox{$\sigma = +1, 0, -1$}
are commonly
\cite{Chandrasekhar:1983}
denoted
$\phi_0$, $\phi_1$, $\phi_2$,
equations~(\ref{chandraF}),
but the notation~(\ref{Fzcomponents})
makes manifest the boost weights of the components.
The propagating component is, equation~(\ref{psir}),
$\Fz_{-1} = \phi_2$ outgoing,
$\Fz_{+1} = \phi_0$ ingoing.

Focus on the right-handed electromagnetic field;
the left-handed field is quite similar.
Define the scaled right-handed electromagnetic field tensor $\hat{F}_{\sigma}$
(with a hat instead of a tilde over the $F$)
by
\begin{equation}
  \Fz_{\sigma}
  =
  f_1 \,
  \hat{F}_{\sigma}
  \ ,
\end{equation}
in accordance with equation~(\ref{psifs}).
In terms of the scaled electromagnetic field $\hat{F}_{kl}$
and the boost and spin raising and lowering operators~(\ref{ethsi}),
Maxwell's equations~(\ref{maxwell}) are
(equations~(\ref{npMaxwellR}) in Appendix~\ref{Dapp}
give these equations in a general spacetime)
\begin{equation}
\label{maxwellcomponents}
  D^k
  \!
  \left(
  \!
  \begin{array}{l}
%1
  \Fz_{kv}
  \\
%2
  \Fz_{ku}
  \\
%3
  \Fz_{k+}
  \\
%4
  \Fz_{k-}
  \end{array}
  \!
  \right)
  =
  {f_1 \over \sqrt{2}\rho}
  \!
  \left(
  \!
  \begin{array}{r}
%1
  \rhom{}^{-1}
  \ethsi{\frac{1}{2}}{v}
  \rhom
  \hat{F}_{0}
  +
  \rhom
  \ethsi{\frac{1}{2}}{-}
  \rhom{}^{-1}
  \hat{F}_{+1}
  \\
%2
  \mp \,
  \rhom{}^{-1}
  \ethsi{\frac{1}{2}}{u}
  \rhom
  \hat{F}_{0}
  -
  \rhom
  \ethsi{\frac{1}{2}}{+}
  \rhom{}^{-1}
  \hat{F}_{-1}
  \\
%3
  \rhom{}^{-1}
  \ethsi{\frac{1}{2}}{+}
  \rhom
  \hat{F}_{0}
  \mp
  \rhom
  \ethsi{\frac{1}{2}}{u}
  \rhom{}^{-1}
  \hat{F}_{+1}
  \\
%4
  \rhom{}^{-1}
  \ethsi{\frac{1}{2}}{-}
  \rhom
  \hat{F}_{0}
  -
  \rhom
  \ethsi{\frac{1}{2}}{v}
  \rhom{}^{-1}
  \hat{F}_{-1}
  \end{array}
  \!
  \right)
  =
  0
  \ ,
\end{equation}
where the $\mp$ sign in the middle two rows is
$-$ outside the horizon ($\Delta_x \,{>}\, 0$),
$+$ inside the horizon ($\Delta_x \,{<}\, 0$).
%with $\rhom$ defined by equation~(\ref{rhopm}).
Equations~(\ref{maxwellcomponents})
show that the three components
of the right-handed electromagnetic field
evolve not independently, but rather in harmony with each other.
Combining equations~(\ref{maxwellcomponents}) in pairs
%so as to eliminate the spin~$0$ component $\hat{F}_{0}$
yields pairs of equations for each of the three components
$\hat{F}_\sigma$,
%for the spin weight $\mp 1$ components
%$\hat{F}_{-1}$,
%and
%$\hat{F}_{+1}$
\begin{subequations}
\label{spinonewave}
\begin{align}
\label{spinonewave1}
  \rhom^{-1}
  \left(
  \ethsi{-1}{v}
  \rhom^{2}
  \ethsi{0}{u}
  \pm
  \ethsi{-1}{+}
  \rhom^{-2}
  \ethsi{0}{-}
  \right)
  \rhom^{-1}
  \hat{F}_{+1}
  &=
  0
  \ ,
\\
\label{spinonewave2}
  \rhom^{-1}
  \left(
  \ethsi{1}{u}
  \rhom^{2}
  \ethsi{0}{v}
  \pm
  \ethsi{1}{-}
  \rhom^{2}
  \ethsi{0}{+}
  \right)
  \rhom^{-1}
  \hat{F}_{-1}
  &=
  0
  \ ,
\\
\label{spinonewave3}
  %\ethsi{0}{v} \, \ethsi{-1}{v}
  \ethsi{}{v}^2
  \hat{F}_{-1}
  +
  %\ethsi{0}{-} \, \ethsi{1}{-}
  \ethsi{}{-}^2
  \hat{F}_{+1}
  &=
  0
  \ ,
\\
\label{spinonewave4}
  %\ethsi{0}{u} \, \ethsi{1}{u}
  \ethsi{}{u}^2
  \hat{F}_{+1}
  +
  %\ethsi{0}{+} \, \ethsi{-1}{+}
  \ethsi{}{+}^2
  \hat{F}_{-1}
  &=
  0
  \ ,
\\
\label{spinonewave5}
  \rhom
  \left(
  \ethsi{-1}{v}
  \rhom^{-2}
  \ethsi{0}{u}
  \pm
  \ethsi{-1}{+}
  \rhom^{-2}
  \ethsi{0}{-}
  \right)
  \rhom
  \hat{F}_{0}
  &=
  0
  \ ,
\\
\label{spinonewave6}
  \rhom
  \left(
  \ethsi{1}{u}
  \rhom^{-2}
  \ethsi{0}{v}
  \pm
  \ethsi{1}{-}
  \rhom^{-2}
  \ethsi{0}{+}
  \right)
  \rhom
  \hat{F}_{0}
  &=
  0
  \ .
\end{align}
\end{subequations}
The top and bottom pairs of equations~(\ref{spinonewave}) can also be written
in terms of the radial and angular wave operators $\square$
defined by equations~(\ref{waveoperatorsraiselower}),
\begin{subequations}
\label{spinonewavesq}
\begin{align}
\label{spinonewavesqp1}
  \left(
  \sq{1}{vu}
  -
  \sq{1}{+-}
  \right)
  \hat{F}_{+1}
  &=
  0
  \ ,
\\
\label{spinonewavesqm1}
  \bigl(
  \sq{-1}{uv}
  -
  \sq{-1}{-+}
  \bigr)
  \hat{F}_{-1}
  &=
  0
  \ ,
\\
\label{spinonewavesq0a}
  \bigl(
  \sq{0}{vu}
  -
  \sq{0}{+-}
  -
  2 \rho^2 \Cz_0
  \bigr)
  \hat{F}_{0}
  &=
  0
  \ ,
\\
\label{spinonewavesq0b}
  \bigl(
  \sq{0}{uv}
  -
  \sq{0}{-+}
  -
  2 \rho^2 \Cz_0
  \bigr)
  \hat{F}_{0}
  &=
  0
  \ ,
\end{align}
\end{subequations}
where $\Cz_0$ is the spin~0 component of the right-handed Weyl tensor,
equation~(\ref{Cz0}).
The top pair of equations~(\ref{spinonewavesq})
establish the correctness of
the claimed wave equation~(\ref{waveequations})
for the case of electromagnetic waves, $s = 1$.
The top pair admit separated solutions for the
boost weight $\pm 1$
components of the right-handed
($\varsigma = \sigma$)
electromagnetic field,
\begin{equation}
\label{Fkl}
  \hat{F}_{\sigma}
  =
  \ee^{- \im (\wel t + m \phi)}
  X_\sigma(x) Y_\varsigma(y)
  \ .
\end{equation}
The corresponding eigenvalues are
$\lambda_{\sigma\varsigma}$.

The spin~$1$ wave operators $\square_{kl}$
can be checked to satisfy the Teukolsky-Starobinsky
\cite{Press:1973,Starobinsky:1973}
identities
\begin{subequations}
\label{spinoneteukolskystarobinski}
\begin{align}
\label{spinoneteukolskystarobinski1}
  \left(
  \square_{uv}
  \,
  %\ethsi{0}{u} \, \ethsi{1}{u}
  \ethsi{}{u}^2
  -
  %\ethsi{0}{u} \, \ethsi{1}{u}
  \ethsi{}{u}^2
  \,
  \square_{vu}
  \right)
  \hat{F}_{+1}
  &=
  0
  \ ,
\\
\label{spinoneteukolskystarobinski2}
  \left(
  \square_{vu}
  \,
  %\ethsi{0}{v} \, \ethsi{-1}{v}
  \ethsi{}{v}^2
  -
  %\ethsi{0}{v} \, \ethsi{-1}{v}
  \ethsi{}{v}^2
  \,
  \square_{uv}
  \right)
  \hat{F}_{-1}
  &=
  0
  \ ,
\\
\label{spinoneteukolskystarobinski3}
  \left(
  \square_{-+}
  \,
  %\ethsi{0}{-} \, \ethsi{1}{-}
  \ethsi{}{-}^2
  -
  %\ethsi{0}{-} \, \ethsi{1}{-}
  \ethsi{}{-}^2
  \,
  \square_{+-}
  \right)
  \hat{F}_{+1}
  &=
  0
  \ ,
\\
\label{spinoneteukolskystarobinski4}
  \left(
  \square_{+-}
  \,
  %\ethsi{0}{+} \, \ethsi{-1}{+}
  \ethsi{}{+}^2
  -
  %\ethsi{0}{+} \, \ethsi{-1}{+}
  \ethsi{}{+}^2
  \,
  \square_{-+}
  \right)
  \hat{F}_{-1}
  &=
  0
  \ .
\end{align}
\end{subequations}
The identities~(\ref{spinoneteukolskystarobinski})
are for the right-handed components;
a similar set of identities holds for the left-handed components.
The first Teukolsky-Starobinski identity~(\ref{spinoneteukolskystarobinski1})
shows that 
$\ethsi{}{u}^2 \hat{F}_{+1}$
is an eigenfunction of $\square_{vu}$
with eigenvalue $\lambda_{v+}$,
while
the fourth Teukolsky-Starobinski identity~(\ref{spinoneteukolskystarobinski4})
shows that
$\ethsi{}{+}^2 \hat{F}_{-1}$
is an eigenfunction of $\square_{+-}$
with eigenvalue $\lambda_{u-}$.
But equation~(\ref{spinonewave4}) shows that
$\ethsi{}{u}^2 \hat{F}_{+1}$
equals
$\ethsi{0}{+}^2 \hat{F}_{-1}$ modulo a minus sign,
so it follows that the eigenvalues must be equal,
$\lambda_{v+} = \lambda_{u-} = \lambda$.

The middle pair of equations~(\ref{spinonewave})
imply that
$Y_{+1}$ lowered twice
is proportional to
$Y_{-1}$,
and that
$Y_{-1}$ raised twice
is proportional to
$Y_{+1}$.
A similar statement holds for
$X_{+1}$ and $X_{-1}$.
Lowering $Y_{+1}$ twice, then raising the result twice,
yields some constant $\mu^2$ times
$Y_{+1}$.
By adjusting the relative normalization of
$Y_{+1}$ and $Y_{-1}$,
the result of $Y_{+1}$ lowered twice can be taken to be $\mu Y_{-1}$,
and $Y_{-1}$ raised twice to be $\mu Y_{+1}$.
%Since the lowering and raising operators
%$\ethsi{}{-}$ and $\ethsi{}{+}$
%applied to $Y_{+1}$ and $Y_{-1}$
%are real, equation~(\ref{ethsiseppm}),
%the constant $\mu^2$ must be real.
The middle pair of equations~(\ref{spinonewave}) imply that the radial
constant of proportionality is minus the angular constant.
Thus
\begin{subequations}
\begin{align}
\label{Xvu}
  %\ethsi{0}{u} \, \ethsi{1}{u} \,
  \ethsi{}{u}^2
  X_{+1}
  =
  - \mu
  X_{-1}
  &
  \ , \quad
  %\ethsi{0}{v} \, \ethsi{-1}{v} \,
  \ethsi{}{v}^2
  X_{-1}
  =
  - \mu
  X_{+1}
  \ ,
\\
\label{Ypm}
  %\ethsi{0}{-} \, \ethsi{1}{-} \,
  \ethsi{}{-}^2
  Y_{+1}
  =
  \mu
  Y_{-1}
  &
  \ , \quad
  %\ethsi{0}{+} \, \ethsi{-1}{+} \,
  \ethsi{}{+}^2
  Y_{-1}
  =
  \mu
  Y_{+1}
  \ .
\end{align}
\end{subequations}
Solving for $\mu^2$ in
\begin{equation}
\label{raiseraiselowerlower}
  \left(
  %\ethsi{0}{+} \, \ethsi{-1}{+}
  \ethsi{}{+}^2
  \,
  %\ethsi{0}{-} \, \ethsi{1}{-}
  \ethsi{}{-}^2
  -
  \mu^2
  \right)
  Y_{+1}
  =
  0
  \ ,
\end{equation}
given that
$\Fz_{+1}$ satisfies
$( \square_{+-} - \lambda ) \Fz_{+1} = 0$,
yields the standard result
\cite[p.~388 eqs.~(35) and (52)]{Chandrasekhar:1983}
\begin{equation}
  \mu^2
  =
  \lambda^2
  -
  4 a \wel ( a \wel + m )
  \ .
\end{equation}
Since the raising and lowering operators are Hermitian conjugates,
equation~(\ref{hermitianconjugatespin}),
the operator on the left hand side of equation~(\ref{raiseraiselowerlower})
can be written
\begin{equation}
  %\ethsi{0}{+} \, \ethsi{-1}{+}
  \ethsi{}{+}^2
  \,
  %\ethsi{0}{-} \, \ethsi{1}{-}
  \ethsi{}{-}^2
  =
  %( \ethsi{0}{-} \, \ethsi{1}{-} )^\dagger
  ( \ethsi{}{-}^2 )^\dagger
  %\ethsi{0}{-} \, \ethsi{1}{-}
  \ethsi{}{-}^2
  \ ,
\end{equation}
which is positive definite
being the product of an operator and its Hermitian conjugate.
It follows that the eigenvalue $\mu^2$ is real and positive,
and consequently $\mu$ is real.

Expressions for the scaled boost weight~$0$ component
$\hat{F}_{0}$ of the electromagnetic field
follow from any of the four equations~(\ref{maxwellcomponents}).
Consider the first of the four equations~(\ref{maxwellcomponents}),
which expresses a certain derivative of $\hat{F}_{0}$
in terms of a certain derivative of
the boost weight~1 component $\hat{F}_{+1}$.
This first of equations~(\ref{maxwellcomponents}) can be solved
for $\hat{F}_{0}$ by expressing the radial part of the separated solution
$\hat{F}_{+1}$ as $X_{+1} = - \mu^{-1} \ethsi{}{v}^2 X_{-1}$,
equation~(\ref{Xvu}),
and by using the relation~(\ref{rhomethsicommutation}).
The first and last of the four equations~(\ref{maxwellcomponents})
redundantly yield one expression for
$\hat{F}_{0}$,
while the second and third equation
redundantly yield a second expression.
The two expressions are,
with
$\hat{F}_{0} \equiv \ee^{- \im (\wel t + m \phi)} \mathring{F}_0 (x,y)$,
%DROP POSSIBLE CONSTANT $c/\rhom$ OF INTEGRATION
\begin{subequations}
\label{F0}
\begin{align}
  &
  \mathring{F}_0 (x,y)
\nonumber
\\
  &
  \mkern-2mu
  =
  \mp
  {1 \over \mu}
  \Bigl[
  \ethsi{1}{u} \, \ethsi{-1}{+} 
  -
  {1 \over \rhom}
  \left(
  \mp
  R^2 \sqrt{| \Delta_x |}
  \,
  \ethsi{-1}{+}
  +
  \im a \sqrt{\Delta_y}
  \,
  \ethsi{1}{u}
  \right)
  \Bigr]
  X_{+1} Y_{-1}
  \ ,
\\
  &
  =
  {1 \over \mu}
  \Bigl[
  \ethsi{-1}{v} \, \ethsi{1}{-} 
  +
  {1 \over \rhom}
  \left(
  \mp
  R^2 \sqrt{| \Delta_x |}
  \,
  \ethsi{1}{-}
  +
  \im a \sqrt{\Delta_y}
  \,
  \ethsi{-1}{v}
  \right)
  \Bigr]
  X_{-1} Y_{+1}
\end{align}
\end{subequations}
where the $\mp$ signs are $-$ outside the horizon, $\Delta_x \,{>}\, 0$,
and $+$ inside the horizon, $\Delta_x \,{<}\, 0$.
Unlike the boost weight $\pm 1$ components,
equation~(\ref{Fkl}),
the boost weight 0 function
$\mathring{F}_0 (x,y)$
is not a separated product of radial and angular coordinates $x$ and $y$.

\section{Spin~$\frac{3}{2}$ waves}
\label{spinthreehalves-sec}

Spin~$\frac{3}{2}$ fields
arise in supersymmetric local gauge theories,
where the generators of the gauge group are taken to be spinors
\cite{vanNieuwenhuizen:1981,vanNieuwenhuizen:2004rh,Ferrara:2017hed}.
%original papers
%\cite{Freedman:1976,Deser:1976eh}
%reviews
%\cite{Duff:2004jj}.
%\cite{Schenkel:2011nv}
%Supergravity http://www.damtp.cam.ac.uk/research/gr/members/gibbons/gwgPartIII_Supergravity.pdf
%Issues with spin~$\frac{3}{2}$
%\cite{Gsponer:2002,Frauendiener:2002qj,Hack:2011yv}
%Non-minimal coupling \cite{Villanueva:2001rh}
The signature feature of supersymmetry is that it transforms
bosonic (integral spin) fields into fermionic (half-integral spin) fields
and vice versa.
Spin~$\tfrac{3}{2}$ fields in $\Lambda$-Kerr-Newman black holes have been
considered by
\cite{Guven:1980,Aichelburg:1981,TorresdelCastillo:1992}.
The supersymmetric gauge connection defines
the gravitino potential $\bARS$,
a vector of spinors,
\begin{equation}
  \bARS
  \equiv
  \ARS_k \bgamma^k
  \equiv
  \ARS_{ka} \bgamma^k \otimes \bepsilon^a
  \ ,
\end{equation}
where $\bgamma^k$ and $\bepsilon^a$
are respectively vector and spinor basis elements,
\S\ref{spinor-sec}.
Each of the 4 vector components of the gravitino potential $\bARS$
is a spinor $\ARS_k$,
\begin{equation}
  \ARS_k
  \equiv
  \ARS_{ka} \bepsilon^a
  \ .
\end{equation}
The gauge-covariant supersymmetric derivative is
\begin{equation}
\label{supersymderiv}
  \bDD + \bARS
  =
  ( \DD_k + \ARS_k ) \bgamma^k
  \ ,
\end{equation}
where $\bDD$, equation~(\ref{bDD}),
is the usual general-relativistic covariant derivative.
A vector-spinor $\bARS$ has $4 \times 4 = 16$ complex components,
whose irreducible parts under Lorentz transformations
comprise a 4-component spin~$\tfrac{1}{2}$ part
and a 12-component spin~$\tfrac{3}{2}$ part.
The spin~$\tfrac{1}{2}$ parts of the vector-spinor $\bARS$ are removed,
leaving only the spin~$\tfrac{3}{2}$ parts of the gravitino,
by imposing the 4 conditions
(here the Dirac $\gamma$-matrices act by matrix multiplication on
the spinors $\ARS_k$)
\begin{equation}
\label{ARS32}
  \bgamma^k
  \ARS_k
  =
  0
  \ .
\end{equation}
A supersymmetric gauge transformation by a spinor $\lambda$
transforms the gravitino potential $\ARS_k$ as
\begin{equation}
\label{ARSgaugetransformation}
  \ARS_k \rightarrow \ARS_k + \DD_k \lambda
  \ .
\end{equation}
Recall that the covariant derivative acting on a spinor is
$\DD_k \lambda = ( \partial_k + \tfrac{1}{2} \bGamma_k ) \lambda$,
equation~(\ref{spinhalfwaveGamma}).
The supersymmetric gauge freedom can be removed
by imposing some gauge condition.
When dealing with waves, a convenient choice of gauge
is the analog of the Lorenz gauge of electromagnetism,
\begin{equation}
\label{ARSLorenz}
  \DD^k \ARS_k = 0
  \ .
\end{equation}
The gauge condition~(\ref{ARSLorenz})
removes 4 of the 12 complex degrees of freedom of
the spin~$\tfrac{3}{2}$ gravitino potential,
leaving it with 8 physical degrees of freedom.
The massless gravitino potential decomposes further into right- and left-handed
chiral parts,
each with 4 physical degrees of freedom
(the tilde on ${\ARSz}$
signifies a right-handed or left-handed chiral component),
\begin{equation}
  {\ARSz}_k
  \equiv
  \tfrac{1}{2} ( 1 \pm \gamma_5 ) \ARS_k
  \equiv
  \tfrac{1}{2} ( 1 \pm \gamma_5 ) \ARS_{ka} \bepsilon^a
  \ ,
\end{equation}
where the $\pm$ sign is $+$ for right-handed, $-$ for left-handed.

The commutator
of the supersymmetric gauge-covariant derivative
defines the curvature,
\begin{equation}
\label{DDsy}
  [ \bDD + \bARS , \bDD + \bARS ]
  \equiv
  [ \DD_k + \ARS_k , \DD_l + \ARS_l ]
  \, \bgamma^k \wedgie \bgamma^l
  \ ,
\end{equation}
implicitly summed over distinct antisymmetric $kl$.
The curvature~(\ref{DDsy}) has both
a bosonic part the Riemann curvature tensor $\bR$,
and a fermionic part the gravitino field $\bFRS$,
\begin{equation}
  [ \bDD + \bARS , \bDD + \bARS ]
  =
  \bR
  +
  \bFRS
  \ .
\end{equation}
The Riemann tensor $\bR$ is a bivector of bivectors,
while the gravitino field $\bFRS$ is a bivector of spinors,
\begin{subequations}
\begin{alignat}{3}
  \bR
  &\equiv
  \bR_{kl}
  \, \bgamma^k \wedgie \bgamma^l
  &&\equiv
  R_{klmn}
  \,
  ( \bgamma^k \wedgie \bgamma^l )
  \otimes
  ( \bgamma^m \wedgie \bgamma^n )
  \ ,
  \\
  \bFRS
  &\equiv
  \FRS_{kl} \, \bgamma^k \wedgie \bgamma^l
  &&\equiv
  \FRS_{kla} ( \bgamma^k \wedgie \bgamma^l ) \otimes \bepsilon^a
  \ ,
\end{alignat}
\end{subequations}
with implicit summation over distinct antisymmetric bivector indices $kl$
(and $mn$).
Compared to its usual general relativistic expression,
the Riemann tensor $\bR$ contains an additional part
$[ \bARS , \bARS ] = ( \{ \ARS_k \ARS_l \} \spinordot ) \, \bgamma^k \wedgie \bgamma^l$
proportional to a square of the gravitino potential $\bARS$.
The factor $\{ \ARS_k \ARS_l \} \spinordot$,
a symmetric outer product of like-handed spinors (in 4 spacetime dimensions)
is dictated by the requirement that the spinor product
transform like a bivector
($\ARS_k \spinordot \equiv \ARS_k^\top \varepsilon$,
with $\varepsilon$ the spinor metric~(\ref{spinormetric}),
denotes the row spinor associated with the column spinor $\ARS_k$).
%see \cite{Hamilton:2022a} for details.
The contribution $[ \bARS , \bARS ]$
plays no role in the present paper,
since the gravitino field vanishes in the background spacetime,
so $[ \bARS , \bARS ]$ is of quadratic order in the gravitino field,
and can be neglected to linear order of wave amplitudes.
Each of the 6 bivector components $\FRS_{kl}$ of the gravitino field $\bFRS$
is a spinor,
\begin{equation}
  \FRS_{kl}
  \equiv
  \DD_k \psi_{l} - \DD_l \psi_{k}
  \equiv
  ( \DD_k \psi_{la} - \DD_l \psi_{ka} ) \bepsilon^a
  \ .
\end{equation}
As usual,
the covariant derivative $\DD_k$ acts on both vector and spinor components.

Like massless fields of other nonzero spin,
massless spin~$\frac{3}{2}$ waves of opposite chirality do not mix.
The right and left-handed chiral components
${\FRSz}_{kl}$
of the gravitino field
(the tilde on ${\FRSz}$
signifies a right- or left-handed chiral component)
are
\begin{equation}
\label{FRSz}
  {\FRSz}_{kl}
  \equiv
  \tfrac{1}{2}
  ( 1 \pm \gamma_5 )
  \left( \FRS_{kl} \mp \im \, \varepsilon_{kl}{}^{mn} \FRS_{mn} \right)
\end{equation}
implicitly summed over distinct antisymmetric indices $mn$,
where the upper and lower signs are respectively right- and left-handed.
It can be shown that
\begin{equation}
  \bgamma^k \FRSz_{kl}
  =
  \bgamma_l \DD^k \ARSz_k
  \ ,
\end{equation}
so imposing the Lorenz gauge condition~(\ref{ARSLorenz}) ensures that
\begin{equation}
\label{FRSLorenz}
  \bgamma^k \FRSz_{kl}
  =
  0
  \ .
\end{equation}
A bivector spinor $\FRS$ has $6 \times 4 = 24$ components,
but the projection~(\ref{FRSz}) projects both spinor and bivector parts
into their chiral components,
leaving each chiral component with 6 complex components.
The Lorenz gauge condition~(\ref{ARSLorenz}) removes 2 components
from each chirality, leaving each with 4 physical components,
which is as it should be.
The right- and left-handed components of the gravitino field
${\FRSz}_{kl}$
subject to the gauge condition~(\ref{ARSLorenz})
are conveniently labeled ${\FRSz}_\sigma$ by their boost weights
$\sigma = +\tfrac{3}{2}$, $+\tfrac{1}{2}$, $-\tfrac{1}{2}$, $-\tfrac{3}{2}$
(spin weights $\varsigma = \sigma$ right-handed,
$\varsigma = -\sigma$ left-handed):
\begin{equation}
\label{FRSzcomponents}
  {\FRSz}_\sigma
  \equiv
  \left(
  \begin{array}{c}
  {\FRSz}_{+3/2}
  \\
  {\FRSz}_{+1/2}
  \\
  {\FRSz}_{-1/2}
  \\
  {\FRSz}_{-3/2}
  \end{array}
  \right)
  \equiv
  \left\{
  \begin{array}{ll}
  \left(
  \begin{array}{c}
  {\FRSz}_{v+\smallUpup}
  \\
  {\FRSz}_{vu\smallUpup}
  \\
  {\FRSz}_{uv\smallDowndown}
  \\
  {\FRSz}_{u-\smallDowndown}
  \end{array}
  \right)
  &
  \mbox{right}
  \\
  \left(
  \begin{array}{c}
  {\FRSz}_{v-\smallUpdown}
  \\
  {\FRSz}_{vu\smallUpdown}
  \\
  {\FRSz}_{uv\smallDownup}
  \\
  {\FRSz}_{u+\smallDownup}
  \end{array}
  \right)
  &
  \mbox{left}
  \end{array}
  \right.
  \ .
\end{equation}
The boost $\pm\tfrac{3}{2}$ components
%${\FRSz}_{\pm 3/2}$
are gauge invariant.
The condition~(\ref{FRSLorenz})
that follows from the Lorenz gauge condition~(\ref{ARSLorenz})
imposes on the boost $\pm\tfrac{1}{2}$ components
%${\FRSz}_{\pm 1/2}$
the conditions
\begin{align}
  {\FRSz}_{1/2}
  &=
  \left\{
  \begin{array}{ll}
  {\FRSz}_{vu\smallUpup}
  =
  {\FRSz}_{v+\smallDowndown}
  &
  \mbox{right}
  \\
  {\FRSz}_{vu\smallUpdown}
  =
  -
  {\FRSz}_{v-\smallDownup}
  &
  \mbox{left}
  \end{array}
  \right.
  \ ,
\nonumber
\\
  {\FRSz}_{-1/2}
  &=
  \left\{
  \begin{array}{ll}
  {\FRSz}_{uv\smallDowndown}
  =
  -
  {\FRSz}_{u-\smallUpup}
  &
  \mbox{right}
  \\
  {\FRSz}_{uv\smallDownup}
  =
  {\FRSz}_{u+\smallUpdown}
  &
  \mbox{left}
  \end{array}
  \right.
  \ .
\end{align}

For brevity, denote the supersymmetric covariant derivative by
${\cal D}_k \equiv \DD_k + \Psi_k$.
Wave equations for the chiral gravitino fields $\bFRSz$
follow from the Jacobi identity
(also known as Bianchi identities)
\begin{equation}
\label{bianchi}
  {\cal D}_{[k} [ {\cal D}_l , {\cal D}_{m]} ]
  =
  [ {\cal D}_{[k} , {\cal D}_l ] {\cal D}_{m]}
  \ .
\end{equation}
The brackets around indices mean antisymmetrize over bracketed indices.
The fermionic part of the Jacobi identity~(\ref{bianchi}) is
\begin{equation}
\label{bianchifermion}
  \DD_{[k} \FRS_{lm]}
  =
  \tfrac{1}{2} \bR_{[kl} \ARS_{m]}
\end{equation}
(the right-hand side of equation~(\ref{bianchifermion})
comes from the Riemann operator acting on the spinor indices $a$
of the gravitino potential $\ARS_{ma}$;
the possible contribution $R_{[klm]n} \ARS^n$
from the Riemann operator acting on the vector indices $m$
of $\ARS_{ma}$ vanishes to linear order
because the vanishing of torsion in the background spacetime implies
$R_{[klm]n} = 0$ in the background).
Applied to the right- and left-handed gravitino fields,
the curl on the left hand side of equation~(\ref{bianchifermion})
can be replaced by a divergence,
\begin{subequations}
\begin{alignat}{3}
  \DD_{[+} \FRSz_{-v]}
  &=
  \mp
  \DD^k \FRSz_{kv}
  \ , \quad
  &
  \DD_{[v} \FRSz_{u+]}
  &=
  \pm
  \DD^k \FRSz_{k+}
  \ ,
\\
  \DD_{[+} \FRSz_{-u]}
  &=
  \pm
  \DD^k \FRSz_{ku}
  \ , \quad
  &
  \DD_{[v} \FRSz_{u-]}
  &=
  \mp
  \DD^k \FRSz_{k-}
  \ ,
\end{alignat}
\end{subequations}
where the upper and lower signs are respectively right- and left-handed.

Focus on the right-handed gravitino field;
the left-handed field is quite similar.
Define the scaled right-handed gravitino field $\hat{\FRS}_{kl}$
(with a hat instead of a tilde over the $\FRS$)
by
\begin{equation}
  {\FRSz}_{kl}
  =
  f_{3/2} \,
  \hat{\FRS}_{kl}
  \ ,
\end{equation}
in accordance with equation~(\ref{psifs}).
The Jacobi identity~(\ref{bianchifermion}) provides
6 independent equations governing the 4 components
of the right-handed gravitino field.
The left hand side of the Jacobi identity~(\ref{bianchifermion}) is
(equations~(\ref{npRSR}) in Appendix~\ref{Dapp}
give these equations in a general spacetime):
\begin{align}
\label{RScomponents}
  &
  \DD^k
  \left(
  \begin{array}{c}
  \hat{\FRS}_{kv\smallUpup}
  \\
  \hat{\FRS}_{k+\smallUpup}
  \\
  %=
  %\hat{\FRS}_{ku\smallUpup}
  \hat{\FRS}_{k-\smallUpup}
  \\
  \hat{\FRS}_{k+\smallDowndown}
  %=
  %\hat{\FRS}_{kv\smallDowndown}
  \\
  \hat{\FRS}_{k-\smallDowndown}
  \\
  \hat{\FRS}_{ku\smallDowndown}
  \end{array}
  \right)
  =
\nonumber
\\
  &
  {f_{3/2} \over \sqrt{2} \rho}
  \left(
  \begin{array}{c}
  \rhom{}^{-2}
  \ethsi{1/2}{v}
  \rhom{}^2
  \hat{\FRS}_{+1/2}
  +
  \rhom{}^2
  \ethsi{3/2}{-}
  \rhom{}^{-2}
  \hat{\FRS}_{+3/2}
  \\
  \rhom{}^{-2}
  \ethsi{1/2}{+}
  \rhom{}^2
  \hat{\FRS}_{+1/2}
  \mp
  \rhom{}^2
  \ethsi{3/2}{u}
  \rhom{}^{-2}
  \hat{\FRS}_{+3/2}
  \\
  \ethsi{-1/2}{v}
  \hat{\FRS}_{-1/2}
  +
  \ethsi{1/2}{-}
  \hat{\FRS}_{+1/2}
  \\
  \mp \,
  \ethsi{1/2}{u}
  \hat{\FRS}_{+1/2}
  +
  \ethsi{-1/2}{+}
  \hat{\FRS}_{-1/2}
  \\
  \rhom{}^{-2}
  \ethsi{-1/2}{-}
  \rhom{}^2
  \hat{\FRS}_{-1/2}
  -
  \rhom{}^2
  \ethsi{-3/2}{v}
  \rhom{}^{-2}
  \hat{\FRS}_{-3/2}
  \\
  \mp \,
  \rhom{}^{-2}
  \ethsi{-1/2}{u}
  \rhom{}^2
  \hat{\FRS}_{-1/2}
  -
  \rhom{}^2
  \ethsi{-3/2}{+}
  \rhom{}^{-2}
  \hat{\FRS}_{-3/2}
  \end{array}
  \right)
  \ ,
\end{align}
where the upper sign of $\pm$ or $\mp$ is outside the horizon, $\Delta_x \,{>}\, 0$,
the lower sign inside the horizon, $\Delta_x \,{<}\, 0$.
Since the radial operators
$\ethsi{}{\vu}$
commute with the angular operators
$\ethsi{}{\pm}$,
equations for each of the four components $\hat{\FRS}_\sigma$
can be obtained by combining equations~(\ref{RScomponents}) in pairs.
Derivatives of the right hand side of the Jacobi identity~(\ref{bianchifermion})
turn the gravitino potential $\ARS$ into the gravitino field $\FRS$,
yielding contributions proportional to chiral components of the Riemann tensor
that are nonvanishing in the background spacetime,
namely the spin~0 component $\Cz_0$ of the Weyl tensor, equation~(\ref{Cz0}),
and the Ricci scalar $R$,
whose only nonvanishing contribution in the conformally separable spacetimes
is from the cosmological constant, where $R = 4 \Lambda$.
The result is 6 equations for the four components $\hat{\FRS}_\sigma$
of the right-handed gravitino field,
\begin{subequations}
\label{spinthreehalfwave}
\begin{align}
  \Bigl(
  \mp
  \rhom^{-2} \ethsi{}{v} \rhom^4 \ethsi{}{u} \rhom^{-2}
  -
  \rhom^{-2} \ethsi{}{+} \rhom^4 \ethsi{}{-} \rhom^{-2}
  &
\\
\nonumber
  +
  \rho^2 ( 2 \Cz_0 - \tfrac{1}{12} R )
  &
  \Bigr)
  \hat{\FRS}_{+3/2}
  =
  0
  \ ,
\\
  \Bigl(
  \mp
  \rhom^{2} \ethsi{}{u} \rhom^{-4} \ethsi{}{v} \rhom^{2}
  -
  \rhom^{2} \ethsi{}{-} \rhom^{-4} \ethsi{}{+} \rhom^{2}
  &
\\
\nonumber
  +
  \rho^2 ( 2 \Cz_0 - \tfrac{1}{12} R )
  &
  \Bigr)
  \hat{\FRS}_{+1/2}
  =
  0
  \ ,
\\
  \Bigl(
  \mp
  \ethsi{}{v} \ethsi{}{u}
  -
  \ethsi{}{+} \ethsi{}{-}
  +
  \rho^2 ( - 4 \Cz_0 - \tfrac{1}{12} R )
  &
  \Bigr)
  \hat{\FRS}_{+1/2}
  =
  0
  \ ,
\\
  \Bigl(
  \mp
  \ethsi{}{u} \ethsi{}{v}
  -
  \ethsi{}{-} \ethsi{}{+}
  +
  \rho^2 ( - 4 \Cz_0 - \tfrac{1}{12} R )
  &
  \Bigr)
  \hat{\FRS}_{-1/2}
  =
  0
  \ ,
\\
  \Bigl(
  \mp
  \rhom^{2} \ethsi{}{v} \rhom^{-4} \ethsi{}{u} \rhom^{2}
  -
  \rhom^{2} \ethsi{}{+} \rhom^{-4} \ethsi{}{-} \rhom^{2}
  &
\\
\nonumber
  +
  \rho^2 ( 2 \Cz_0 - \tfrac{1}{12} R )
  &
  \Bigr)
  \hat{\FRS}_{-1/2}
  =
  0
  \ ,
\\
  \Bigl(
  \mp
  \rhom^{-2} \ethsi{}{u} \rhom^4 \ethsi{}{v} \rhom^{-2}
  -
  \rhom^{-2} \ethsi{}{-} \rhom^4 \ethsi{}{+} \rhom^{-2}
  &
\\
\nonumber
  +
  \rho^2 ( 2 \Cz_0 - \tfrac{1}{12} R )
  &
  \Bigr)
  \hat{\FRS}_{-3/2}
  =
  0
  \ .
\end{align}
\end{subequations}
Equations~(\ref{spinthreehalfwave}) can be recast
in terms of the wave operators defined by~(\ref{waveoperatorsraiselower}) as
\begin{subequations}
\label{spinthreehalfwavesq}
\begin{align}
\label{spinthreehalfwavesqp32}
  \left(
  \sq{3/2}{vu}
  -
  \sq{3/2}{+-}
  -
  \tfrac{1}{3} \rho^2 \Lambda
  \right)
  \hat{\FRS}_{+3/2}
  &=
  0
  \ ,
\\
\label{spinthreehalfwavesqp12a}
  \bigl(
  \sq{1/2}{uv}
  -
  \sq{1/2}{-+}
  -
  4 \rho^2 \Cz_0
  -
  \tfrac{1}{3} \rho^2 \Lambda
  \bigr)
  \hat{\FRS}_{+1/2}
  &=
  0
  \ ,
\\
\label{spinthreehalfwavesqp12b}
  \bigl(
  \sq{1/2}{vu}
  -
  \sq{1/2}{+-}
  -
  4 \rho^2 \Cz_0
  -
  \tfrac{1}{3} \rho^2 \Lambda
  \bigr)
  \hat{\FRS}_{+1/2}
  &=
  0
  \ ,
\\
\label{spinthreehalfwavesqm12a}
  \bigl(
  \sq{-1/2}{uv}
  -
  \sq{-1/2}{-+}
  -
  4 \rho^2 \Cz_0
  -
  \tfrac{1}{3} \rho^2 \Lambda
  \bigr)
  \hat{\FRS}_{-1/2}
  &=
  0
  \ ,
\\
\label{spinthreehalfwavesqm12b}
  \bigl(
  \sq{-1/2}{vu}
  -
  \sq{-1/2}{+-}
  -
  4 \rho^2 \Cz_0
  -
  \tfrac{1}{3} \rho^2 \Lambda
  \bigr)
  \hat{\FRS}_{-1/2}
  &=
  0
  \ ,
\\
\label{spinthreehalfwavesqm32}
  \bigl(
  \sq{-3/2}{uv}
  -
  \sq{-3/2}{-+}
  -
  \tfrac{1}{3} \rho^2 \Lambda
  \bigr)
  \hat{\FRS}_{-3/2}
  &=
  0
  \ .
\end{align}
\end{subequations}
The wave equations~(\ref{spinthreehalfwavesqp32})
and~(\ref{spinthreehalfwavesqm32})
for $\hat{\FRS}_{+3/2}$ and $\hat{\FRS}_{-3/2}$
are separable,
after the adjustment~(\ref{waveoperatorsLambda}) to the wave operators
if the cosmological constant $\Lambda$ is nonvanishing.
The separated solutions can be written~(\ref{psis}),
and the corresponding eigenvalues are $\lambda_{\sigma\varsigma}$.

The Teukolsky-Starobinski identities for the gravitino field
do not work out quite so nicely
in the conformally separable spacetimes as they do in vacuum spacetimes.
The Teukolsky-Starobinski identities for the gravitino field are
\begin{subequations}
\label{spinthreehalvesteukolskystarobinski}
\begin{align}
\label{spinthreehalvesteukolskystarobinski1}
  \left(
  \square_{uv}
  \,
  %\ethsi{-\frac{1}{2}}{u} \, \ethsi{\frac{1}{2}}{u} \, \ethsi{3/2}{u}
  \ethsi{}{u}^3
  -
  %\ethsi{-\frac{1}{2}}{u} \, \ethsi{\frac{1}{2}}{u} \, \ethsi{3/2}{u}
  \ethsi{}{u}^3
  \,
  \square_{vu}
  \right)
  \hat{\FRS}_{+3/2}
  &=
  \tfrac{1}{6} | \Delta_x |^{3/2}
  K_x
  \hat{\FRS}_{+3/2}
  \ ,
\\
\label{spinthreehalvesteukolskystarobinski2}
  \left(
  \square_{vu}
  \,
  %\ethsi{\frac{1}{2}}{v} \, \ethsi{-\frac{1}{2}}{v} \, \ethsi{-3/2}{v}
  \ethsi{}{v}^3
  -
  %\ethsi{\frac{1}{2}}{v} \, \ethsi{-\frac{1}{2}}{v} \, \ethsi{-3/2}{v}
  \ethsi{}{v}^3
  \,
  \square_{uv}
  \right)
  \hat{\FRS}_{-3/2}
  &=
  -
  \tfrac{1}{6} | \Delta_x |^{3/2}
  K_x
  \hat{\FRS}_{-3/2}
  \ ,
\\
\label{spinthreehalvesteukolskystarobinski3}
  \left(
  \square_{-+}
  \,
  %\ethsi{-\frac{1}{2}}{-} \, \ethsi{\frac{1}{2}}{-} \, \ethsi{3/2}{-}
  \ethsi{}{-}^3
  -
  %\ethsi{-\frac{1}{2}}{-} \, \ethsi{\frac{1}{2}}{-} \, \ethsi{3/2}{-}
  \ethsi{}{-}^3
  \,
  \square_{+-}
  \right)
  \hat{\FRS}_{+3/2}
  &=
  \tfrac{1}{6} \Delta_y^{3/2}
  K_y
  \hat{\FRS}_{+3/2}
  \ ,
\\
\label{spinthreehalvesteukolskystarobinski4}
  \left(
  \square_{+-}
  \,
  %\ethsi{\frac{1}{2}}{+} \, \ethsi{-\frac{1}{2}}{+} \, \ethsi{-3/2}{+}
  \ethsi{}{+}^3
  -
  %\ethsi{\frac{1}{2}}{+} \, \ethsi{-\frac{1}{2}}{+} \, \ethsi{-3/2}{+}
  \ethsi{}{+}^3
  \,
  \square_{-+}
  \right)
  \hat{\FRS}_{-3/2}
  &=
  -
  \tfrac{1}{6} \Delta_y^{3/2}
  K_y
  \hat{\FRS}_{-3/2}
  \ ,
\end{align}
\end{subequations}
where
$K_x$ and $K_y$ are radial and angular functions defined by
\begin{equation}
\label{Kxy}
  K_x
  \equiv
%  {\ddi{5} \Delta_x \over \dd x^5}
%  +
%  20 a^2
%  {\ddi{3} \Delta_x \over \dd x^3}
%  +
%  64 a^4
%  {\dd \Delta_x \over \dd x}
  R^4 {\ddi{5} R^4 \Delta_x \over \dd r^5}
  \ , \quad
  K_y
  \equiv
  {\ddi{5} \Delta_y \over \dd y^5}
  \ .
\end{equation}
The radial and angular functions
$K_x$ and $K_y$
vanish in $\Lambda$-Kerr(-Newman) spacetimes,
where $R^4 \Delta_x$ and $\Delta_y$
are quartic in respectively $r$ and $y$,
but they do not vanish in the conformally separable spacetimes.
The wave operators $\square$ in
the Teukolsky-Starobinski identities~(\ref{spinthreehalvesteukolskystarobinski})
are for the wave operators defined by~(\ref{waveoperatorsraiselower}),
{\em not\/} adjusted for a cosmological constant
per the modification~(\ref{waveoperatorsLambda}).
We have not found a comparably simple set of identities
if the wave operators are adjusted for a cosmological constant
per~(\ref{waveoperatorsLambda}).
For this reason,
the cosmological constant is taken to be zero, $\Lambda = 0$,
in the remainder of this section~\ref{spinthreehalves-sec}.

For Kerr spacetime (without a charge and without a cosmological constant),
the Teukolsky-Starobinski identities~(\ref{spinthreehalvesteukolskystarobinski1})
show that 
$X_{+3/2}$ lowered three times (with $\ethsi{}{u}^3$)
is proportional to $X_{-3/2}$,
and
$X_{-3/2}$ raised three times (with $\ethsi{}{v}^3$)
is proportional to $X_{+3/2}$;
and similarly
$Y_{+3/2}$ lowered three times (with $\ethsi{}{-}^3$)
is proportional to $Y_{-3/2}$,
and
$Y_{-3/2}$ raised three times (with $\ethsi{}{+}^3$)
is proportional to $Y_{+3/2}$.
With a convenient choice of relative normalization,
the eigenfunctions are related by
\begin{subequations}
\begin{align}
\label{Xthreehalf}
  \ethsi{}{u}^3
  X_{+3/2}
  =
  -
  \mu_x
  X_{-3/2}
  &
  \ , \ \ 
  \ethsi{}{v}^3
  X_{-3/2}
  =
  \sgn(\Delta_x) \mu_x
  X_{+3/2}
  \ ,
\\
\label{Ythreehalf}
  \ethsi{}{-}^3
  Y_{+3/2}
  =
  \mu_y
  Y_{-3/2}
  &
  \ , \ \ 
  \ethsi{}{+}^3
  Y_{-3/2}
  =
  \mu_y
  Y_{+3/2}
  \ .
\end{align}
\end{subequations}
Solving for $\mu_x^2$ and $\mu_y^2$ in
\begin{subequations}
\label{raiseraiseraiselowerlowerlower}
\begin{align}
  \left(
  \ethsi{}{v}^3
  \,
  \ethsi{}{u}^3
  +
  \sgn(\Delta_x)
  \mu_x^2
  \right)
  X_{+3/2}
  &=
  0
  \ ,
\\
  \left(
  \ethsi{}{+}^3
  \,
  \ethsi{}{-}^3
  -
  \mu_y^2
  \right)
  Y_{+3/2}
  &=
  0
  \ ,
\end{align}
\end{subequations}
given that
$\FRS_{+3/2}$ satisfies
\mbox{$( \square_{vu} - \lambda ) \FRS_{+3/2}$}
=
\mbox{$( \square_{+-} - \lambda ) \FRS_{+3/2}$} = 0,
shows that $\mu_x$ and $\mu_y$ are related to the eigenvalue $\lambda$ by,
for Kerr,
\begin{align}
  &
  \mu_x = \mu_y
  =
\\
\nonumber
  &
  \lambda^3
  -
  \lambda
  \left[
  \tfrac{1}{12}
  +
  4 a \wel ( a \wel + m )
  \right]
  +
  \tfrac{1}{108}
  +
  \tfrac{2}{3}
  a \wel ( 4 a \wel + m )
  \ .
\end{align}
%\begin{subequations}
%\begin{align}
%  \mu_y^2
%  &=
%  \lambda^3
%  -
%  \lambda
%  \left[
%  \tfrac{1}{12}
%  +
%  4 a \wel ( a \wel + m )
%  +
%  \tfrac{1}{18}
%  a^2 \Lambda ( - 7 + \tfrac{1}{6} a^2 \Lambda )
%  \right]
%\nn
%  &\quad
%  + \,
%  \tfrac{1}{108}
%  +
%  \tfrac{2}{3}
%  a \wel ( 4 a \wel + m )
%\nn
%  &\quad
%  - \,
%  \tfrac{1}{108}
%  a^2 \Lambda
%  \bigl[
%  -
%  11
%  +
%  24 ( a \wel + m ) ( 4 a \wel + 3 m )
%\nn
%  &\quad
%  + \,
%  \tfrac{1}{3}
%  a^2 \Lambda
%  ( 11 + \tfrac{1}{9} a^2 \Lambda )
%  \bigr]
%  \ ,
%\\
%  \mu_x^2
%  &=
%  \mu_y^2
%  %-
%  %\tfrac{1}{3}
%  %\lambda \Qelecbh^2 \Lambda
%  -
%  \tfrac{1}{6}
%  \Mbh^2 \Lambda
%%\nn
%%  &\quad
%%  - \,
%%  \Qelecbh^2
%%  \left[
%%  2 \wel^2
%%  +
%%  \tfrac{1}{9}
%%  \Lambda
%%  \left(
%%  1
%%  -
%%  \tfrac{1}{3}
%%  a^2 \Lambda
%%  \right)
%%  \right]
%  \ .
%\end{align}
%\end{subequations}
The eigenvalues agree with those in
\cite[eqs.~(12) and~(15)]{TorresdelCastillo:1992}
with the substitutions
(there $\leftrightarrow$ here)
$\lambdabar = 2 \lambda - \frac{1}{3}$
and $\sigma = \wel$.

\section{Spin~2 waves}
\label{spintwo-sec}

Wave equations for massless spin~2 waves,
gravitational waves,
follow from the Bianchi identities,
which imply the Weyl evolution equations
\begin{equation}
\label{weylevolution}
  D^k C_{klmn}
  =
  J_{lmn}
  \ ,
\end{equation}
where $C_{klmn}$ is the Weyl tensor, the traceless part of the Riemann tensor,
and $J_{lmn}$ is the Weyl current,
defined in terms of the Einstein tensor $G_{mn}$ and its trace $G$ by
\begin{equation}
\label{weylcurrent}
  J_{lmn}
  \equiv
  \tfrac{1}{2}
  ( D_m G_{ln} - D_n G_{lm} )
  -
  \tfrac{1}{6}
  ( \gamma_{ln} D_m G - \gamma_{lm} D_n G )
  \ .
\end{equation}
Like the Weyl tensor,
the Weyl current $J_{lmn}$ is traceless and satisfies the cyclic symmetry
$J_{[lmn]} = 0$.
It satisfies the conservation law
\begin{equation}
\label{weylcurrentconservation}
  \DD^l J_{lmn}
  =
  0
  \ ,
\end{equation}
which can be thought of as the gravitational analog of Maxwell's equations.

Like massless fields of other nonzero spin,
massless spin~$2$ waves of opposite chirality do not mix.
The right and left-handed chiral components
$\Cz_{klmn}$
of the gravitational field constitute the complex self-dual Weyl tensor
(the tilde on ${\Cz}$
signifies a right- or left-handed chiral component)
\begin{equation}
\label{Cz}
  \Cz_{klmn}
  \equiv
  \tfrac{1}{4}
  \left(
  \delta_k^p \delta_l^q \mp \im \, \varepsilon_{kl}{}^{pq}
  \right)
  \left(
  \delta_m^r \delta_n^s \mp \im \, \varepsilon_{mn}{}^{rs}
  \right)
  C_{pqrs}
  \ ,
\end{equation}
implicitly summed over distinct antisymmetric indices $pq$ and $rs$,
the upper and lower signs being respectively right- and left-handed.
The wave equations for the right- and left-handed fields
follow from the Weyl evolution equations~(\ref{weylevolution}),
\begin{equation}
\label{weylevolutionz}
  D^k \Cz_{klmn}
  =
  \Jz_{lmn}
  \ ,
\end{equation}
where $\Jz_{lmn}$ is the complex Weyl current
\begin{equation}
  \Jz_{lmn}
  \equiv
  \tfrac{1}{2}
  \left(
  \delta_m^r \delta_n^s \mp \im \, \varepsilon_{mn}{}^{rs}
  \right)
  J_{lrs}
  \ ,
\end{equation}
the upper and lower signs being again respectively right- and left-handed.

The complex Weyl tensor $\Cz_{klmn}$
is a traceless, symmetric, complex $3 \times 3$ matrix of bivectors,
with five complex degrees of freedom.
It is convenient to label the components $\Cz_\sigma$ by their boost weights
$\sigma = +2$, $+1$, 0, $-1$, $-2$
(spin weights $\varsigma = \sigma$ right-handed,
$\varsigma = -\sigma$ left-handed):
\begin{equation}
\label{Czcomponent}
  \Cz_\sigma
  \equiv
  \left(
  \begin{array}{l}
  \Cz_{+2} \\
  \Cz_{+1} \\
  \Cz_{0} \\
  \Cz_{-1} \\
  \Cz_{-2}
  \end{array}
  \right)
  \equiv
  \left\{
  \begin{array}{ll}
  \left(
  \begin{array}{l}
  \Cz_{v+ v+} \\
  \Cz_{vuv+} \\
  \Cz_{vuvu} \\
  \Cz_{uvu-} \\
  \Cz_{u- u-}
  \end{array}
  \right)
  & \mbox{right}
  \ ,
  \\
  \left(
  \begin{array}{l}
  \Cz_{v- v-} \\
  \Cz_{vuv-} \\
  \Cz_{vuvu} \\
  \Cz_{uvu+} \\
  \Cz_{u+ u+}
  \end{array}
  \right)
  & \mbox{left}
  \ .
  \end{array}
  \right.
\end{equation}
The five components~(\ref{Czcomponent}) are commonly
\cite{Chandrasekhar:1983}
denoted (minus) $\Psi_0$, $\Psi_1$, $\Psi_2$, $\Psi_3$, and $\Psi_4$,
equations~(\ref{chandraC}),
but the notation~(\ref{Czcomponent})
makes manifest their transformation properties.
The propagating component is, equation~(\ref{psir}),
$\Cz_{-2} = -\Psi_4$ outgoing,
$\Cz_{+2} = -\Psi_0$ ingoing.

Focus on the right-handed gravitational field;
the left-handed field is quite similar.
Define the scaled Weyl tensor $\hat{C}_\sigma$
and scaled Weyl current $\hat{J}_{lmn}$
(with hat instead of tildes over the $C$ and $J$)
by
\begin{equation}
  \Cz_\sigma
  =
  f_2 \,
  \hat{C}_\sigma
  \ , \quad
  \Jz_{lmn}
  =
  f_2 \,
  \hat{J}_{lmn}
  \ ,
\end{equation}
in accordance with equation~(\ref{psifs}).
The left hand side of the complex Weyl evolution equation~(\ref{weylevolutionz})
has 8 nonvanishing components
(equations~(\ref{npWeylR}) in Appendix~\ref{Dapp}
give these equations in a general spacetime):
\begin{align}
\label{DCz}
  &
  D^k
  \left(
  \begin{array}{c}
%5
  \hat{C}_{kvv+}
  \\
%7
  \hat{C}_{k+ v+}
  \\
%1
  \hat{C}_{k- v+}
  \\
%3
  \hat{C}_{kuv+}
  \\
%2
  \hat{C}_{k+ u-}
  \\
%4
  \hat{C}_{kvu-}
  \\
%6
  \hat{C}_{kuu-}
  \\
%8
  \hat{C}_{k- u-}
  \end{array}
  \right)
  =
  \left(
  \begin{array}{c}
%5
  - \, 3 \Gamma_{+ vv} \hat{C}_{0}
  \\
%7
  - \, 3 \Gamma_{+ v+} \hat{C}_{0}
  \\
%1
  %- \, 2 \Gamma_{+ vv} \hat{C}_{-1} + \, \Gamma_{- u-} \hat{C}_{+2}
  0
  \\
%3
  %- \, 2 \Gamma_{+ v+} \hat{C}_{-1} + \Gamma_{- uu} \hat{C}_{+2}
  0
  \\
%2
  %- \, 2 \Gamma_{- uu} \hat{C}_{+1} + \Gamma_{+ v+} \hat{C}_{-2}
  0
  \\
%4
  %- \, 2 \Gamma_{- u-} \hat{C}_{+1} + \Gamma_{+ vv} \hat{C}_{-2}
  0
  \\
%6
  - \, 3 \Gamma_{- uu} \hat{C}_{0}
  \\
%8
  - \, 3 \Gamma_{- u-} \hat{C}_{0}
  \end{array}
  \right)
\nonumber
\\
  &\quad
  +
  {f_2 \over \sqrt{2} \rho}
  \left(
  \begin{array}{c}
%5
  \rhom{}^{-3}
  \ethsi{1}{v}
  \rhom{}^3
  \hat{C}_{+1}
  +
  \rhom{}^3
  \ethsi{2}{-}
  \rhom{}^{-3}
  \hat{C}_{+2}
  \\
%7
  \rhom{}^{-3}
  \ethsi{1}{+}
  \rhom{}^3
  \hat{C}_{+1}
  \mp
  \rhom{}^3
  \ethsi{2}{u}
  \rhom{}^{-3}
  \hat{C}_{+2}
  \\
%1
  \rhom{}^{-1}
  \ethsi{0}{v}
  \rhom{}
  \hat{C}_{0}
  +
  \rhom{}
  \ethsi{1}{-}
  \rhom{}^{-1}
  \hat{C}_{+1}
  \\
%3
  \rhom{}^{-1}
  \ethsi{0}{+}
  \rhom{}
  \hat{C}_{0}
  \mp
  \rhom{}
  \ethsi{1}{u}
  \rhom{}^{-1}
  \hat{C}_{+1}
  \\
%2
  \pm \,
  \rhom{}^{-1}
  \ethsi{0}{u}
  \rhom{}
  \hat{C}_{0}
  -
  \rhom{}
  \ethsi{-1}{+}
  \rhom{}^{-1}
  \hat{C}_{-1}
  \\
%4
  - \,
  \rhom{}^{-1}
  \ethsi{0}{-}
  \rhom{}
  \hat{C}_{0}
  -
  \rhom{}
  \ethsi{-1}{v}
  \rhom{}^{-1}
  \hat{C}_{-1}
  \\
%6
  \pm \,
  \rhom{}^{-3}
  \ethsi{-1}{u}
  \rhom{}^3
  \hat{C}_{-1}
  -
  \rhom{}^3
  \ethsi{-2}{+}
  \rhom{}^{-3}
  \hat{C}_{-2}
  \\
%8
  - \,
  \rhom{}^{-3}
  \ethsi{-1}{-}
  \rhom{}^3
  \hat{C}_{-1}
  -
  \rhom{}^3
  \ethsi{-2}{v}
  \rhom{}^{-3}
  \hat{C}_{-2}
  \end{array}
  \right)
  %=
  %\hat{J}_{lmn}
  \ .
\end{align}
The non-derivative terms to the immediate right of the equals sign
in equations~(\ref{DCz}) involve products
of boost and spin weight $\pm 2$ components of Lorentz connections,
also known as shears,
with the boost and spin weight 0 Weyl tensor $\Cz_0$, equation~(\ref{Cz0}).
As \cite{Chandrasekhar:1983} has emphasized,
although the shears vanish in the conformally separable background,
their derivatives,
equations~(\ref{Dshears}),
yield boost/spin weight $\pm 2$ Weyl components $\Cz_{\pm 2}$
that contribute to the wave equations for those components.
For example,
the difference of
$\rhom{}^{-3} \ethsi{1}{+} \rhom{}^{3}$ times the first row
of equations~(\ref{DCz}) and
$\rhom{}^{-3} \ethsi{1}{v} \rhom{}^{3}$ times the second row
%eliminates the $\hat{C}_{+1}$ component, yielding
yields
\begin{align}
\label{DCz2}
  \left(
  \rhom{}^{-3}
  \ethsi{1}{v}
  \rhom{}^{6}
  \ethsi{2}{u}
  \rhom{}^{-3}
  \pm
  \rhom{}^{-3}
  \ethsi{1}{+}
  \rhom{}^{6}
  \ethsi{2}{-}
  \rhom{}^{-3}
  +
  6 \rho^2 \Cz_0
  \right)
  \hat{C}_{+2}
  &
\\
\nonumber
  =
  \rhom{}^{-3}
  \left(
  \ethsi{1}{v}
  \rhom{}^{3}
  \hat{J}_{+ v+}
  -
  \ethsi{1}{+}
  \rhom{}^{3}
  \hat{J}_{vv+}
  \right)
  &
  \ ,
\end{align}
the $6 \rho^2 \Cz_0$ term coming from the difference of derivatives
of the shears
$\Gamma_{+vv}$ and $\Gamma_{+v+}$,
top row of equation~(\ref{DshearsR}) with $\psi_{s\sigma\varsigma} = \Cz_0$.
Remarkably, the combination of derivatives of Weyl currents $\hat{J}_{lmn}$
on the right hand side of equation~(\ref{DCz2})
vanishes for the conformally separable solutions,
despite the fact that neither of the Weyl currents vanishes individually.
This proves to be true for all of the wave equations derived from
combining equations in pairs~(\ref{DCz}):
in all cases, the combinations of derivatives of Weyl currents
from the right hand sides of the Weyl evolution equations~(\ref{weylevolutionz})
vanish for the conformally separable spacetimes,
despite the fact that none of the Weyl currents vanishes individually.
The shear derivatives contribute precisely what is needed to make
the wave equations for the boost/weight $\pm 2$ components separable,
equations~(\ref{spintwowavesqp2}) and~(\ref{spintwowavesqm2}),
but the same shear derivatives complicate the wave equations for the
boost/weight spin $\pm 1$ components that follow from combining
respectively the top and bottom pairs of equations~(\ref{DCz}),
so these equations are omitted from equations~(\ref{spintwowave}) that follow.

Combining equations~(\ref{DCz}) in pairs
so as to eliminate one of the two spin components on each row
yields 6 second-order differential equations:
\begin{subequations}
\label{spintwowave}
\begin{align}
  \bigl(
  \mp
  \rhom^{-3} \ethsi{}{v} \rhom^6 \ethsi{}{u} \rhom^{-3}
  -
  \rhom^{-3} \ethsi{}{+} \rhom^6 \ethsi{}{-} \rhom^{-3}
  +
  6 \rho^2
  &
  \Cz_0
  \bigr)
  \hat{C}_{+2}
  =
  0
  \ ,
%\\
%  \bigl(
%  \mp
%  \rhom^{3} \ethsi{}{u} \rhom^{-6} \ethsi{}{v} \rhom^{3}
%  -
%  \rhom^{3} \ethsi{}{-} \rhom^{-6} \ethsi{}{+} \rhom^{3}
%  \bigr)
%  \hat{C}_{+1}
%  &=
%  0
%  \ ,
\\
  \bigl(
  \mp
  \rhom^{-1}
  \ethsi{-1}{v}
  \rhom^{2}
  \ethsi{0}{u}
  \rhom^{-1}
  -
  \rhom^{-1}
  \ethsi{-1}{+}
  \rhom^{2}
  \ethsi{0}{-}
  \rhom^{-1}
  \bigr)
  \hat{C}_{+1}
  &=
  0
  \ ,
\\
  \bigl(
  \mp
  \rhom
  \ethsi{-1}{v}
  \rhom^{-2}
  \ethsi{0}{u}
  \rhom
  -
  \rhom
  \ethsi{-1}{+}
  \rhom^{-2}
  \ethsi{0}{-}
  \rhom
  \bigr)
  \hat{C}_{0}
  &=
  0
  \ ,
\\
  \bigl(
  \mp
  \rhom
  \ethsi{1}{u}
  \rhom^{-2}
  \ethsi{0}{v}
  \rhom
  -
  \rhom
  \ethsi{1}{-}
  \rhom^{-2}
  \ethsi{0}{+}
  \rhom
  \bigr)
  \hat{C}_{0}
  &=
  0
  \ ,
\\
  \bigl(
  \mp
  \rhom^{-1}
  \ethsi{1}{u}
  \rhom^{2}
  \ethsi{0}{v}
  \rhom^{-1}
  -
  \rhom^{-1}
  \ethsi{1}{-}
  \rhom^{2}
  \ethsi{0}{+}
  \rhom^{-1}
  \bigr)
  \hat{C}_{-1}
  &=
  0
  \ ,
%\\
%  \bigl(
%  \mp
%  \rhom^{3} \ethsi{}{v} \rhom^{-6} \ethsi{}{u} \rhom^{3}
%  -
%  \rhom^{3} \ethsi{}{+} \rhom^{-6} \ethsi{}{-} \rhom^{3}
%  \bigr)
%  \hat{C}_{-1}
%  &=
%  0
%  \ ,
\\
  \bigl(
  \mp
  \rhom^{-3} \ethsi{}{u} \rhom^6 \ethsi{}{v} \rhom^{-3}
  -
  \rhom^{-3} \ethsi{}{-} \rhom^6 \ethsi{}{+} \rhom^{-3}
  +
  6 \rho^2
  &
  \Cz_0
  \bigr)
  \hat{C}_{-2}
  =
  0
  \ .
\end{align}
\end{subequations}
Equations~(\ref{spintwowave}) can be recast
in terms of the wave operators defined by~(\ref{waveoperatorsraiselower}) as
\begin{subequations}
\label{spintwowavesq}
\begin{align}
\label{spintwowavesqp2}
  \left(
  \sq{3/2}{vu}
  -
  \sq{3/2}{+-}
  \right)
  \hat{C}_{+2}
  &=
  0
  \ ,
%\\
%  \bigl(
%  \sq{1/2}{uv}
%  -
%  \sq{1/2}{-+}
%  -
%  12 \rho^2 \Cz_0
%  \bigr)
%  \hat{C}_{+1}
%  &=
%  0
%  \ ,
\\
\label{spintwowavesqp1}
  \bigl(
  \sq{1/2}{vu}
  -
  \sq{1/2}{+-}
  \bigr)
  \hat{C}_{+1}
  &=
  0
  \ ,
\\
\label{spintwowavesq0a}
  \bigl(
  \sq{1/2}{uv}
  -
  \sq{1/2}{-+}
  -
  2 \rho^2 \Cz_0
  \bigr)
  \hat{C}_{0}
  &=
  0
  \ ,
\\
\label{spintwowavesq0b}
  \bigl(
  \sq{1/2}{vu}
  -
  \sq{1/2}{+-}
  -
  2 \rho^2 \Cz_0
  \bigr)
  \hat{C}_{0}
  &=
  0
  \ ,
\\
\label{spintwowavesqm1}
  \bigl(
  \sq{-1}{uv}
  -
  \sq{-1}{-+}
  \bigr)
  \hat{C}_{-1}
  &=
  0
  \ ,
%\\
%  \bigl(
%  \sq{-1}{vu}
%  -
%  \sq{-1}{+-}
%  -
%  12 \rho^2 \Cz_0
%  \bigr)
%  \hat{C}_{-1}
%  &=
%  0
%  \ ,
\\
\label{spintwowavesqm2}
  \bigl(
  \sq{-2}{uv}
  -
  \sq{-2}{-+}
  \bigr)
  \hat{C}_{-2}
  &=
  0
  \ .
\end{align}
\end{subequations}
Note the similarity of the gravitational wave
equations~(\ref{spintwowavesqm1})--(\ref{spintwowavesqp1})
with boost weights $\pm 1$ and 0
to the electromagnetic wave equations~(\ref{spinonewavesq}).
The gravitational wave equations for not only the boost weight $\pm 2$
components but also the boost weight $\pm 1$ components are separable.
\cite{Chandrasekhar:1983} discusses this on p.~435,
where he says that the separability of the boost weight $\pm 1$ wave equations
is a gauge choice associated with the freedom of Lorentz transformations
of the tetrad frame;
but in the present case the tetrad frame is chosen to be aligned
with the principal null directions,
and that choice leads to separable wave equations for $\Cz_{\pm 1}$.

The Teukolsky-Starobinski identities for the boost weight $\pm 1$
components $\Cz_{\pm 1}$ of the gravitational field are the same as
those~(\ref{spinoneteukolskystarobinski}) for the electromagnetic field,
and the relations between separated factors $X_{\pm 1}$ and $Y_{\pm 1}$
for the gravitational field are the same as those~(\ref{raiseraiselowerlower})
for the electromagnetic field.

The Teukolsky-Starobinski identities for the boost weight $\pm 2$
components $\Cz_{\pm 2}$ of the gravitational field are
\begin{subequations}
\label{spintwoteukolskystarobinski}
\begin{align}
\label{spintwoteukolskystarobinski1}
  &\left(
  \square_{uv}
  \,
  %\ethsi{-1}{u} \, \ethsi{0}{u} \, \ethsi{1}{u} \, \ethsi{2}{u}
  \ethsi{}{u}^4
  -
  %\ethsi{-1}{u} \, \ethsi{0}{u} \, \ethsi{1}{u} \, \ethsi{2}{u}
  \ethsi{}{u}^4
  \,
  \square_{vu}
  \right)
  \hat{C}_{+2}
\nn
  &\quad=
  - K_x^{1/2} R^3 | \Delta_x |^{3/2}
  \ethsi{2}{u}
  \left(
  K_x^{1/2} R^{-3}
  \hat{C}_{+2}
  \right)
  \ ,
\\
\label{spintwoteukolskystarobinski2}
  &\left(
  \square_{vu}
  \,
  %\ethsi{1}{v} \, \ethsi{0}{v} \, \ethsi{-1}{v} \, \ethsi{-2}{v}
  \ethsi{}{v}^4
  -
  %\ethsi{1}{v} \, \ethsi{0}{v} \, \ethsi{-1}{v} \, \ethsi{-2}{v}
  \ethsi{}{v}^4
  \,
  \square_{uv}
  \right)
  \hat{C}_{-2}
\nn
  &\quad=
  K_x^{1/2} R^3 | \Delta_x |^{3/2}
  \ethsi{-2}{v}
  \left(
  K_x^{1/2} R^{-3}
  \hat{C}_{-2}
  \right)
  \ ,
\\
\label{spintwoteukolskystarobinski3}
  &\left(
  \square_{-+}
  \,
  %\ethsi{-1}{-} \, \ethsi{0}{-} \, \ethsi{1}{-} \, \ethsi{2}{-}
  \ethsi{}{-}^4
  -
  \ethsi{}{-}^4
  \,
  \square_{+-}
  \right)
  \hat{C}_{+2}
\nn
  &\quad=
  - K_y^{1/2} \Delta_y^{3/2}
  \ethsi{2}{-}
  \left(
  K_y^{1/2}
  \hat{C}_{+2}
  \right)
  \ ,
\\
\label{spintwoteukolskystarobinski4}
  &\left(
  \square_{+-}
  \,
  %\ethsi{1}{+} \, \ethsi{0}{+} \, \ethsi{-1}{+} \, \ethsi{-2}{+}
  \ethsi{}{+}^4
  -
  %\ethsi{1}{+} \, \ethsi{0}{+} \, \ethsi{-1}{+} \, \ethsi{-2}{+}
  \ethsi{}{+}^4
  \,
  \square_{-+}
  \right)
  \hat{C}_{-2}
\nn
  &\quad=
  K_y^{1/2} \Delta_y^{3/2}
  \ethsi{-2}{+}
  \left(
  K_y^{1/2}
  \hat{C}_{-2}
  \right)
  \ ,
\end{align}
\end{subequations}
where $R$ is the radial coordinate~(\ref{Rdef})
(not the Ricci scalar),
and
$K_x$ and $K_y$ are the radial and angular functions
defined by equations~(\ref{Kxy}).
As remarked following equations~(\ref{Kxy}),
the functions $K_x$ and $K_y$ vanish in $\Lambda$-Kerr(-Newman) spacetimes,
but they do not vanish in the conformally separable spacetimes.
For the remainder of this section~\ref{spintwo-sec},
the spacetime is taken to be $\Lambda$-Kerr.

For $\Lambda$-Kerr spacetimes,
with a convenient choice of relative normalization,
the boost and spin weight $\pm 2$ eigenfunctions are related by
\begin{subequations}
\begin{align}
\label{Xtwo}
  \ethsi{}{u}^4
  X_{+2}
  =
  \mu_x
  X_{-2}
  &
  \ , \ \ 
  \ethsi{}{v}^4
  X_{-2}
  =
  \mu_x
  X_{+2}
  \ ,
\\
\label{Ytwo}
  \ethsi{}{-}^4
  Y_{+2}
  =
  \mu_y
  Y_{-2}
  &
  \ , \ \ 
  \ethsi{}{+}^4
  Y_{-2}
  =
  \mu_y
  Y_{+2}
  \ .
\end{align}
\end{subequations}
Solving for $\mu_x^2$ and $\mu_y^2$ in
\begin{subequations}
\label{raiseraiseraiseraiselowerlowerlowerlower}
\begin{align}
  \left(
  \ethsi{}{v}^4
  \,
  \ethsi{}{u}^4
  -
  \mu_x^2
  \right)
  X_{+2}
  &=
  0
  \ ,
\\
  \left(
  \ethsi{}{+}^4
  \,
  \ethsi{}{-}^4
  -
  \mu_y^2
  \right)
  Y_{+2}
  &=
  0
  \ ,
\end{align}
\end{subequations}
given that
$\Cz_{2}$ satisfies
\mbox{$( \square_{vu} - \lambda ) \Cz_{+2}$}
=
\mbox{$( \square_{+-} - \lambda ) \Cz_{+2}$} = 0,
shows that $\mu_x$ and $\mu_y$ are related to the eigenvalue $\lambda$ by,
for $\Lambda$-Kerr,
\begin{subequations}
\label{muxytwo}
\begin{align}
  \mu_x^2
  =& \,
  \mu_y^2
  +
  144 \wel^2 \Mbh^2
  %- 96 \wel^2 \Qelecbh^2 ( \lambda - 1 )
%  +
%  8 \Qelecbh^2 \Lambda
%  \bigl[
%  - \lambda^2
%  + 1
%\nn
%  &
%  + \,
%  4 a \wel ( 4 a \wel + 3 m )
%  - \tfrac{14}{3} a^2 \Lambda
%  + 2 \Qelecbh^2 \Lambda
%  + \tfrac{1}{9} a^4 \Lambda^2
%  \bigr]
  \ ,
\\
  \mu_y^2
  =
  &
  \left[ \lambda^2 - 1
  - 4 a \wel ( a \wel + m )
  -
  \tfrac{1}{3}
  a^2 \Lambda
  \left(
  14
  +
  \tfrac{1}{3}
  a^2 \Lambda
  \right)
  \right]
\nn
  &
  \times
  \left[ \lambda^2 - 1
  - 36 a \wel ( a \wel + m )
  +
  \tfrac{1}{3}
  a^2 \Lambda
  \left(
  42
  -
  \tfrac{1}{3}
  a^2 \Lambda
  \right)
  \right]
\nn
  &
  + \,
  32
  \left( \lambda - 1 + \tfrac{1}{3} a^2 \Lambda \right)
  \left[
  a \wel ( 4 a \wel + m )
  \right.
\nn
  &
  \left.
  \qquad
  - \,
  \tfrac{1}{3} a^2 \Lambda
  ( a \wel + m ) ( 4 a \wel + 3 m )
  \right]
\nn
  &
  - \,
  \tfrac{16}{3} a^2 \Lambda
  \left[
  8 a \wel ( a \wel + m )
  -
  \tfrac{49}{3} a^2 \Lambda
  \right]
  \ .
\end{align}
\end{subequations}
Equations~(\ref{muxytwo})
agree with
\cite[p.~440 eq.~(61)]{Chandrasekhar:1983},
who gives the case $\Lambda = 0$,
with the translations
(there $\leftrightarrow$ here)
$\lambdabar = \lambda - \frac{1}{2}$
and
$\mathscr{C} = \mu_x$,
$D = \mu_y$,
$\sigma = \wel$.

\section{Conclusions}

The wave equations in the conformally separable solutions of
\cite{Hamilton:2010a,Hamilton:2010b}
for accreting, rotating, uncharged black holes are solved
for massless fields of spin 0, $\tfrac{1}{2}$, 1, $\tfrac{3}{2}$, and 2,
resulting in the generalized
Teukolsky wave equations~(\ref{altwaveoperatorscoordinate})
with the potentials~(\ref{altpotential}),
generalizing the well-known Teukolsky wave equations for stationary black holes
\cite{Teukolsky:1972my,Teukolsky:1973ha,Teukolsky:2015,Chandrasekhar:1983,Staicova:2015}.
As is well known,
massless waves resolve into independently evolving right- and left-handed
chiralities.
A wave of given chirality and spin $s$
has $2s+1$ components, with boost weights
$\sigma = -s , -s{+}1 , ... , s$,
and spin weights $\varsigma$ either equal
($\varsigma = \sigma$, right-handed chirality)
or opposite
($\varsigma = -\sigma$, left-handed chirality)
to the boost weight.
The $2s+1$ different components are coupled by equations of motion,
so are not independent of each other,
but rather oscillate in harmony.
The propagating components of a wave have boost weight $\sigma = -s$
for outgoing waves, and $\sigma = +s$ for ingoing waves.
The wave equations for components with $\sigma = \pm s$,
which include the propagating components,
are separable in the conformally separable solutions,
as they are in stationary solutions for black holes.
In addition, the wave equations for boost weight $\pm 1$ components
of gravitational waves ($s = 2$) are separable.

The Teukolsky-Starobinsky identities
\cite{Press:1973,Starobinsky:1973}
carry through essentially unchanged (with a modified horizon function)
for fields of spin $\tfrac{1}{2}$ and 1,
but are more complicated for fields of spin $\tfrac{3}{2}$ and 2.

\begin{acknowledgements}
This research was supported in part by FQXI mini-grant FQXI-MGB-1626.
The equations in this paper were checked with Mathematica.
\end{acknowledgements}

%\section*{References}

%\bibliographystyle{apsrev4-2}
\bibliography{bh}

\appendix

\onecolumngrid

\section{Wave equations in a general spacetime}
\label{Dapp}

This Appendix takes a deeper dive into the derivation of wave equations
in a general spacetime.
Subsections~\ref{scalar-app}--\ref{spintwo-app}
give the equations of motion that lead to wave equations
for spins $s = 0$, $\tfrac{1}{2}$, 1, $\tfrac{3}{2}$, and 2.
Subsection~\ref{shear-app}
gives a general expression for derivatives of shear
needed in the wave equations for gravitational waves.

Massless waves are described by independently evolving chiral fields
with all-right-handed or all-left-handed bivector and spinor indices.
For example, spin~$\tfrac{3}{2}$ waves have a bivector index and a spinor
index, and those indices are either both right-handed or both left-handed,
equation~(\ref{FRSz}).
Similarly, spin~2 waves have two bivector indices,
and those indices are either both right-handed or both left-handed,
equation~(\ref{Cz}).
Equations of motion
%(or Jacobi identites)
yield linear differential equations that relate components
of adjacent boost (and spin) weight to each other.
The right- and left-handed differential equations involve
purely right- or purely left-handed Lorentz connections.
Right- and left-handed Lorentz connections are defined by
(the tilde on $\Gammaz$ signifies a right- or left-handed chiral component)
\begin{equation}
  \Gammaz_{klp}
  \equiv
  \tfrac{1}{2}
  \left( \Gamma_{klp} \mp \im \, \varepsilon_{kl}{}^{mn} \Gamma_{mnp} \right)
  \ .
\end{equation}
In a completely general spacetime
(not necessarily a conformally separable spacetime),
the most general right- or left-handed
linear differential operators with the properties that
(1) they involve purely right- or left-handed Lorentz connections, and
(2) they raise and lower the boost/spin weight by one,
are the radial $\DSi{s\sigma}{\vu}$ and angular $\DSi{s\varsigma}{\pm}$
operators defined by
\begin{subequations}
\label{Deltas}
\begin{align}
  \DSi{s\sigma}{\vu}
  &\equiv
  \partial_\vu
  \pm
  \sigma \, \Gamma_{\overset{{\scriptstyle u v v}}{{\scriptstyle v u u}}}
  +
  \left\{
  \begin{array}{ll}
  \mp
  \sigma \, \Gamma_{\overset{{\scriptstyle \smallminus \smallplus v}}{{\scriptstyle \smallplus \smallminus u}}}
  +
  ( s \pm \sigma + 1 ) \, \Gamma_{\overset{{\scriptstyle \smallplus v \smallminus}}{{\scriptstyle \smallminus u \smallplus}}}
  &
  \mbox{right}
  \\
  \mp
  \sigma \, \Gamma_{\overset{{\scriptstyle \smallplus \smallminus v}}{{\scriptstyle \smallminus \smallplus u}}}
  +
  ( s \pm \sigma + 1 ) \, \Gamma_{\overset{{\scriptstyle \smallminus v \smallplus}}{{\scriptstyle \smallplus u \smallminus}}}
  &
  \mbox{left}
  \end{array}
  \right.
  \ ,
\\
  \DSi{s\varsigma}{\pm}
  &\equiv
  \partial_\pm
  \mp
  \varsigma \, \Gamma_{\overset{\smallminus \smallplus \smallplus}{\smallplus \smallminus \smallminus}}
  +
  \left\{
  \begin{array}{ll}
  \pm
  \varsigma \, \Gamma_{\overset{{\scriptstyle u v \smallplus}}{{\scriptstyle v u \smallminus}}}
  +
  ( s \pm \varsigma + 1 ) \, \Gamma_{\overset{{\scriptstyle \smallplus v u}}{{\scriptstyle \smallminus u v}}}
  &
  \mbox{right}
  \\
  \pm
  \varsigma \, \Gamma_{\overset{{\scriptstyle u v \smallplus}}{{\scriptstyle v u \smallminus}}}
  +
  ( s \pm \varsigma + 1 ) \, \Gamma_{\overset{{\scriptstyle \smallplus u v}}{{\scriptstyle \smallminus v u}}}
  &
  \mbox{left}
  \end{array}
  \right.
  \ .
\end{align}
\end{subequations}
%In a general spacetime, the differential operators in the equations
%relating components of adjacent boost (and spin) weight to each other
%are of the form~(\ref{Deltas}).
Note that the boost and spin weights of the various terms of each operator
agree, as they must, per equations~(\ref{boostspinrules}).

Explicit calculation shows that
the index $s$ of the operators~(\ref{Deltas}) denotes the spin of the field,
with $s = 0, \tfrac{1}{2} , 1 , \tfrac{3}{2} , 2$ for respectively
scalar, spinor, electromagnetic, gravitino, and gravitational fields.
The index $\sigma$ denotes the boost weight of the field,
which ranges over the $2s+1$ components $-s , -s{+}1 , ... , s$,
while $\varsigma = \pm\sigma$ denotes the spin weight of the field,
with $+$ right-handed, $-$ left-handed.
For the most part, the spin $s$ and boost/spin weight $\sigma$ and $\varsigma$
indices can be suppressed,
because they equal the spin and boost/spin weight
of the field they are acting on.
Equations for the various spins are~(\ref{npscalarconfsep}),
(\ref{npDiracR}) \&~(\ref{npDiracL}),
(\ref{npMaxwellR}) \&~(\ref{npMaxwellL}),
(\ref{npRSR}) \&~(\ref{npRSL}),
and
(\ref{npWeylR}) \&~(\ref{npWeylL}).

The operators $\DSi{}{}$ defined by equations~(\ref{Deltas})
do not commute with each other,
but (primed) operators defined by
\begin{subequations}
\label{Dpeltas}
\begin{align}
  \DpSi{s\sigma}{\vu}
  &\equiv
  \DSi{s\sigma}{\vu}
  +
  \left\{
  \begin{array}{ll}
  -
  \Gamma_{\overset{{\scriptstyle \smallminus \smallplus v}}{\smallplus \smallminus {\scriptstyle u}}}
  +
  \Gamma_{\overset{{\scriptstyle \smallminus v \smallplus}}{{\scriptstyle \smallplus u \smallminus}}}
  &
  \mbox{right}
  \\
  -
  \Gamma_{\overset{{\scriptstyle \smallplus \smallminus v}}{\smallminus \smallplus {\scriptstyle u}}}
  +
  \Gamma_{\overset{{\scriptstyle \smallplus v \smallminus}}{{\scriptstyle \smallminus u \smallplus}}}
  &
  \mbox{left}
  \end{array}
  \right.
  \ ,
\\
  \DpSi{s\varsigma}{\pm}
  &\equiv
  \DSi{s\varsigma}{\pm}
  +
  \left\{
  \begin{array}{ll}
  \Gamma_{\overset{{\scriptstyle u v \smallplus}}{{\scriptstyle v u \smallminus}}}
  -
  \Gamma_{\overset{{\scriptstyle \smallplus u v}}{{\scriptstyle \smallminus v u}}}
  &
  \mbox{right}
  \\
  \Gamma_{\overset{{\scriptstyle v u \smallplus}}{{\scriptstyle u v \smallminus}}}
  -
  \Gamma_{\overset{{\scriptstyle \smallplus v u}}{{\scriptstyle \smallminus u v}}}
  &
  \mbox{left}
  \end{array}
  \right.
  \ ,
\end{align}
\end{subequations}
have the property that,
in the conformally separable spacetimes,
the radial and angular
$\Dsi{}{}$
and
$\Dpsi{}{}$
operators do commute,
\begin{equation}
\label{DpDcommute}
  \DpSi{s\varsigma}{\pm}
  \,
  \DSi{s\sigma}{\vu}
  -
  \DpSi{s\sigma}{\vu}
  \,
  \DSi{s\varsigma}{\pm}
  =
  %2 ( 1 + s \pm 2 \sigma )
  %\Cz_{\overset{\scriptstyle{v u v \smallplus}}{\scriptstyle{u v u \smallminus}}}
  %\Cz_{\pm 1}
  0
  \ ,
\end{equation}
in both right- and left-handed versions.
The existence of radial/angular operators $\DpSi{}{}$
that commute with angular/radial operators $\DSi{}{}$
is crucial to forming wave equations for individual components
of particular boost and spin weight.
In the conformally separable spacetimes,
the 
$\Dsi{s\sigma}{}$
and
$\Dpsi{s\sigma}{}$
operators are related by
\begin{equation}
  \DpSi{s\sigma}{k}
  =
  \rho^{-1}
  \DSi{s\sigma}{k}
  \, \rho
\end{equation}
in both right- and left-handed versions,
with $\rho$ the conformal factor, equation~(\ref{rho}).

In conformally separable spacetimes,
the raising and lowering operators
$\ethSi{s\sigma}{k}$ defined by equations~(\ref{ethsi})
are related to the
$\DSi{s\sigma}{k}$
operators by, in the right-handed case,
\begin{equation}
\label{Deth}
  \sqrt{2} \rho \,
  \DSi{s\sigma}{k}
%  =
%  (\pm)
%  {1 \over
%  \rhoi^{1 + s}
%  \rhom^{1 + s \pm 2\sigma}}
%  \,
%  \ethSi{s\sigma}{k}
%  \,
%  \rhoi^{1 + s}
%  \rhom^{1 + s \pm 2\sigma}
  =
  (\pm)
  f_s
  \,
  \rhom^{\mp 2\sigma - 1}
  \,
  \ethSi{\sigma}{k}
  \,
  \rhom^{\pm 2\sigma + 1}
  \,
  f_s^{-1}
  \ ,
\end{equation}
where $f_s$ is given by equation~(\ref{fs}),
and the initial $(\pm)$ sign is:
$+$ for $k = v$ or $+$, or $k = u$ outside the horizon;
$-$ for $k = -$, or $k = u$ inside the horizon;
while the $\pm$ sign multiplying $2\sigma$ is
$+$ for indices $k = v$ or $+$, and $-$ for $k = u$ or $-$.
The left-handed relation is the complex conjugate
of the right-handed relation~(\ref{Deth}),
obtained by replacing $f_s \rightarrow f_s^\ast$ and $\rhom \rightarrow \rhom^\ast$.

Wave equations are obtained by taking the differential equations
linear in the operators $\Dsi{}{}$, equations~(\ref{Deltas}),
that follow from the equations of motion,
applying operators $\Dpsi{}{}$, equations~(\ref{Dpeltas}),
and differencing the resulting second order differential equations
in such a way that the commutation~(\ref{DpDcommute})
leads to cancellation of terms.
The resulting equations can be expressed in terms of
the difference
of the radial and angular wave operators $\square$
defined by equations~(\ref{waveoperatorsraiselower}).
For arbitrary spins $s$ and boost weights $\sigma$,
the difference
$\sqvu - \sqpm$
of radial and angular wave operators,
expressed in terms of the $\Dsi{}{}$ and $\Dpsi{}{}$ operators,
and then in terms of the $\ethsi{}{}$ raising and lower operators~(\ref{ethsi})
are, for right-handed ($\varsigma = +\sigma$) modes,
\begin{align}
\label{sqsR}
  \left(
  \Sqvu{\sigma} - \Sqpm{\sigma}
  \right)
  \hatpsi_{\sigma}
  &=
  2
  \rho^2
  f_s^{-1}
  \left[
  \left(
  - \,
  \DpSi{s,\sigma\mp1}{\vu}
  \,
  \DSi{s,\sigma}{\uv}
  +
  \DpSi{s,\sigma\mp1}{\pm}
  \,
  \DSi{s,\sigma}{\mp}
  \right)
  +
  2
  ( \sigma \mp \tfrac{1}{2} ) ( \sigma \mp 1 ) \Cz_{0}
  \right]
  f_s
  \hatpsi_{\sigma}
\nonumber
\\
  &=
  \Bigl[
  \rhom^{-(\pm 2\sigma-1)}
  \left(
  - \sgn(\Delta_x)
  \,
  \ethSi{\sigma\mp1}{\vu}
  \,
  \rhom^{2(\pm 2\sigma-1)}
  \,
  \ethSi{\sigma}{\uv}
  -
  \ethSi{\sigma\mp1}{\pm}
  \,
  \rhom^{2(\pm 2\sigma-1)}
  \,
  \ethSi{\sigma}{\mp}
  \right)
  \rhom^{-(\pm 2\sigma-1)}
\nonumber
\\
  &
  \quad\ \ 
  +
  4 \rho^2 ( \sigma \mp \tfrac{1}{2} ) ( \sigma \mp 1 ) \Cz_{0}
  \Bigr]
  \hatpsi_{\sigma}
  \ .
\end{align}
The equivalent result for left-handed modes ($\varsigma = -\sigma$)
is obtained by flipping angular indices $+ \leftrightarrow -$,
and complex-conjugating
$f_s \rightarrow f_s^\ast$ and $\rhom \rightarrow \rhom^\ast$,
\begin{align}
\label{sqsL}
  \left(
  \Sqvu{\sigma} - \Sqmp{\sigma}
  \right)
  \hatpsi_{\sigma}
  &=
  2
  \rho^2
  ( f_s^\ast )^{-1}
  \left[
  \left(
  - \,
  \DpSi{s,\sigma\mp1}{\vu}
  \,
  \DSi{s,\sigma}{\uv}
  +
  \DpSi{s,\sigma\pm1}{\mp}
  \,
  \DSi{s,\sigma}{\pm}
  \right)
  +
  2
  ( \sigma \mp \tfrac{1}{2} ) ( \sigma \mp 1 ) \Cz_{0}
  \right]
  f_s^\ast
  \hatpsi_{\sigma}
\nonumber
\\
  &=
  \Bigl[
  (\rhom^\ast)^{-(\pm 2\sigma-1)}
  \left(
  - \sgn(\Delta_x)
  \,
  \ethSi{\sigma\mp1}{\vu}
  \,
  (\rhom^\ast)^{2(\pm 2\sigma-1)}
  \,
  \ethSi{\sigma}{\uv}
  -
  \ethSi{\sigma\pm1}{\mp}
  \,
  (\rhom^\ast)^{2(\pm 2\sigma-1)}
  \,
  \ethSi{\sigma}{\pm}
  \right)
  (\rhom^\ast)^{-(\pm 2\sigma-1)}
\nonumber
\\
  &
  \quad\ \ 
  +
  4 \rho^2 ( \sigma \mp \tfrac{1}{2} ) ( \sigma \mp 1 ) \Cz_{0}
  \Bigr]
  \hatpsi_{\sigma}
  \ .
\end{align}

\subsection{Spin~0 waves}
\label{scalar-app}

In a general spacetime, the d'Alembertian operator that goes in
the scalar wave equation~(\ref{DD0square}) is
%\begin{equation}
%  \DD^k \DD_k \varphi
%  =
%  \left(
%  - \ 
%  \DpSi{1,-1}{v} \ 
%  \DSi{-1,0}{u}
%  \,-\,
%  \DpSi{1,+1}{u} \ 
%  \DSi{-1,0}{v}
%  \,+\,
%  \DpSi{1,-1}{+} \ 
%  \DSi{-1,0}{-}
%  \,+\,
%  \DpSi{1,+1}{-} \ 
%  \DSi{-1,0}{+}
%  \right)
%  \varphi
%  \ ,
%\end{equation}
\begin{equation}
\label{npscalar}
  \DD^k \DD_k \varphi
  =
  \left(
  - \,
  \DpSi{1,-1}{v} \,
  \partial_u
  -
  \DpSi{1,+1}{u} \,
  \partial_v
  +
  \DpSi{1,-1}{+} \,
  \partial_-
  +
  \DpSi{1,+1}{-} \,
  \partial_+
  \right)
  \varphi
  \ ,
\end{equation}
which holds true for both right- and left-handed versions of
the $\Dpsi{s\sigma}{}$ operators~(\ref{Dpeltas}).
In conformally separable spacetimes, equation~(\ref{npscalar})
can be recast as
\begin{equation}
\label{npscalarconfsep}
  \DD^k \DD_k \varphi
  =
  2
  \left(
  - \
  \DpSi{0,\mp 1}{\vu} \
  \DSi{0,0}{\uv}
  \,+\,
  \DpSi{0,\mp 1}{\pm} \ 
  \DSi{0,0}{\mp}
  \,+\,
  \Cz_0
  \,+\,
  \tfrac{1}{12}
  R
  \right)
  \varphi
  \ ,
\end{equation}
which again holds true for both right- and left-handed versions of the
$\Dsi{s\sigma}{}$ and $\Dpsi{s\sigma}{}$ operators.
Note that whereas the index $s$ on $\DpSi{s\sigma}{k}$ is $s = 1$
in the general equation~(\ref{npscalar}),
the index is $s = 0$
on $\DSi{s\sigma}{k}$ and $\DpSi{s\sigma}{k}$
in the conformally separable version~(\ref{npscalarconfsep}).

\subsection{Spin~$\tfrac{1}{2}$ waves}
\label{spinhalf-app}

In a general spacetime,
the Dirac equation~(\ref{spinhalfwave})
for the components $\psi_{\sigma}$ of the spinor field,
equation~(\ref{Fzcomponents}),
expressed with respect to a Newman-Penrose tetrad
in terms of the differential operators defined by equations~(\ref{Deltas}),
are, for the right-handed components,
\begin{subequations}
\label{npDiracR}
\begin{alignat}{3}
\label{npDiracRa}
  {1 \over \sqrt{2}}
  ( \bDD \psi )_{\overset{\smallUpdown}{\smallDownup}}
  &=
  \DSi{\mp 1/2}{\vu} \, \psi_{\overset{\smallDowndown}{\smallUpup}}
  -
  \DSi{\pm 1/2}{\mp} \, \psi_{\overset{\smallUpup}{\smallDowndown}}
  &&=
  0
  \ ,
\end{alignat}
\end{subequations}
and for the left-handed components,
\begin{subequations}
\label{npDiracL}
\begin{alignat}{3}
\label{npDiracLa}
  {1 \over \sqrt{2}}
  ( \bDD \psi )_{\overset{\smallUpup}{\smallDowndown}}
  &=
  - \,
  \DSi{\mp 1/2}{\vu} \, \psi_{\overset{\smallDownup}{\smallUpdown}}
  -
  \DSi{\mp 1/2}{\pm} \, \psi_{\overset{\smallUpdown}{\smallDownup}}
  &&=
  0
  \ .
\end{alignat}
\end{subequations}

\subsection{Spin~1 waves}
\label{spinone-app}

In a general spacetime,
Maxwell's equations~(\ref{maxwell})
for the components $\Fz_{\sigma}$ of the electromagnetic field,
equation~(\ref{Fzcomponents}),
expressed with respect to a Newman-Penrose tetrad
in terms of the differential operators defined by equations~(\ref{Deltas}),
are, for the right-handed components ($\varsigma = +\sigma$),
\begin{subequations}
\label{npMaxwellR}
\begin{alignat}{3}
\label{npMaxwellRa}
  \DD^k \Fz_{\overset{{\scriptstyle k} {\scriptstyle v}}{k {\scriptstyle u}}}
  &=
  \Gamma_{\overset{\smallplus {\scriptstyle v} {\scriptstyle v}}{\smallminus {\scriptstyle u} {\scriptstyle u}}} \, \Fz_{\mp 1}
  \, \pm \, \DSi{0}{\vu} \, \Fz_0
  - \DSi{\pm 1}{\mp} \, \Fz_{\pm 1}
  &&=
  \jz_\vu
  \ ,
\\
\label{npMaxwellRb}
  \DD^k \Fz_{\overset{{\scriptstyle k} \smallplus}{k \smallminus}}
  &=
  \Gamma_{\overset{\smallplus {\scriptstyle v} \smallplus}{\smallminus {\scriptstyle u} \smallminus}} \, \Fz_{\mp 1}
  \, \pm \, \DSi{0}{\pm} \, \Fz_{0}
  - \DSi{\pm 1}{\uv} \, \Fz_{\pm 1}
  &&=
  \jz_\pm
  \ ,
\end{alignat}
\end{subequations}
and for the left-handed components ($\varsigma = -\sigma$),
\begin{subequations}
\label{npMaxwellL}
\begin{alignat}{3}
\label{npMaxwellLa}
  \DD^k \Fz_{\overset{{\scriptstyle k} {\scriptstyle v}}{k {\scriptstyle u}}}
  &=
  \Gamma_{\overset{\smallminus {\scriptstyle v} {\scriptstyle v}}{\smallplus {\scriptstyle u} {\scriptstyle u}}} \, \Fz_{\mp 1}
  \, \pm \,
  \DSi{0}{\vu} \, \Fz_0
  -
  \DSi{\mp 1}{\pm} \, \Fz_{\pm 1}
  &&=
  \jz_\vu
  \ ,
\\
\label{npMaxwellLb}
  \DD^k \Fz_{\overset{{\scriptstyle k} \smallminus}{k \smallplus}}
  &=
  \Gamma_{\overset{\smallminus {\scriptstyle v} \smallminus}{\smallplus {\scriptstyle u} \smallplus}} \, \Fz_{\mp 1}
  \, \pm \,
  \DSi{0}{\mp} \, \Fz_{0}
  -
  \DSi{\pm 1}{\uv} \, \Fz_{\pm 1}
  &&=
  \jz_\pm
  \ .
\end{alignat}
\end{subequations}
The units of the currents $j_k$ are Heaviside;
in Gaussian units the currents would be multiplied by $4\pi$.
Conservation of electric current is expressed by
\begin{equation}
\label{Dj}
  \DD^k j_k
  =
  - \,
  \DpSi{+1}{u} \,
  j_v
  -
  \DpSi{-1}{v} \,
  j_u
  +
  \DpSi{-1}{+} \,
  j_\smallminus
  +
  \DpSi{+1}{-} \,
  j_\smallplus
  =
  0
  \ .
\end{equation}

\subsection{Spin~$\tfrac{3}{2}$ waves}
\label{spinthreehalf-app}

Wave equations for the spin~$\tfrac{3}{2}$ gravitino field $\bFRS$ follow from
the Jacobi identity~(\ref{bianchifermion}).
In a general spacetime,
the left hand sides of
the Jacobi identity~(\ref{bianchifermion})
in terms of the differential operators defined by equations~(\ref{Deltas}),
are, for the right-handed components,
\begin{subequations}
\label{npRSR}
\begin{alignat}{3}
\label{npRSRa}
  \frac{1}{\sqrt{2}}
  \bigl(
  \bDD \FRSz_{\overset{{\scriptstyle v} {\scriptstyle u}}{{\scriptstyle u} {\scriptstyle v}}}
  \bigr)_{\overset{\smallUpdown}{\smallDownup}}
  =
  \DD^k \FRSz_{\overset{{\scriptstyle k} {\scriptstyle v} \smallDowndown}{{\scriptstyle k} {\scriptstyle u} \smallUpup}}
  =
  \DD^k \FRSz_{\overset{{\scriptstyle k} \smallminus \smallDowndown}{{\scriptstyle k} \smallplus \smallUpup}}
  &=
  \Gamma_{\overset{\smallplus {\scriptstyle v} {\scriptstyle v}}{\smallminus {\scriptstyle u} {\scriptstyle u}}} \, \FRSz_{\mp 3/2}
  \pm
  \DSi{\mp 1/2}{\vu} \, \FRSz_{\mp 1/2}
  \mp
  \DSi{\pm 1/2}{\mp} \, \FRSz_{\pm 1/2}
  +
  \Gamma_{\overset{\smallminus {\scriptstyle u} \smallminus}{\smallplus {\scriptstyle v} \smallplus}} \, \FRSz_{\pm 3/2}
  \ ,
\\
\label{npRSRc}
  \frac{1}{\sqrt{2}}
  \bigl(
  \bDD \FRSz_{\overset{{\scriptstyle v} \smallplus}{{\scriptstyle u} \smallminus}}
  \bigr)_{\overset{\smallUpdown}{\smallDownup}}
  =
  \DD^k \FRSz_{\overset{{\scriptstyle k} {\scriptstyle v} \smallUpup}{{\scriptstyle k} {\scriptstyle u} \smallDowndown}}
  &=
  \mp \,
  2 \Gamma_{\overset{\smallplus {\scriptstyle v} {\scriptstyle v}}{\smallminus {\scriptstyle u} {\scriptstyle u}}} \, \FRSz_{\mp 1/2}
  \pm
  \DSi{\pm 1/2}{\vu} \, \FRSz_{\pm 1/2}
  -
  \DSi{\pm 3/2}{\mp} \, \FRSz_{\pm 3/2}
  \ ,
\\
\label{npRSRd}
  - \frac{1}{\sqrt{2}}
  \bigl(
  \bDD \FRSz_{\overset{{\scriptstyle v} \smallplus}{{\scriptstyle u} \smallminus}}
  \bigr)_{\overset{\smallDownup}{\smallUpdown}}
  =
  \DD^k \FRSz_{\overset{{\scriptstyle k} \smallplus \smallUpup}{{\scriptstyle k} \smallminus \smallDowndown}}
  &=
  \mp \,
  2 \Gamma_{\overset{\smallplus {\scriptstyle v} \smallplus}{\smallminus {\scriptstyle u} \smallminus}} \, \FRSz_{\mp 1/2}
  \pm
  \DSi{\pm 1/2}{\pm} \, \FRSz_{\pm 1/2}
  -
  \DSi{\pm 3/2}{\uv} \, \FRSz_{\pm 3/2}
  \ ,
\end{alignat}
\end{subequations}
and for the left-handed components,
\begin{subequations}
\label{npRSL}
\begin{alignat}{3}
\label{npRSLa}
  - \frac{1}{\sqrt{2}}
  \bigl(
  \bDD \FRSz_{\overset{{\scriptstyle v} {\scriptstyle u}}{{\scriptstyle u} {\scriptstyle v}}}
  \bigr)_{\overset{\smallUpup}{\smallDowndown}}
  =
  \DD^k \FRSz_{\overset{{\scriptstyle k} {\scriptstyle v} \smallDownup}{{\scriptstyle k} {\scriptstyle u} \smallUpdown}}
  =
  \pm
  \DD^k \FRSz_{\overset{{\scriptstyle k} \smallplus \smallUpdown}{{\scriptstyle k} \smallminus \smallDownup}}
  &=
  \Gamma_{\overset{\smallminus {\scriptstyle v} {\scriptstyle v}}{\smallplus {\scriptstyle u} {\scriptstyle u}}} \, \FRSz_{\mp 3/2}
  \pm
  \DSi{\mp 1/2}{\vu} \, \FRSz_{\mp 1/2}
  \pm
  \DSi{\mp 1/2}{\pm} \, \FRSz_{\pm 1/2}
  -
  \Gamma_{\overset{\smallplus {\scriptstyle u} \smallplus}{\smallminus {\scriptstyle v} \smallminus}} \, \FRSz_{\pm 3/2}
  \ ,
\\
\label{npRSLc}
  \frac{1}{\sqrt{2}}
  \bigl(
  \bDD \FRSz_{\overset{{\scriptstyle v} \smallminus}{{\scriptstyle u} \smallplus}}
  \bigr)_{\overset{\smallUpup}{\smallDowndown}}
  =
  \DD^k \FRSz_{\overset{{\scriptstyle k} {\scriptstyle v} \smallUpdown}{{\scriptstyle k} {\scriptstyle u} \smallDownup}}
  &=
  \pm \,
  2 \Gamma_{\overset{\smallminus {\scriptstyle v} {\scriptstyle v}}{\smallplus {\scriptstyle u} {\scriptstyle u}}} \, \FRSz_{\mp 1/2}
  \pm
  \DSi{\pm 1/2}{\vu} \, \FRSz_{\pm 1/2}
  -
  \DSi{\mp 3/2}{\pm} \, \FRSz_{\pm 3/2}
  \ ,
\\
\label{npRSLd}
  \frac{1}{\sqrt{2}}
  \bigl(
  \bDD \FRSz_{\overset{{\scriptstyle v} \smallminus}{{\scriptstyle u} \smallplus}}
  \bigr)_{\overset{\smallDowndown}{\smallUpup}}
  =
  \DD^k \FRSz_{\overset{{\scriptstyle k} \smallminus \smallUpdown}{{\scriptstyle k} \smallplus \smallDownup}}
  &=
  \pm \,
  2 \Gamma_{\overset{\smallminus {\scriptstyle v} \smallminus}{\smallplus {\scriptstyle u} \smallplus}} \, \FRSz_{\mp 1/2}
  \pm
  \DSi{\mp 1/2}{\mp} \, \FRSz_{\pm 1/2}
  -
  \DSi{\pm 3/2}{\uv} \, \FRSz_{\pm 3/2}
  \ .
\end{alignat}
\end{subequations}

\subsection{Spin~2 waves}
\label{spintwo-app}

In a general spacetime,
the Weyl evolution equations~(\ref{weylevolutionz})
in terms of the differential operators defined by equations~(\ref{Deltas}),
are,
for right-handed components,
\begin{subequations}
\label{npWeylR}
\begin{alignat}{3}
  \DD^k\Cz_{\overset{{\scriptstyle k} \smallminus {\scriptstyle v} \smallplus}{{\scriptstyle k} \smallplus {\scriptstyle u} \smallminus}}
  &=
  - \, 2 \, \Gamma_{\overset{\smallplus {\scriptstyle v} {\scriptstyle v}}{\smallminus {\scriptstyle u} {\scriptstyle u}}} \, \Cz_{\mp 1}
  + \DSi{0}{\vu} \, \Cz_{0}
  - \DSi{\pm 1}{\mp} \, \Cz_{\pm 1}
  + \Gamma_{\overset{\smallminus {\scriptstyle u} \smallminus}{\smallplus {\scriptstyle v} \smallplus}} \, \Cz_{\pm 2}
  &&=
  \Jz_{\overset{\smallminus {\scriptstyle v} \smallplus}{\smallplus {\scriptstyle u} \smallminus}}
  \ ,
\\
  \DD^k\Cz_{\overset{{\scriptstyle k} {\scriptstyle u} {\scriptstyle v} \smallplus}{{\scriptstyle k} {\scriptstyle v} {\scriptstyle u} \smallminus}}
  &=
  - \, 2 \, \Gamma_{\overset{\smallplus {\scriptstyle v} \smallplus}{\smallminus {\scriptstyle u} \smallminus}} \, \Cz_{\mp 1}
  + \DSi{0}{\pm} \, \Cz_{0}
  - \DSi{\pm 1}{\uv} \, \Cz_{\pm 1}
  + \Gamma_{\overset{\smallminus {\scriptstyle u u}}{\smallplus {\scriptstyle v v}}} \, \Cz_{\pm 2}
  &&=
  \Jz_{\overset{{\scriptstyle u} {\scriptstyle v} \smallplus}{{\scriptstyle v} {\scriptstyle u} \smallminus}}
  \ ,
\\
  \DD^k\Cz_{\overset{{\scriptstyle k} {\scriptstyle v} {\scriptstyle v} \smallplus}{{\scriptstyle k} {\scriptstyle u} {\scriptstyle u} \smallminus}}
  &=
  - \, 3 \, \Gamma_{\overset{\smallplus {\scriptstyle v} {\scriptstyle v}}{\smallminus {\scriptstyle u} {\scriptstyle u}}} \, \Cz_{0}
  + \DSi{\pm 1}{\vu} \, \Cz_{\pm 1}
  - \DSi{\pm 2}{\mp} \, \Cz_{\pm 2}
  &&=
  \Jz_{\overset{{\scriptstyle v} {\scriptstyle v} \smallplus}{{\scriptstyle u} {\scriptstyle u} \smallminus}}
  \ ,
\\
  \DD^k\Cz_{\overset{{\scriptstyle k} \smallplus {\scriptstyle v} \smallplus}{{\scriptstyle k} \smallminus {\scriptstyle u} \smallminus}}
  &=
  - \, 3 \, \Gamma_{\overset{\smallplus {\scriptstyle v} \smallplus}{\smallminus {\scriptstyle u} \smallminus}} \, \Cz_{0}
  + \DSi{\pm 1}{\pm} \, \Cz_{\pm 1}
  - \DSi{\pm 2}{\uv} \, \Cz_{\pm 2}
  &&=
  \Jz_{\overset{\smallplus {\scriptstyle v} \smallplus}{\smallminus {\scriptstyle u} \smallminus}}
  \ ,
\end{alignat}
\end{subequations}
and for left-handed components,
\begin{subequations}
\label{npWeylL}
\begin{alignat}{3}
  \DD^k\Cz_{\overset{{\scriptstyle k} \smallplus {\scriptstyle v} \smallminus}{k \smallminus {\scriptstyle u} \smallplus}}
  &=
  2 \, \Gamma_{\overset{\smallminus {\scriptstyle v} {\scriptstyle v}}{\smallplus {\scriptstyle u} {\scriptstyle u}}} \, \Cz_{\mp 1}
  + \DSi{0}{\vu} \, \Cz_{0}
  + \DSi{\mp 1}{\pm} \, \Cz_{\pm 1}
  + \Gamma_{\overset{\smallplus {\scriptstyle u} \smallplus}{\smallminus {\scriptstyle v} \smallminus}} \, \Cz_{\pm 2}
  &&=
  \Jz_{\overset{\smallplus {\scriptstyle v} \smallminus}{\smallminus {\scriptstyle u} \smallplus}}
  \ ,
\\
  \DD^k\Cz_{\overset{{\scriptstyle k} {\scriptstyle u} {\scriptstyle v} \smallminus}{k {\scriptstyle v} {\scriptstyle u} \smallplus}}
  &=
  2 \, \Gamma_{\overset{\smallminus {\scriptstyle v} \smallminus}{\smallplus {\scriptstyle u} \smallplus}} \, \Cz_{\mp 1}
  + \DSi{0}{\mp} \, \Cz_{0}
  + \DSi{\pm 1}{\uv} \, \Cz_{\pm 1}
  + \Gamma_{\overset{\smallplus {\scriptstyle u u}}{\smallminus {\scriptstyle v v}}} \, \Cz_{\pm 2}
  &&=
  \Jz_{\overset{{\scriptstyle u} {\scriptstyle v} \smallminus}{{\scriptstyle v} {\scriptstyle u} \smallplus}}
  \ ,
\\
  \DD^k\Cz_{\overset{{\scriptstyle k} {\scriptstyle v} {\scriptstyle v} \smallminus}{k {\scriptstyle u} {\scriptstyle u} \smallplus}}
  &=
  - \, 3 \, \Gamma_{\overset{\smallminus {\scriptstyle v} {\scriptstyle v}}{\smallplus {\scriptstyle u} {\scriptstyle u}}} \, \Cz_{0}
  - \DSi{\pm 1}{\vu} \, \Cz_{\pm 1}
  - \DSi{\mp 2}{\pm} \, \Cz_{\pm 2}
  &&=
  \Jz_{\overset{{\scriptstyle v} {\scriptstyle v} \smallminus}{{\scriptstyle u} {\scriptstyle u} \smallplus}}
  \ ,
\\
  \DD^k\Cz_{\overset{{\scriptstyle k} \smallminus {\scriptstyle v} \smallminus}{k \smallplus {\scriptstyle u} \smallplus}}
  &=
  - \, 3 \, \Gamma_{\overset{\smallminus {\scriptstyle v} \smallminus}{\smallplus {\scriptstyle u} \smallplus}} \, \Cz_{0}
  - \DSi{\mp 1}{\mp} \, \Cz_{\pm 1}
  - \DSi{\pm 2}{\uv} \, \Cz_{\pm 2}
  &&=
  \Jz_{\overset{\smallminus {\scriptstyle v} \smallminus}{\smallplus {\scriptstyle u} \smallplus}}
  \ .
\end{alignat}
\end{subequations}
%Conservation of Weyl current is expressed by
%\begin{align}
%\label{DJ}
%  \DD^k
%  \Jz_{kuv}
%  &=
%  \DpSi{+1}{u} \,
%  \Jz_{vvu}
%  -
%  \DpSi{-1}{v} \,
%  \Jz_{uuv}
%  -
%  \DpSi{+1}{-} \,
%  \Jz_{+vu}
%  +
%  \DpSi{-1}{+} \,
%  \Jz_{-uv}
%  -
%  \Gamma_{-uu}
%  \Jz_{vv+}
%  +
%  \Gamma_{+vv}
%  \Jz_{uu-}
%  +
%  \Gamma_{-u-}
%  \Jz_{+v+}
%  -
%  \Gamma_{+v+}
%  \Jz_{-u-}
%  \ ,
%\\
%  \DD^k
%  \Jz_{kv+}
%  &=
%  \DpSi{0}{+} \,
%  \Jz_{vvu}
%  -
%  \DpSi{0}{v} \,
%  \Jz_{+vu}
%  -
%  \DpSi{+2}{u} \,
%  \Jz_{vv+}
%  +
%  \DpSi{+2}{-} \,
%  \Jz_{+v+}
%  -
%  2
%  \Gamma_{+vv}
%  \Jz_{uuv}
%  +
%  2
%  \Gamma_{+v+}
%  \Jz_{-uv}
%  \ ,
%\\
%  \DD^k
%  \Jz_{ku-}
%  &=
%  \DpSi{0}{-} \,
%  \Jz_{uuv}
%  -
%  \DpSi{0}{u} \,
%  \Jz_{-uv}
%  -
%  \DpSi{-2}{v} \,
%  \Jz_{uu-}
%  +
%  \DpSi{-2}{+} \,
%  \Jz_{-u-}
%  -
%  2
%  \Gamma_{-uu}
%  \Jz_{vvu}
%  +
%  2
%  \Gamma_{-u-}
%  \Jz_{+vu}
%  \ .
%\end{align}

\subsection{Derivatives of shear}
\label{shear-app}

The Newman-Penrose formalism makes a 2+2 split of the tangent space of spacetime
into a radial subspace (null indices $v$ and $u$)
and an angular subspace (angular indices $+$ and $-$).
Conformally separable black-hole spacetimes are shear-free,
meaning that the eight Lorentz connections of the form
$\Gamma_{aza}$,
where $a$ and $z$ are from opposite spaces,
are all zero.
Although the shears all vanish in the unperturbed background,
their derivatives yield spin~$\pm 2$ components of the Riemann tensor
(compare 
\cite{Chandrasekhar:1983} p.~431, last of eqs.~(3) and~(4)):
\begin{subequations}
\label{Dshears}
\begin{alignat}{3}
\label{DshearsR}
  \DpSi{s,\varsigma\pm 1}{\pm}
  \left(
  \Gamma_{\overset{\smallplus {\scriptstyle v} {\scriptstyle v}}{\smallminus {\scriptstyle u} {\scriptstyle u}}}
  \, \psi_{s\sigma\varsigma}
  \right)
  -
  \DpSi{s,\sigma \pm 1}{\vu}
  \left(
  \Gamma_{\overset{\smallplus {\scriptstyle v} \smallplus}{\smallminus {\scriptstyle u} \smallminus}}
  \, \psi_{s\sigma\varsigma}
  \right)
  -
  \Gamma_{\overset{\smallplus {\scriptstyle v} {\scriptstyle v}}{\smallminus {\scriptstyle u} {\scriptstyle u}}}
  \, \DSi{s\varsigma}{\pm}
  \, \psi_{s\sigma\varsigma}
  +
  \Gamma_{\overset{\smallplus {\scriptstyle v} \smallplus}{\smallminus {\scriptstyle u} \smallminus}}
  \, \DSi{s\sigma}{\vu}
  \, \psi_{s\sigma\varsigma}
  &=
  R_{\overset{{\scriptstyle v} \smallplus {\scriptstyle v} \smallplus}{{\scriptstyle u} \smallminus {\scriptstyle u} \smallminus}}
  \, \psi_{s\sigma\varsigma}
  &&\quad
  \mbox{right}
  \ ,
\\
\label{DshearsL}
  \DpSi{s,\varsigma \mp 1}{\mp}
  \left(
  \Gamma_{\overset{\smallminus {\scriptstyle v} {\scriptstyle v}}{\smallplus {\scriptstyle u} {\scriptstyle u}}}
  \, \psi_{s\sigma\varsigma}
  \right)
  -
  \DpSi{s,\sigma \pm 1}{\vu}
  \left(
  \Gamma_{\overset{\smallminus {\scriptstyle v} \smallminus}{\smallplus {\scriptstyle u} \smallplus}}
  \, \psi_{s\sigma\varsigma}
  \right)
  -
  \Gamma_{\overset{\smallminus {\scriptstyle v} {\scriptstyle v}}{\smallplus {\scriptstyle u} {\scriptstyle u}}}
  \, \DSi{s\varsigma}{\mp}
  \, \psi_{s\sigma\varsigma}
  +
  \Gamma_{\overset{\smallminus {\scriptstyle v} \smallminus}{\smallplus {\scriptstyle u} \smallplus}}
  \, \DSi{s\sigma}{\vu}
  \, \psi_{s\sigma\varsigma}
  &=
  R_{\overset{{\scriptstyle v} \smallminus {\scriptstyle v} \smallminus}{{\scriptstyle u} \smallplus {\scriptstyle u} \smallplus}}
  \, \psi_{s\sigma\varsigma}
  &&\quad
  \mbox{left}
  \ .
\end{alignat}
\end{subequations}
Equations~(\ref{Dshears}) are valid in an arbitrary spacetime
for arbitrary spin $s$ and boost/spin weight $\sigma$ and $\varsigma$.
The case relevant to gravitational waves, equation~(\ref{DCz2}), has
$\psi_{s\sigma\varsigma} = \Cz_0$,
for which $s = \sigma = \varsigma = 0$.

\section{A relation among differential operators}
\label{relation-app}

In the conformally separable spacetimes considered in this paper,
the differential operators $\ethsi{}{k}$ defined by equations~(\ref{ethsi}),
acting on any arbitrary (not necessarily separable) function of the coordinates
$\{ x,t,y,\phi \}$, satisfy the following cubic relations:
\begin{subequations}
\label{rhomethsicommutation}
\begin{align}
  \rhom \ethsi{}{l} \rhom^{-1}
  \ethsi{}{k}^2
  &=
  \rhom^{-1} \ethsi{}{k} \rhom
  \left[
  \ethsi{}{k} \ethsi{}{l}
  -
  \frac{1}{\rhom}
  \left(
  \pm_k \sgn(\Delta_x)
  R^2 \sqrt{|\Delta_x|} \ethsi{}{l}
  \pm_l
  \im a \sqrt{\Delta_y} \ethsi{}{k}
  \right)
  \right]
  \ ,
\\
  \rhom \ethsi{}{k} \rhom^{-1}
  \ethsi{}{l}^2
  &=
  \rhom^{-1} \ethsi{}{l} \rhom
  \left[
  \ethsi{}{k} \ethsi{}{l}
  -
  \frac{1}{\rhom}
  \left(
  \pm_k \sgn(\Delta_x)
  R^2 \sqrt{|\Delta_x|} \ethsi{}{l}
  \pm_l
  \im a \sqrt{\Delta_y} \ethsi{}{k}
  \right)
  \right]
  \ ,
\end{align}
\end{subequations}
where the index $k$ is either of $v$ or $u$,
and the index $l$ is either of $+$ or $-$.
The $\pm_k$ sign is
$+$ or $-$ as $k = v$ or $u$,
while the $\pm_l$ sign is
$+$ or $-$ as $l = +$ or $-$.
The relations~(\ref{rhomethsicommutation}) lead to the expressions~(\ref{F0})
for the boost weight~0 component $\Fz_0$ of the electromagnetic field.

\section{Comparison to Chandrasekhar \cite{Chandrasekhar:1983} notation}
\label{chandra-app}

Chandrasekhar's null directions $\bl$, $\bn$, $\bmm$, $\bar{\bmm}$,
p.~41 eq.~(283), correspond to
\begin{equation}
\label{chandralnmm}
  \bl = \bgamma_v
  \ , \quad
  \bn = \bgamma_u
  \ , \quad
  \bmm = \bgamma_+
  \ , \quad
  \bar{\bmm} = \bgamma_-
  \ .
\end{equation}
Chandrasekhar's null indices 1,2,3,4 are
\begin{equation}
\label{chandra1234}
  1 = v
  \ , \quad
  2 = u
  \ , \quad
  3 = +
  \ , \quad
  4 = -
  \ .
\end{equation}
Chandrasekhar's spin coefficients $\gamma_{klm}$, p.~37 eq.~(253),
coincide with the Lorentz connections $\Gamma_{klm}$,
\begin{equation}
  \gamma_{klm} = \Gamma_{klm}
  \ .
\end{equation}
Chandrasekhar's electromagnetic fields $\phi_i$ are, p.~51 eq.~(324),
\begin{subequations}
\label{chandraF}
\begin{alignat}{3}
  \phi_0
  &=
  F_{+1}
  &&
  \equiv
  \Fz_{v+}
  \ ,
\\
  \phi_1
  &=
  \Fz_0
  &&
  \equiv
  \tfrac{1}{2} ( F_{vu} - F_{+-} )
  \ ,
\\
  \phi_2
  &=
  \Fz_{-1}
  &&
  \equiv
  F_{u-}
  \ ,
\end{alignat}
\end{subequations}
and his gravitational fields $\Psi_i$ are, p.~43 eq~(294),
\begin{subequations}
\label{chandraC}
\begin{alignat}{3}
  -
  \Psi_0
  &=
  \Cz_{+2}
  &&
  \equiv
  C_{v+v+}
  \ ,
\\
  -
  \Psi_1
  &=
  \Cz_{+1}
  &&
  \equiv
  C_{vuv+}
  =
  C_{-+v+}
  \ ,
\\
  -
  \Psi_2
  &=
  \Cz_{0}
  &&
  \equiv
  C_{v+-u}
  =
  \tfrac{1}{2} ( C_{vuvu} - C_{vu+-} )
  =
  \tfrac{1}{2} ( C_{+-+-} - C_{vu+-} )
  \ ,
\\
  -
  \Psi_3
  &=
  \Cz_{-1}
  &&
  \equiv
  C_{uvu-}
  =
  C_{+-u-}
  \ ,
\\
  -
  \Psi_4
  &=
  \Cz_{-2}
  &&
  \equiv
  C_{u-u-}
  \ .
\end{alignat}
\end{subequations}
%I'LL GUESS SIGN TRACES TO PENROSE
%Chandrasekhar offers no explanation for the $-$ sign.
Chandrasekhar's operators
$\mathscr{D}_n$, $\mathscr{D}^\prime_n$,
$\mathscr{L}_n$, $\mathscr{L}^\prime_n$,
p.~383 eq.~(3),
are related to the raising and lowering operators~(\ref{ethsi})
of the present paper by
\begin{equation}
\label{chandraDL}
  \mathscr{D}_n
  =
  {1 \over R^2 \sqrt{\Delta_x}}
  \,
  \ethSi{-2n}{v}
  \ , \quad
  \mathscr{D}^\prime_n
  =
  - {1 \over R^2 \sqrt{\Delta_x}}
  \,
  \ethSi{2n}{u}
  \ , \quad
  \mathscr{L}_n
  =
  -
  \,
  \ethSi{n}{-}
  \ , \quad
  \mathscr{L}^\prime_n
  =
  \ethSi{-n}{+}
  \ .
\end{equation}

\end{document}